\documentclass[aps, prd, superscriptaddress, twocolumn,a4paper]{revtex4-1}
\usepackage{hyperref}
\usepackage{graphicx}
\usepackage{amsmath}
\usepackage{color}
\newcommand{\be}{\begin{equation}}
\newcommand{\ee}{\end{equation}}
\newcommand{\ba}{\begin{eqnarray}}
\newcommand{\ea}{\end{eqnarray}}
\newcommand{\nn}{\nonumber\\}
\title{A model of the effect of collisions on QCD plasma instabilities}
\begin{document}
\title{Collective excitations of a hot anisotropic QCD medium with Bhatnagar-Gross-Krook collisional kernel within an  effective description}
\author{Avdhesh Kumar}
\email{avdhesh@prl.res.in}
\affiliation{Physical Research Laboratory, Navrangpura, Ahmedabad 380 009, Gujarat, India}
\author{M. Yousuf Jamal}
\email{mohammad.yousuf@iitgn.ac.in}
\author{Vinod Chandra}
\email{vchandra@iitgn.ac.in}
\affiliation{Indian Institute of Technology Gandhinagar,  Gandhinagar-382355, Gujarat, India}
\author{Jitesh R. Bhatt}
\email{jeet@prl.res.in}
\affiliation{Physical Research Laboratory, Navrangpura, Ahmedabad 380 009, Gujarat, India}

\begin{abstract}
Collective modes of an anisotropic hot QCD
medium have been studied within the semi-classical transport theory employing 
Bhatnagar-Gross-Krook (BGK) collisional kernel.  
The modeling of the isotropic  medium is primarily based
on a recent quasi-particle description of hot QCD equation of state
where the medium effects have been encoded in effective gluon and quark/anti-quark 
momentum distributions that posses non-trivial energy dispersions.  The anisotropic 
distribution functions are obtained in a straightforward the way by stretching
or squeezing the isotropic ones along one of the directions. The gluon self-energy 
is computed using these distribution functions in a linearized 
transport equation with  Bhatnagar-Gross-Krook (BGK) collisional kernel. Further, the 
tensor decomposition of gluon self-energy  leads to the structure functions which eventually
controls the dispersion relations and the collective mode structure of the
medium. It has been seen that both the medium effects and collisions induce appreciable
modifications to the collective modes and plasma excitations in the hot QCD medium. 
\end{abstract}
\maketitle

\section{Introduction}
The matter at extreme conditions of temperature/energy density behaves more like a near perfect fluid 
with the  smallest value of the shear viscosity to entropy ratio among almost all the known fluids in nature. This fact is strongly 
supported  by  the experimental observations at RHIC, BNL \cite{expt_rhic}, and LHC, CERN \cite{expt_lhc}.  In particular, 
 the robust collective flow phenomenon and strong jet quenching both at RHIC and LHC indicate towards the strongly coupled nature of 
 this medium which is commonly termed as quark-gluon-plasma (QGP).  Also, theoretical investigations based on various approaches starting from 
kinetic theory to holographic theories also hint towards a tiny value for the 
shear viscosity to entropy density ratio for the  QGP/hot QCD matter~\cite{Ryu, Denicol1}.

Among a few other interesting observations, suppression of quarkonia yields at high transverse momentum
observed in these experiments highlight the plasma aspects of the medium such a color screening~\cite{Chu:1988wh},
landau damping~\cite{Landau:1984} and energy loss~\cite{Koike:1991mf}. In a hot QCD medium/QGP, such
 aspects can be explored  in terms of  gluon polarization tensor or self-energy of the medium that
helps in exploring  the spectrum of the collective excitations of the medium.  The collective excitations
carry crucial information about the  equilibrated QGP and also provide an  information on the temporal
evolution of the non (near)-equilibrated one. In the present manuscript,  a collisional anisotropic 
QGP is considered where the collisions are governed by BGK kernel. The prime reason to consider
the  anisotropy (momentum) is due to the fact that it has been there in all the stages of heavy-ion collisions.
The BGK collision kernel ensures the local number/current conservation. Therefore, it is always
wise to consider BGK over the RTA (relaxation-time -approximation) while incorporating the collision in the theory.

The collective excitations of the hot QCD medium have
been studied by several groups~\cite{Mrowczynski:1993qm, Mrowczynski:1994xv, Mrowczynski:1996vh,Jamal:2017dqs,avdhesh}.
The results are obtained either by employing linearized semi-classical transport theory or Hard-Thermal-Loop (HTL) effective
theory upto one-loop in weak coupling limit. The two approaches reached to the same
results~\cite{dm_rev1,dm_rev2,Mrowczynski:2000ed,Mrowczynski:2004kv}  for the 
gluon self-energy and collective modes. 
Apart from that there have been a few works where the collisional aspects of the QGP have
been included either with RTA~\cite{Akamatsu:2013pjd} or 
BGK collision term~\cite{Bhatnagar:1954,Jiang:2016dkf,Schenke:2006xu}
while considering isotropic as well as anisotropic aspects of the medium.
In either of the approaches, the dispersion equations are obtained from the gluon polarization tensor 
depicting the conditions that reflects the existence of solutions of the homogeneous equation of motion (the Yang-Mills equations).

It is important to note that the hot QCD plasma (in the abelian limit) do possess collective
excitations ~\cite{Weibel:1959zz} that could be seen as a straightforward generalization of hot
QED plasma ( the difference is only in the effective coupling constant). The collective
excitations (gluonic collective modes) that are commonly termed as plasmons have been 
investigated in isotropic/anisotropic hot QCD medium in Refs. ~\cite{Romatschke:2003ms, Romatschke:2004jh, Mrowczynski:2005ki, Arnold:2003rq}.
The distribution functions for partons employed to explore different aspects of QGP are
briefly discussed in~\cite{Attems:2012js, Florkowski:2012as, Dumitru:2007hy, Martinez:2008di, Schenke:2006yp}. 
In most of these studied, QGP has been considered as the ultra-relativistic non-interacting gas of quarks,
anti-quarks, and gluons which is certainly not desired because of the strongly coupled nature of QGP in experiments. 

The present analysis is the extension of the work in Ref.~\cite{Jamal:2017dqs}, where
the collective excitations of collision-less anisotropic plasma for the different equations of state(EoSs) 
have been studied. We shall focus on the effect of collisions by incorporating the BGK collisional kernel. 
It is important to mention that in Ref.~\cite{Schenke:2006xu} the author also have studied the collective modes of anisotropic QGP in presence
of BGK collisional kernel. However, they have analyzed only the those modes which are propagating 
in the direction of anisotropy vector. In their analysis of zeros of the propagator they found two modes of propagation,
out of which one can be unstable. The main purpose of our analysis is to study this situation with more generality.
We allow the modes to propagate in all possible directions with respect to anisotropy vector. Apart from that, we also consider the 
effective  description of the hot QCD medium within the framework of effective fugacity quasi-particle model (EQPM).
In the small anisotropy limit, while considering all the possible directions of propagation with respect to anisotropy vector,
we find that there exists three modes. Out of these modes, two can be unstable. 
We also showed that the effect of lattice and HTL inspired non-ideal EoSs can also significantly change the dispersion characteristic. 
Whenever possible, we have compared our results with those of in Ref.~\cite{Schenke:2006xu}.

The main work  of the present manuscript  includes, (i) detailed analysis of the stable as well as unstable modes and their 
dependencies on wave vector, strength of anisotropy and collisional frequency,  (ii) studying small
anisotropy with full angular dependence,  (iii) the critical dependence of unstable modes on the 
collisional frequency,  the angle between the propagation vector and anisotropy direction, wave 
vector has been investigated by obtaining their maximum allowed values  while fixing the other two respectively for
various values for anisotropy parameter.

The paper is organized as follows. In section ~\ref{GSE}, We shall give a 
brief derivation of gluon self-energy while considering the BGK collisional
kernel. The modeling of hot QCD medium for the isotropic case as 
well as the anisotropic case is shown in the sub-section~\ref{QPD}. 
A simple decomposition of gluon self-energy in terms of structure functions 
and their forms in weak anisotropy limit will be presented in 
the different sub-sections~\ref{CGSE}. 
In section~\ref{PP}, we shall give a brief mathematical structure of the
dispersion relations which we used to 
study different collective modes. Section \ref{RD} contains the discussions of results. 
In section \ref{SC}, we offer the summary and conclusions of the present work as well as the possible future aspects.

\section{Gluon self-energy/polarization tensor in QCD plasma with BGK collisional kernel}
\label{GSE}
Gluon self-energy/polarization tensor ($\Pi^{\mu\nu}$) carries the
information of QCD medium as it describes the interactions term in
the effective action of QCD. We
are interested here in obtaining the expression for $\Pi^{\mu\nu}$ in the presence of collisions.     
To start our calculation, we shall focus on the physics at
soft scale, $ k \sim gT\ll T$, $g$
is the strong coupling constant. At this scale, we can assume the strength of field
fluctuations, $A$ to be O($\sqrt{g} T$), and the
derivatives $\partial_{x}$ of O($g T$). Applying this
power counting scheme we can restrict ourself to the
abelian limit by neglecting the non-abelian term. This
is because in the field strength tensor,   
$F^{\mu\nu}=\partial^{\mu }A^{\nu}-\partial^{\nu }A^{\mu }-i g\left[A^{\mu },A^{\nu }\right]$,
the order of the non-abelian term is O($g^2$) which is smaller than the order 
O($g^{3/2}$) of first two term in $F^{\mu\nu}(x)$.

In the abelian limit, the linearized semi-classical
transport equations, also given in
refs.\cite{Mrowczynski:1993qm}-\cite{Romatschke:2003ms},
 can be written separately for each color
channel\cite{Jiang:2016dkf,Schenke:2006xu} as,
\ba
v^{\mu}\partial_{\mu} \delta f^{i}_a(p,X) + g \theta_{i}
v_{\mu}F^{\mu\nu}_a(X)\partial_{\nu}^{(p)}f^{i}(\mathbf{p})=\mathcal{C}^{i}_a(p,X),\nn
\label{transportequation}
\ea
where, $x^{\mu}=(t,\mathbf{x}) = X$ and $v^{\mu}=(1,\mathbf{v}) = V$, are the four
space-time coordinate and the velocity of the plasma particle, respectively with
$\mathbf{v}=\mathbf{p}/|\mathbf{p}|$.  $\theta_{i}\in\{\theta_g,\theta_q,\theta_{\bar{q}}\}$ and have the values 
$\theta_{g}=\theta_{q}=1$ and $\theta_{\bar{q}}=-1$.   
$\partial_{\mu}$, $\partial_{\nu}^{(p)}$ are the partial
four derivatives corresponding to space and momentum
respectively. $\mathcal{C}^{i}_a(p,X)$ is the collision
term which describes the effects of collisions between
hard particles in a hot QCD medium.
We consider $\mathcal{C}^{i}_a(p,X)$ to be BGK-type  collision
term \cite{Bhatnagar:1954,Jiang:2016dkf,Schenke:2006xu,Carrington:2004}, given as follows, 
\ba
\mathcal{C}^{i}_a(p,X)=-\nu\left[f^{i}_a(p,X)-\frac{N^{i}_a(X)}{N^{i}_{\text{eq}}}f^{i}_{\text{eq}}(|\mathbf{p}|)\right]\,\text{,}\label{collision}
\ea
where, 
\ba f^{i}_a(p,X)=f^{i}(\mathbf{p})+\delta f^{i}_a(p,X),
\ea
are the distribution functions of
quarks, anti-quarks and gluons, $f^{i}(\mathbf{p})$ is
equilibrium part while $\delta f^{i}_a(p,X)$
perturbed part of the distribution function.
The particle number $N^{i}_a(X)$ and its equilibrium value $N^{i}_{\text{eq}}$ are defined as follows,
\ba
\label{particlenumber1}
N^{i}_a(X)=\int \frac{d^{3}p}{(2\pi)^3} f^{i}_a(p,X)\text{ , ~} \\ N^{i}_{\text{eq}} = \int \frac{d^{3}p}{(2\pi)^3} f^{i}_{\text{eq}}(|\mathbf{p}|) = \int \frac{d^{3}p}{(2\pi)^3} f^{i}(\mathbf{p})\text{,}\label{particlenumber2}
\ea
$\nu$ is the collision frequency. The BGK collision term \cite{Bhatnagar:1954} describes
equilibration of the system due to the collisions in a  time proportional to $\nu^{-1}$.
We consider the collision frequency $\nu$ to be independent of momentum and
particle species. Note that if we take the ratio $\frac{N^{i}_a(X)}{N^{i}_{\text{eq}}}$ to be one we can
see that collision term is the same as in the relaxation time approximation (RTA). 
BGK kernel is important in the sense that it can conserve the particle number
instantaneously in contrast to RTA kernel. This implies that,
\ba
\int \frac{d^{3}p}{(2\pi)^3}\mathcal{C}^{i}_a(p,X)=0.
\ea
The induced current is given by
\cite{Mrowczynski:2000ed, Romatschke:2003ms,Jiang:2016dkf,Schenke:2006xu}
\ba
J_{ind,a}^{\mu}&=&g\int\frac{d^{3}p}{(2\pi)^3} V^{\mu}\{2N_c \delta f^{g}_a(p,X)+N_{f}[\delta f^{q}_a(p,X) \nn
&-&\delta f^{\bar{q}}_a(p,X)]\}\label{indcurrent}.
\ea
Now the using eqs.(\ref{particlenumber1}),  (\ref{particlenumber2}) and (\ref{collision}) 
we write the linearized transport equation (\ref{transportequation}) as follows,
\ba
v^{\mu}\partial_{\mu} \delta f^{i}_a(p,X) + g \theta_{i}
v_{\mu}F^{\mu\nu}_a(X)\partial_{\nu}^{(p)}f^{i}(\mathbf{p})=\nn \\ \nu \left(f_{\text{eq}}^i(\left|\mathbf{p}\right|)- f^i(\mathbf{p})\right)-\nu{\delta f}_a^i(p,X)\nn \\+\frac{\nu  f_{\text{eq}}^i(\left| \mathbf{p}\right|) }{N_{\text{eq}}^i}\left(\int \frac{d^3{p'}}{(2\pi )^3}\delta {f}_a^i(p',X)\right)\label{transportequation1}\text{.}
\ea
Now taking the Fourier transform of above equation we can get,
\begin{widetext}
\ba
\delta f^{i}_a(p,K)=\frac{-ig\theta_iv_{\mu}F_a^{\mu\nu}(K)\partial_{\nu}^{(p)}f^{i}(\mathbf{p})+i\nu(f^{i}_{\text{eq}}(|\mathbf{p}|)-f^{i}(\mathbf{p}))+i\nu f^{i}_{\text{eq}}(|\mathbf{p}|)\left( \int\frac{d^{3}p}{(2\pi)^3}\delta f_a^{i}(p^{\prime},K) \right)/N_{\text{eq}}}{\omega-\mathbf{v}\cdot\mathbf{k}+i\nu}\text{,} \label{induceddistribution}
\ea
\end{widetext}
where, $\delta f^{i}(p,K)$ and $F^{\mu\nu}(K)$ are the
Fourier transforms of $\delta f^{i}(p,X)$ and
$F^{\mu\nu}(X)$, respectively. Note that we use
definition of Fourier transform of a function 
$F(X)=\int _{-\infty-i\sigma} ^{\infty+i\sigma}\frac{d\omega }{2\pi}\int \frac{d^3{\mathbf{k}}}{(2\pi)^3}e^{-iK\cdot X}F(K)$.
Where $K = k^{\mu} = (\omega,{\bf k})$
Now taking the Fourier transform of the induced current
and substituting the the value of $\delta f^{g}_a(p,X)$,
$\delta f^{q}_a(p,X)$ and $\delta f^{\bar{q}}(p,X)$ from the
Eq.(\ref{induceddistribution}) one can get the induced
color current to be of the form,
\begin{widetext}
\ba
J^{\mu}_{\text{ind}\,a}(K)&=&g^2 \int\frac{d^{3}p}{(2\pi)^3}v^{\mu}\partial_{\nu}^{(p)}
f(\mathbf{p})\mathcal{M}^{\nu\alpha}(K,V)D^{-1}(K,\mathbf{v},\nu)A_{\alpha{a}}
+  g i \nu \{2 N_c \mathcal{S}^{g}(K,\nu)+N_f (\mathcal{S}^{q}(K,\nu)-\mathcal{S}^{\bar{q}}(K,\nu))\}\nn
&&+ g (i \nu) \int \frac{d\Omega}{4\pi}v^{\mu}D^{-1}(K,\mathbf{v},\nu)
\int\frac{d^{3}p^{\prime}}{(2\pi)^3}\Big[g \partial_{\nu}^{(p^{\prime})}f(\mathbf{p^{\prime}})
\mathcal{M}^{\nu\alpha}(K,V^{\prime})D^{-1}(K,\mathbf{v^{\prime}},\nu)\mathcal{W}^{-1}(K,\nu)
A_{\alpha{a}}\nn
&&+i\nu(f_{\text{eq}}(|\mathbf{p^{\prime}}|)-f(\mathbf{p^{\prime}}))D^{-1}(K,\mathbf{v^{\prime}},\nu)\Big] \mathcal{W}^{-1}(K,\nu)\,\text{,}\label{fullcurrent}
\ea
\end{widetext}
where,
\begin{equation}
f(p)=2N_c f^g(\mathbf{p})+N_f\left[f^q(\mathbf{p})+f^{\bar{q}}(\mathbf{p})\right]\text{,}
\end{equation}
\begin{equation}
f_{\text{eq}}(|\mathbf{p^{\prime}}|)=2N_c f_{\text{eq}}^{g}(|\mathbf{p^{\prime}}|)+N_f\left[f_{\text{eq}}^q(|\mathbf{p^{\prime}}|)+f_{\text{eq}}^{\bar{q}}(|\mathbf{p^{\prime}}|)\right]\,\text{,}
\end{equation}
\begin{equation}
\mathcal{M}^{\nu\alpha}(K,V)=g^{\nu\alpha}(\omega-\mathbf{k}\cdot\mathbf{v})-K^{\nu}v^{\alpha}\,\text{,}
\end{equation}
\begin{equation}
D(K,\mathbf{v},\nu)=\omega+i\nu-\mathbf{k}\cdot\mathbf{v}\,\text{,}
\end{equation}
\begin{equation}
\mathcal{S}^{i}(K,\nu)=-\int\frac{d^{3}p}{(2\pi)^3} v^{\mu}[f^{i}(\mathbf{p})-f^{i}_{\text{eq}}(|\mathbf{p}|)]D^{-1}(K,\mathbf{v},\nu)\,\text{,}
\end{equation}
\begin{equation}
\mathcal{W}(K,\nu)=1-i \nu \int \frac{d\Omega}{4\pi}D^{-1}(K,\mathbf{v},\nu)\,\text{.}
\end{equation}
Using the relation  $\Pi^{\mu\nu}_{ab}(K)=\frac{\delta J^{\mu}_{\text{ind}\,a}(K)}{\delta
    A_{\nu}^b(K)}\,$ one can obtain the polarization tensor as follows,

\ba
\Pi^{\mu\nu}_{ab}(K)&=&\delta_{ab} g^2 \int\frac{d^{3}p}{(2\pi)^3} v^{\mu}
\partial_{\beta}^{(p)}f(\mathbf{p})\mathcal{M}^{\beta\nu}(K,V)\nn
&& \times D^{-1}(K,\mathbf{v},\nu)+\delta_{ab} g^2 (i \nu)\int \frac{d\Omega}{4\pi}v^{\mu}\nn
&&\times D^{-1}(K,\mathbf{v},\nu)\int\frac{d^{3}p^{\prime}}{(2\pi)^3}\partial_{\beta}^{(p^{\prime})}f(\mathbf{p}^{\prime})\nn
&&\mathcal{M}^{\beta\nu}(K,V^{\prime})D^{-1}(K,\mathbf{v}^{\prime},\nu)\mathcal{W}^{-1}(K,\nu).\nn
\label{selfenergy}
\ea

Now the Maxwell's equation can be written as,
\ba
-i k_{\nu}F^{\nu\mu}(K)=J^{\mu}_{ind}(K)+J^{\mu}_{ext}(K). 
\label{Maxwell:1}
\ea
Here $J^{\mu}_{ext}(K)$ is the external current.  The induced current $J^{\mu}_{ind}(K)$ can be expressed in terms 
of self-energy $\Pi^{\mu\nu}(K)$ as follows,
\ba
J^{\mu}_{ind}(K)=\Pi^{\mu\nu}(K)A_{\nu}(K).
\label{linresponse}
\ea
The Eq.( \ref{Maxwell:1}) can also be written as,
\ba
[K^{2}g^{\mu\nu}-k^{\mu}k^{\nu}+\Pi^{\mu\nu}(K)]A_{\nu}(K)=-J^{\mu}_{ext}(K). 
\ea
Now we make a choice of temporal gauge $A_{0}=0$. In
this case we can write the above equation in terms of a
physical electric field as,
\ba
[\Delta^{-1}(K)]^{ij}E^{j}&=&[({k^2}-{\omega}^{2})\delta^{ij}-k^{i}k^{j}+\Pi^{ij}(K)]E^{j}\nn&=&
i\omega{J^{i}_{ext}}(k), 
\label{eqinA0gauge}
\ea
where,
\ba
[\Delta^{-1}(K)]^{ij}=({k^2}-{\omega}^{2})\delta^{ij}-k^{i}k^{j}+\Pi^{ij}(K),
\label{invprop}
\ea
is the inverse of the propagator. The dispersion equations for collective modes can be obtained by finding the 
 poles of propagator $[\Delta(K)]^{ij}$.
Next, we will discuss the Quasi-particle picture of the hot  isotropic medium.

\begin{figure*}    
    \includegraphics[height=5cm,width=5.8cm]{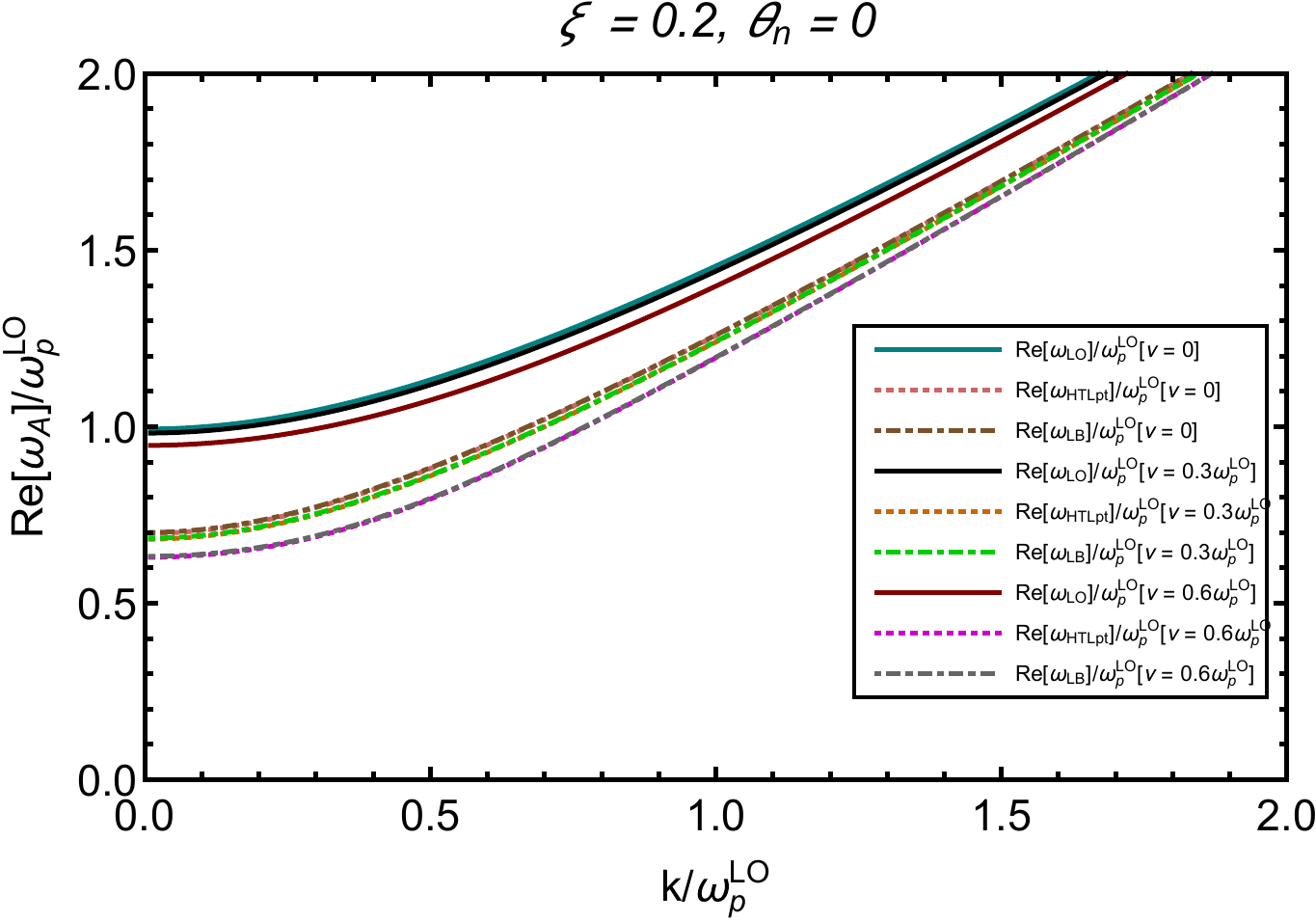}
    \hspace{-1mm}
    \includegraphics[height=5cm,width=5.8cm]{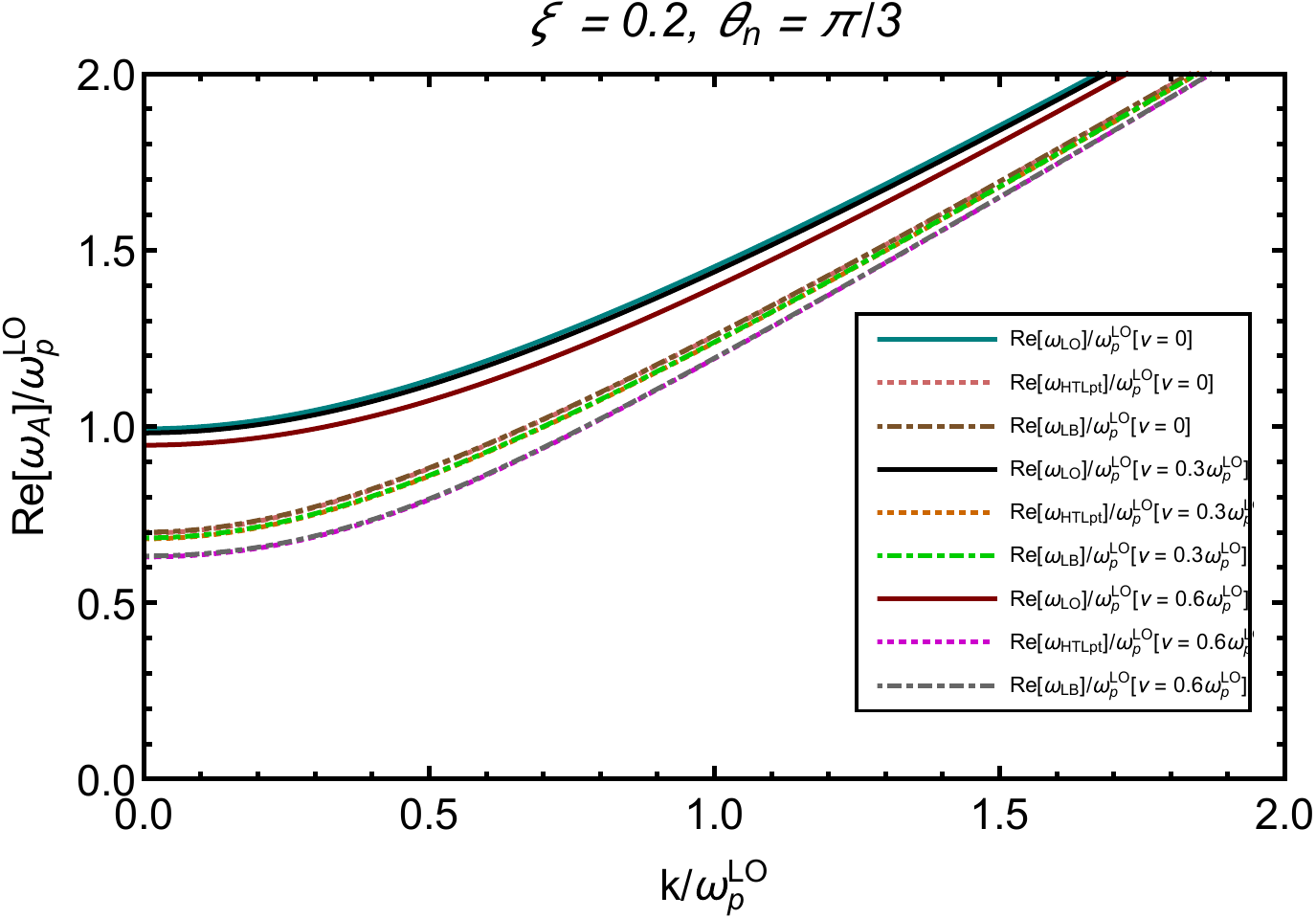}
    \hspace{-1mm}
    \includegraphics[height=5cm,width=5.8cm]{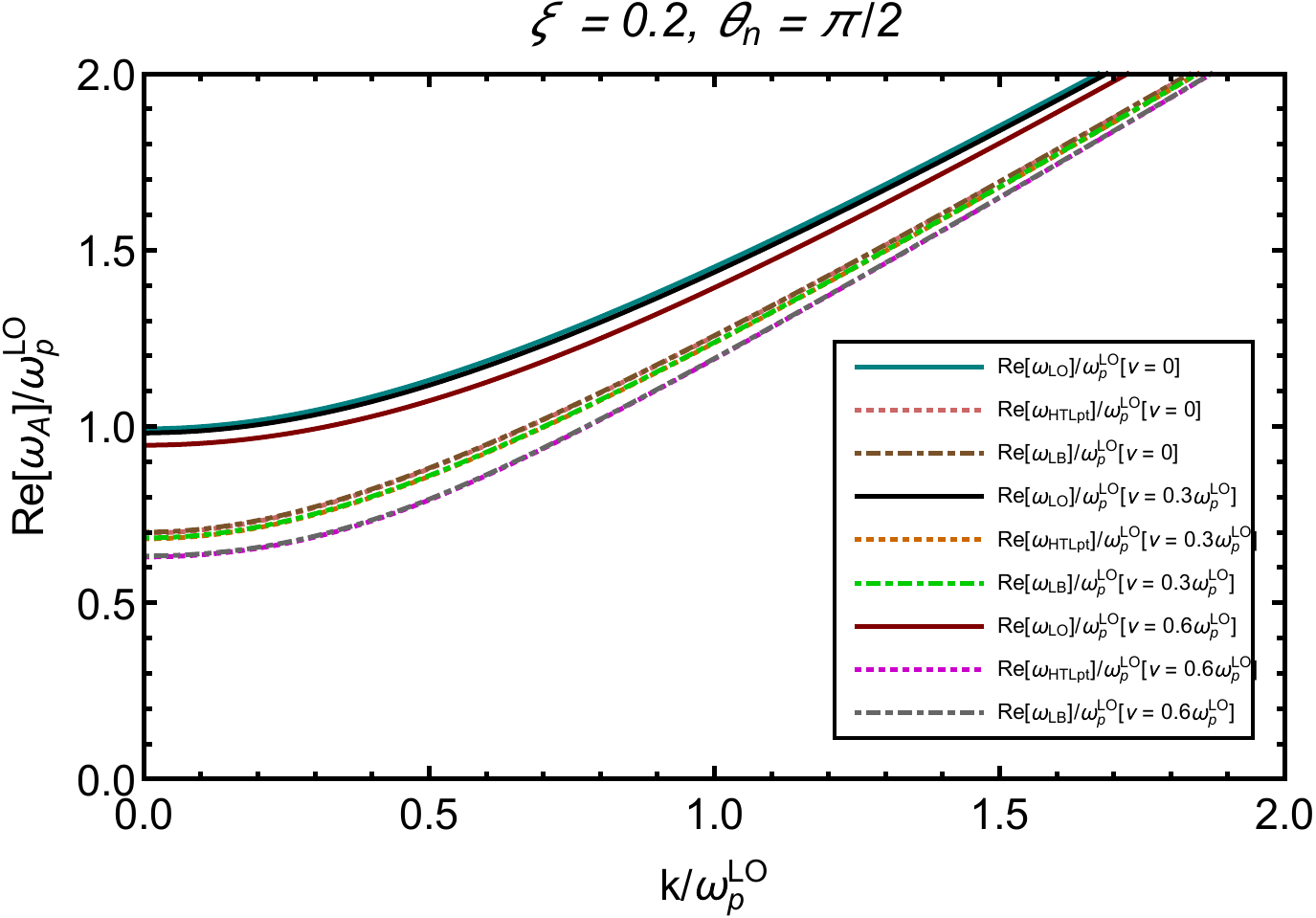}    
    \caption{Real A-mode dispersion curve for various EoSs at $\xi = 0.2$, $T_{c} = 0.17GeV$ and $T = 0.25GeV$ at different $\nu$.}
    \label{fig:Stable_A_modes_Real}
\end{figure*}
\begin{figure*}    
    \includegraphics[height=5cm,width=5.8cm]{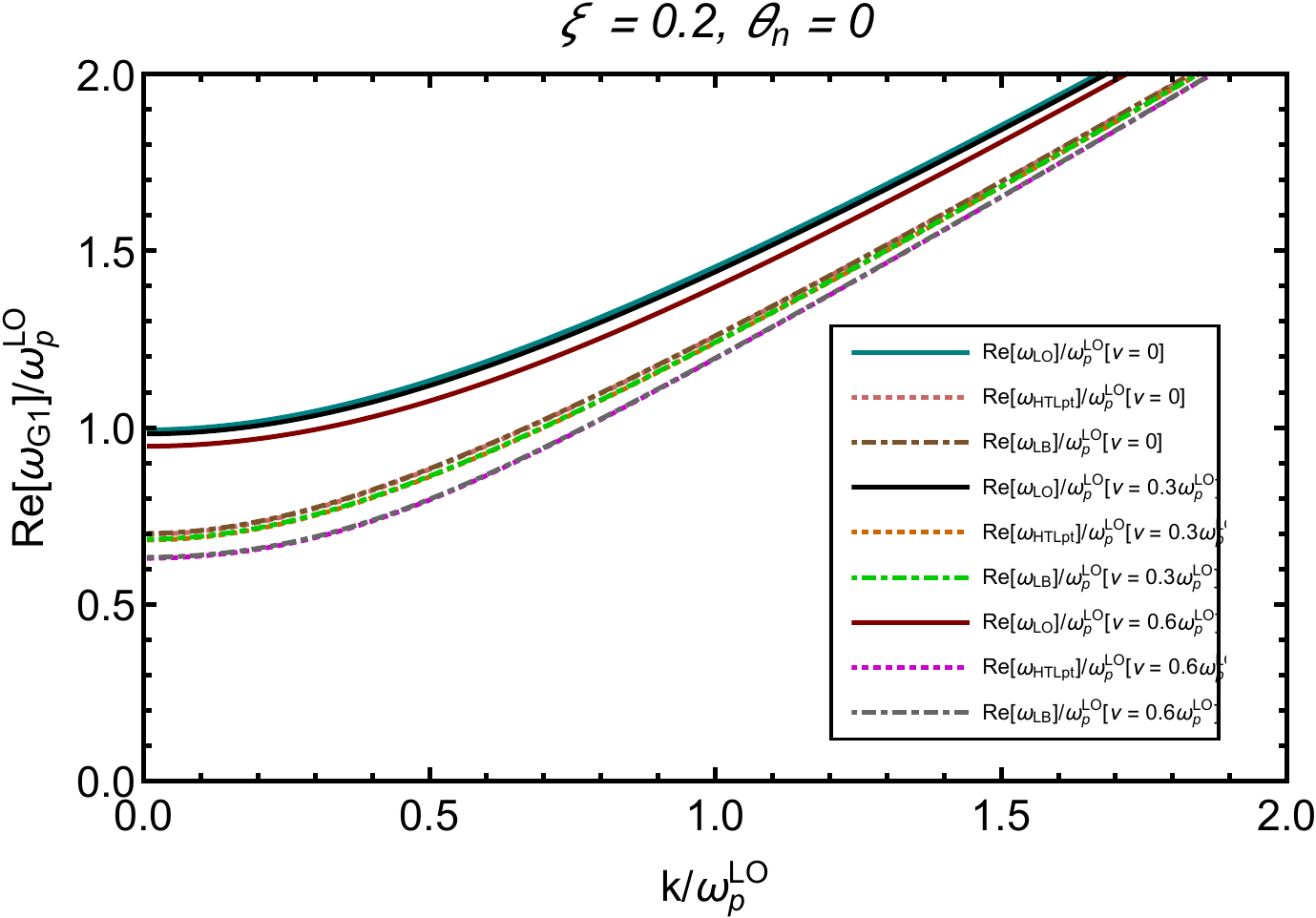}
    \hspace{-1mm}
    \includegraphics[height=5cm,width=5.8cm]{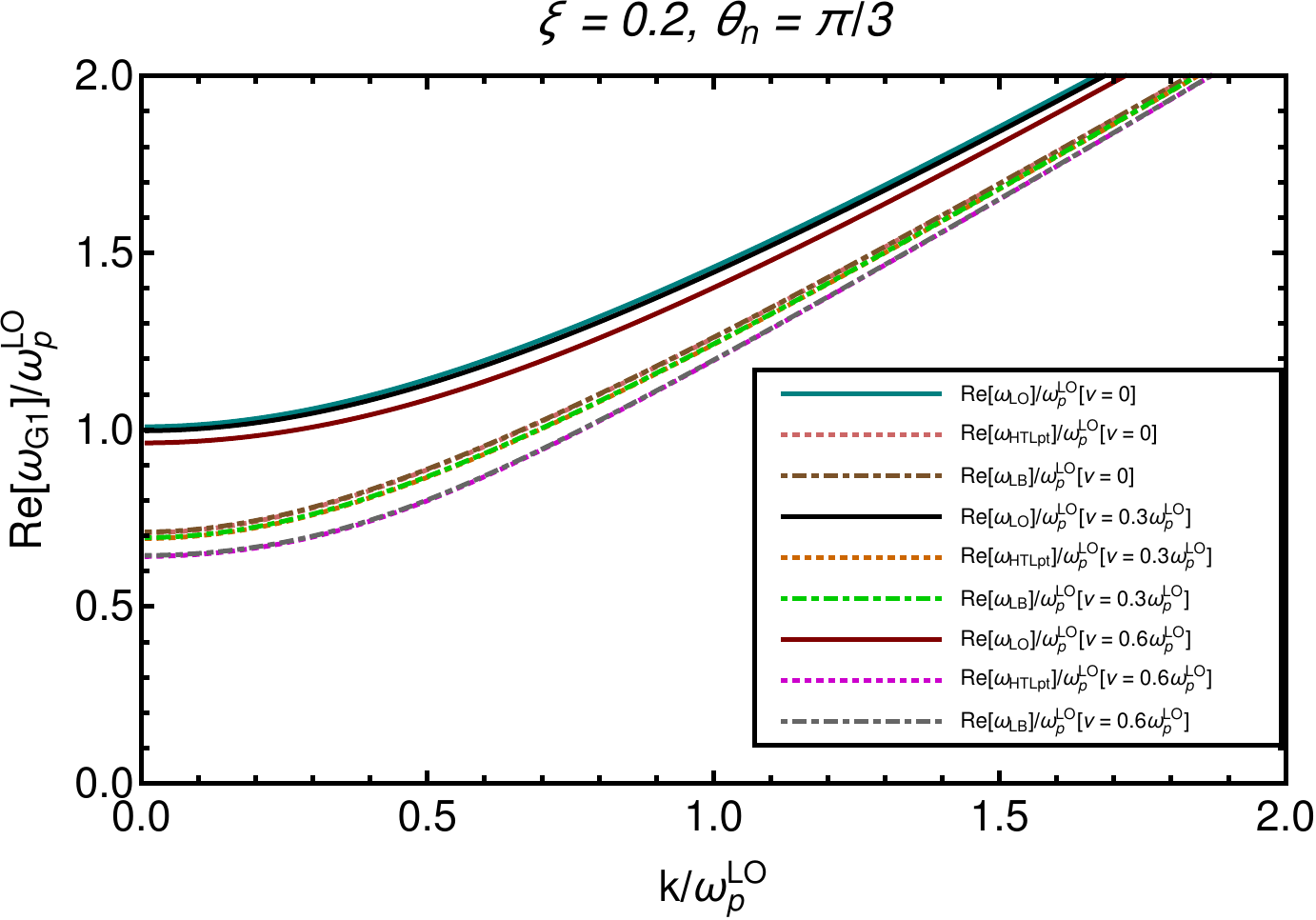}
    \hspace{-1mm}
    \includegraphics[height=5cm,width=5.8cm]{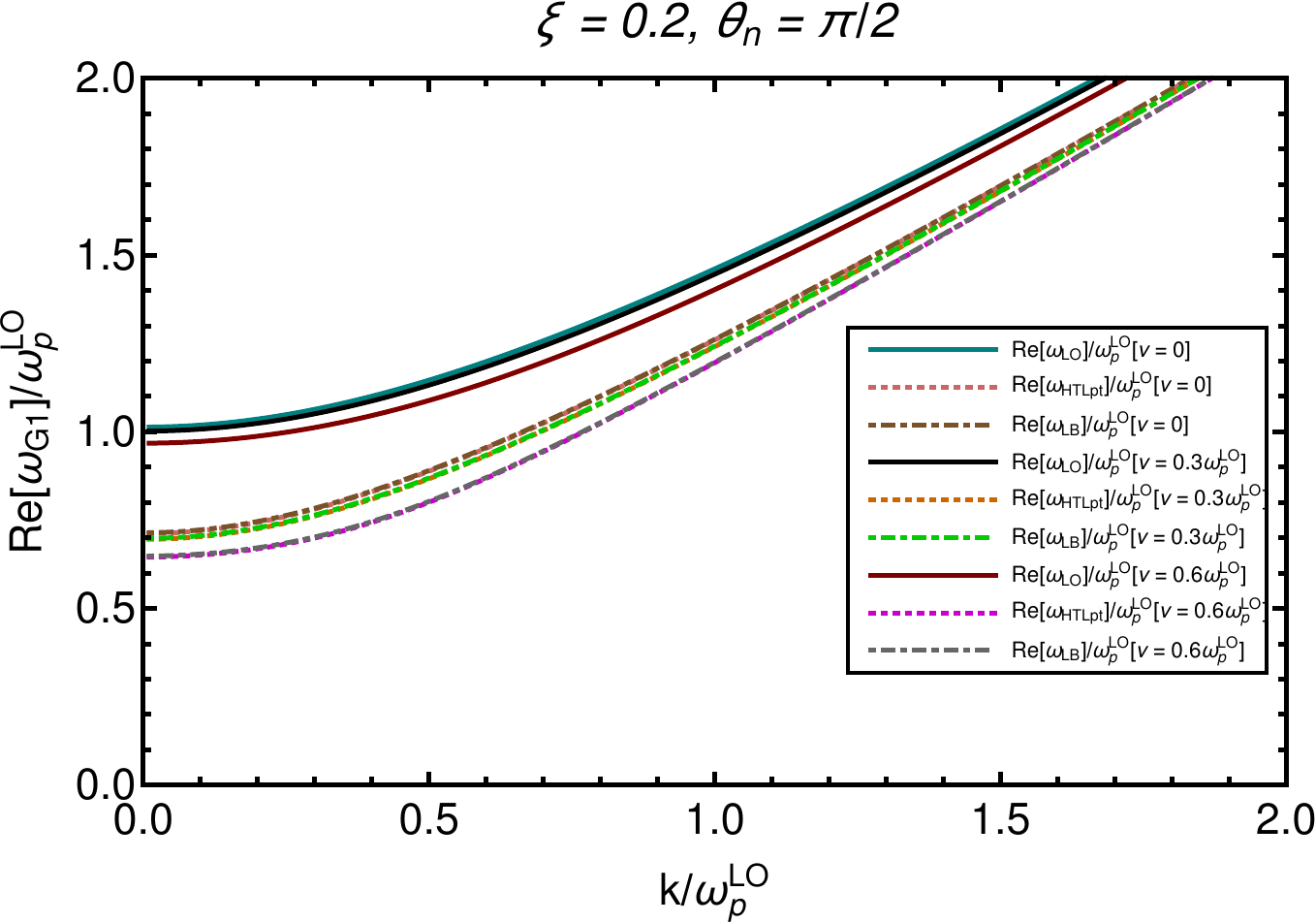}
    \caption{Real G1-mode dispersion curve for various EoSs at $\xi = 0.2$, $T_{c} = 0.17GeV$ and $T = 0.25GeV$ at different $\nu$.}
    \label{fig:Stable_G1_modes_Real}
\end{figure*}
\begin{figure*}    
    \includegraphics[height=5cm,width=5.8cm]{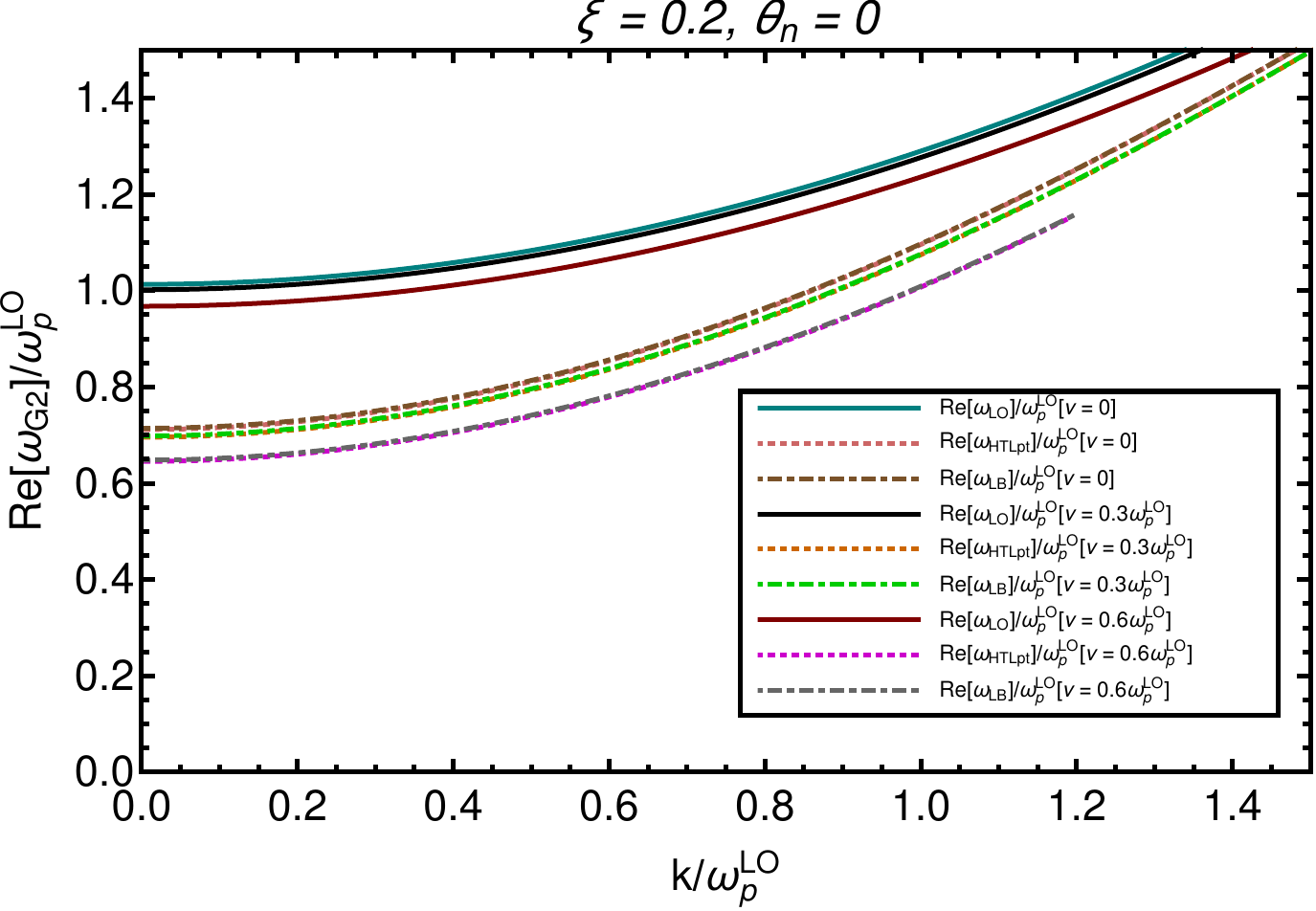}
    \hspace{-1mm}
    \includegraphics[height=5cm,width=5.8cm]{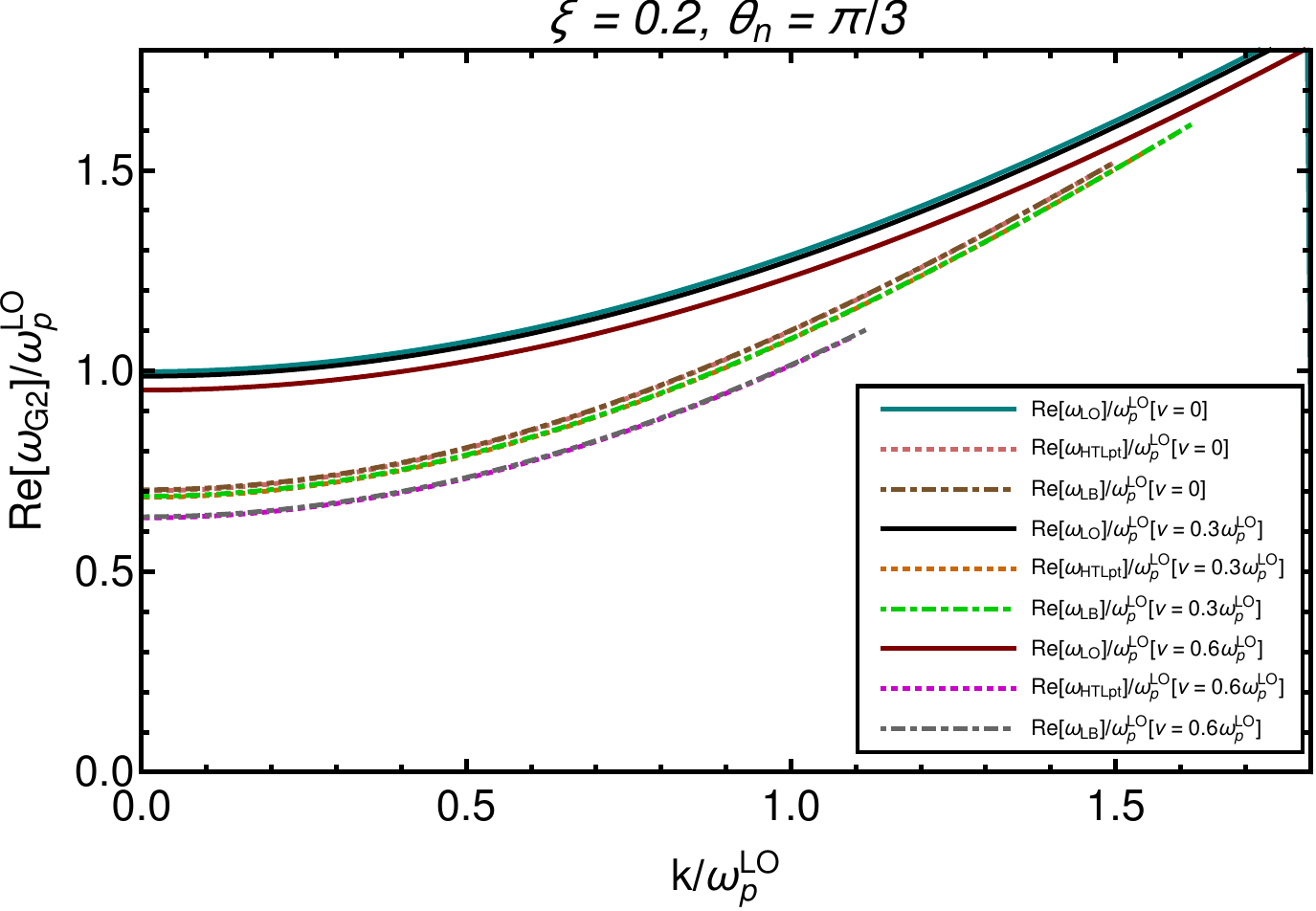}
    \hspace{-1mm}
    \includegraphics[height=5cm,width=5.8cm]{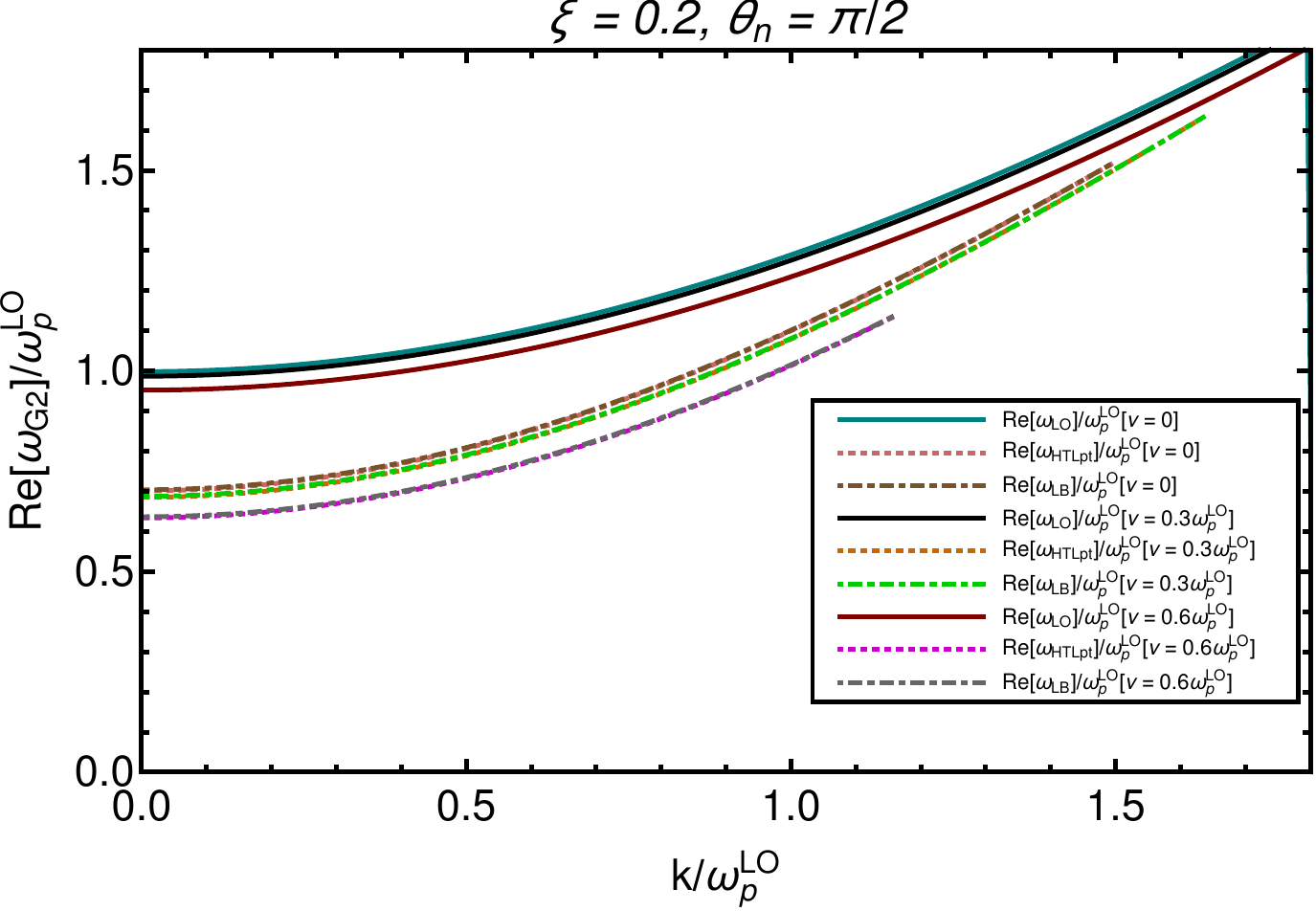}    
    \caption{Real G2-mode dispersion curve for various EoSs at $\xi = 0.2$, $T_{c} = 0.17GeV$ and $T = 0.25GeV$ at different $\nu$.}
    \label{fig:Stable_G2_modes_Real}
\end{figure*}
\begin{figure*}    %
    \includegraphics[height=5cm,width=5.8cm]{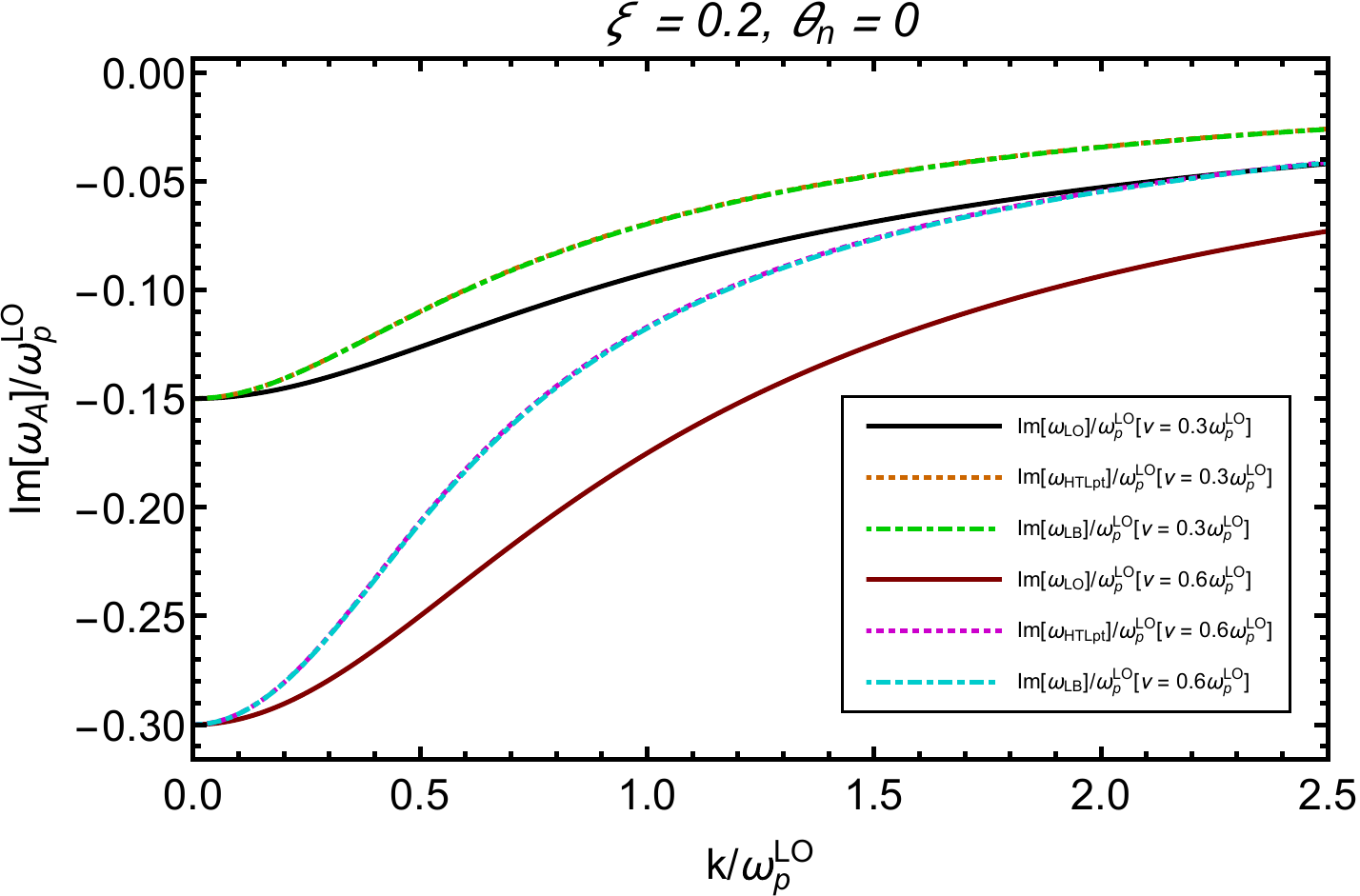}
    \hspace{-1mm}
    \includegraphics[height=5cm,width=5.8cm]{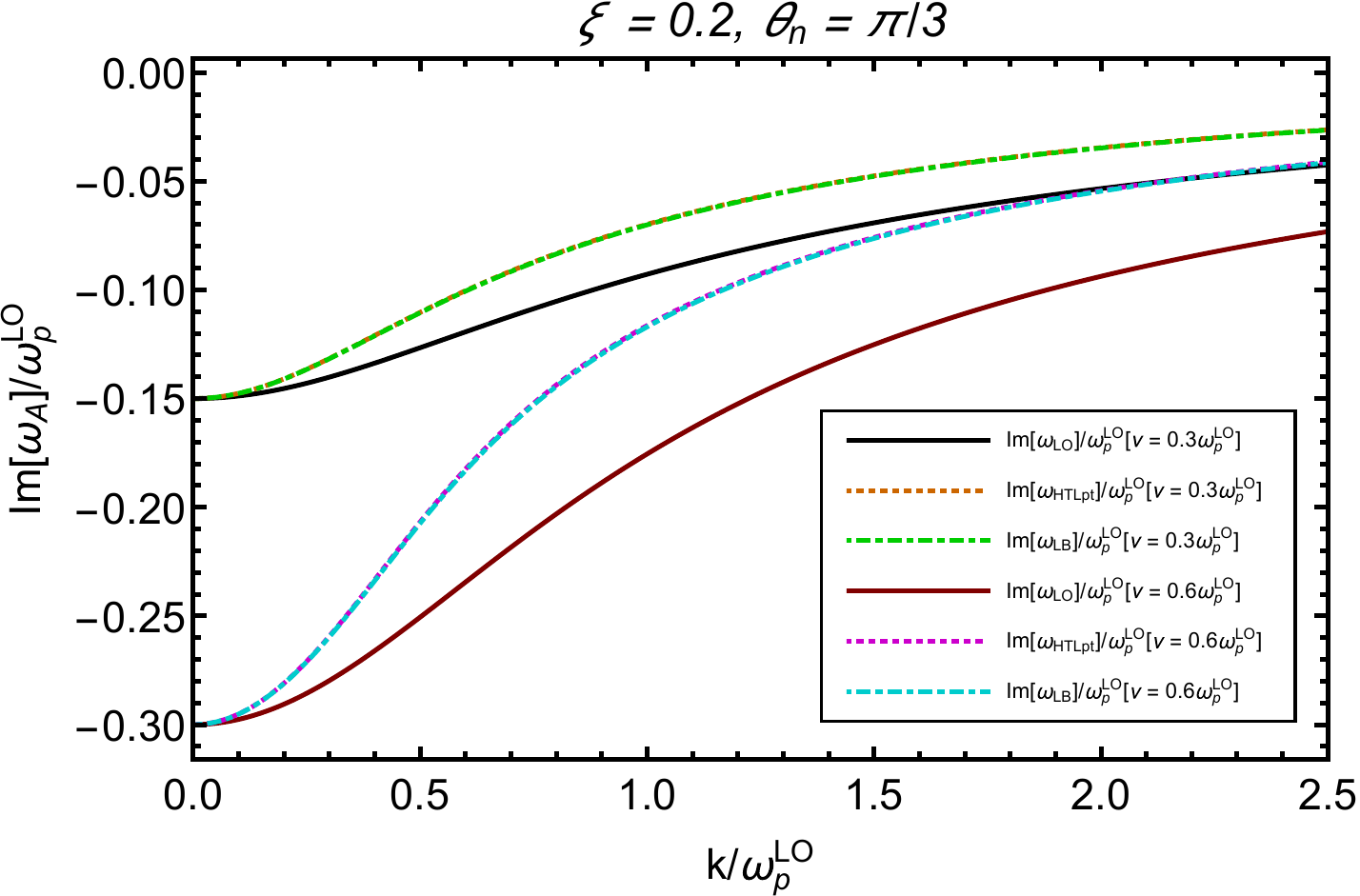}
    \hspace{-1mm}
    \includegraphics[height=5cm,width=5.8cm]{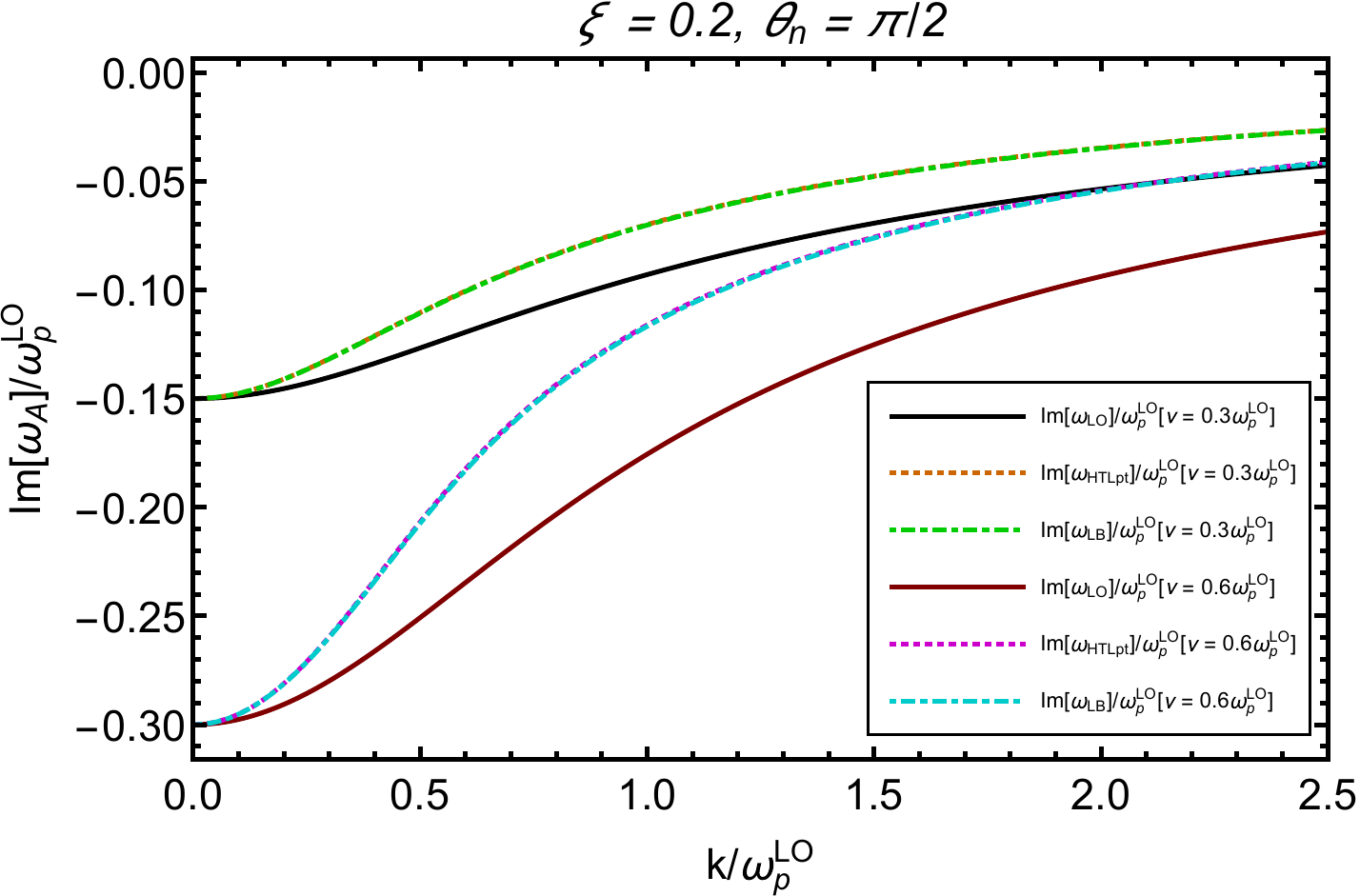}
    \caption{Imaginary A-mode dispersion curves for various EoSs at $\xi = 0.2$, $T_{c} = 0.17GeV$ and $T = 0.25GeV$ at different $\nu$.}
    \label{fig:Stable_A_modes_Imaginary}
\end{figure*}
\begin{figure*}    
    \includegraphics[height=5cm,width=5.8cm]{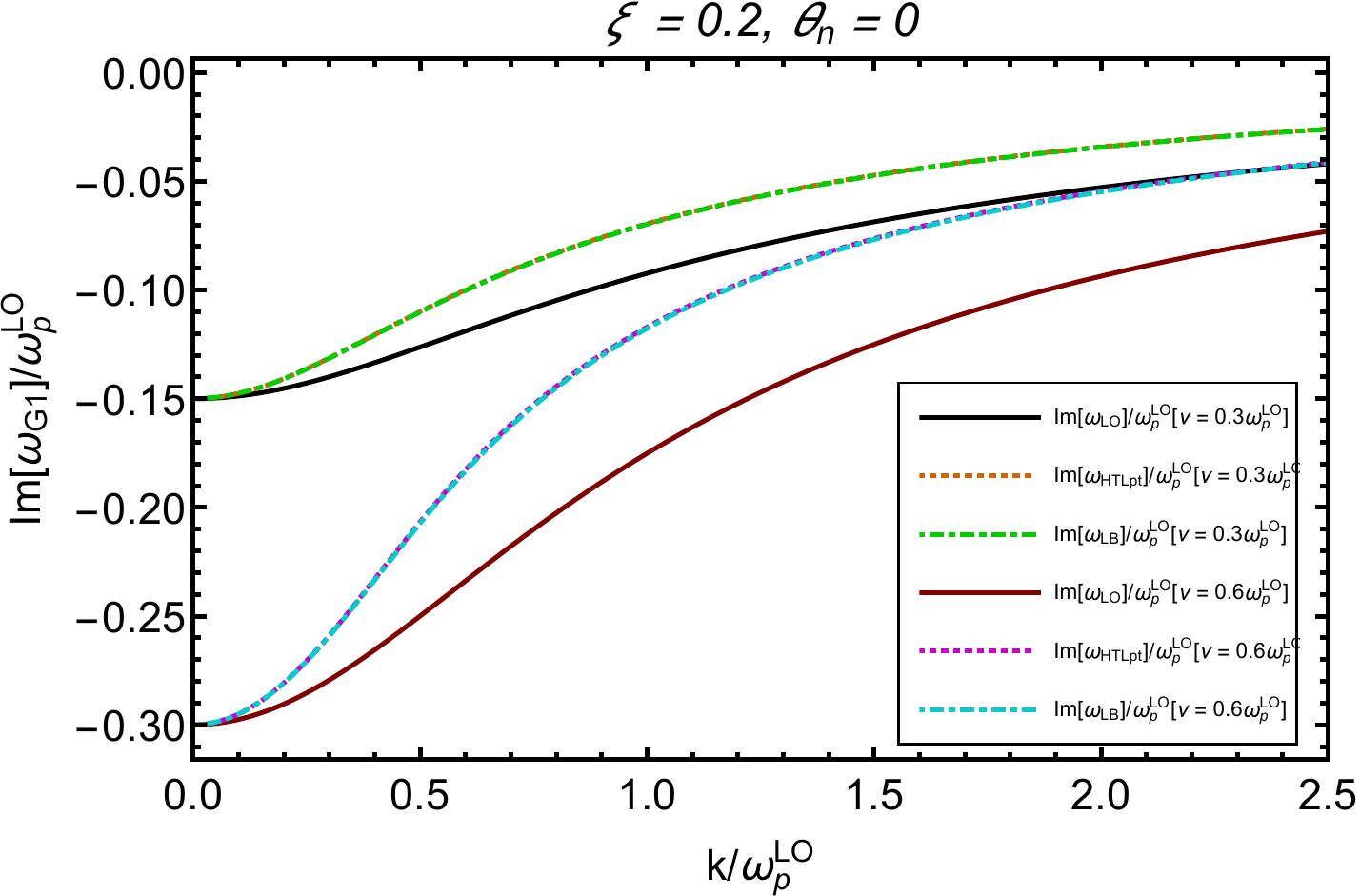}
    \hspace{-1mm}
    \includegraphics[height=5cm,width=5.8cm]{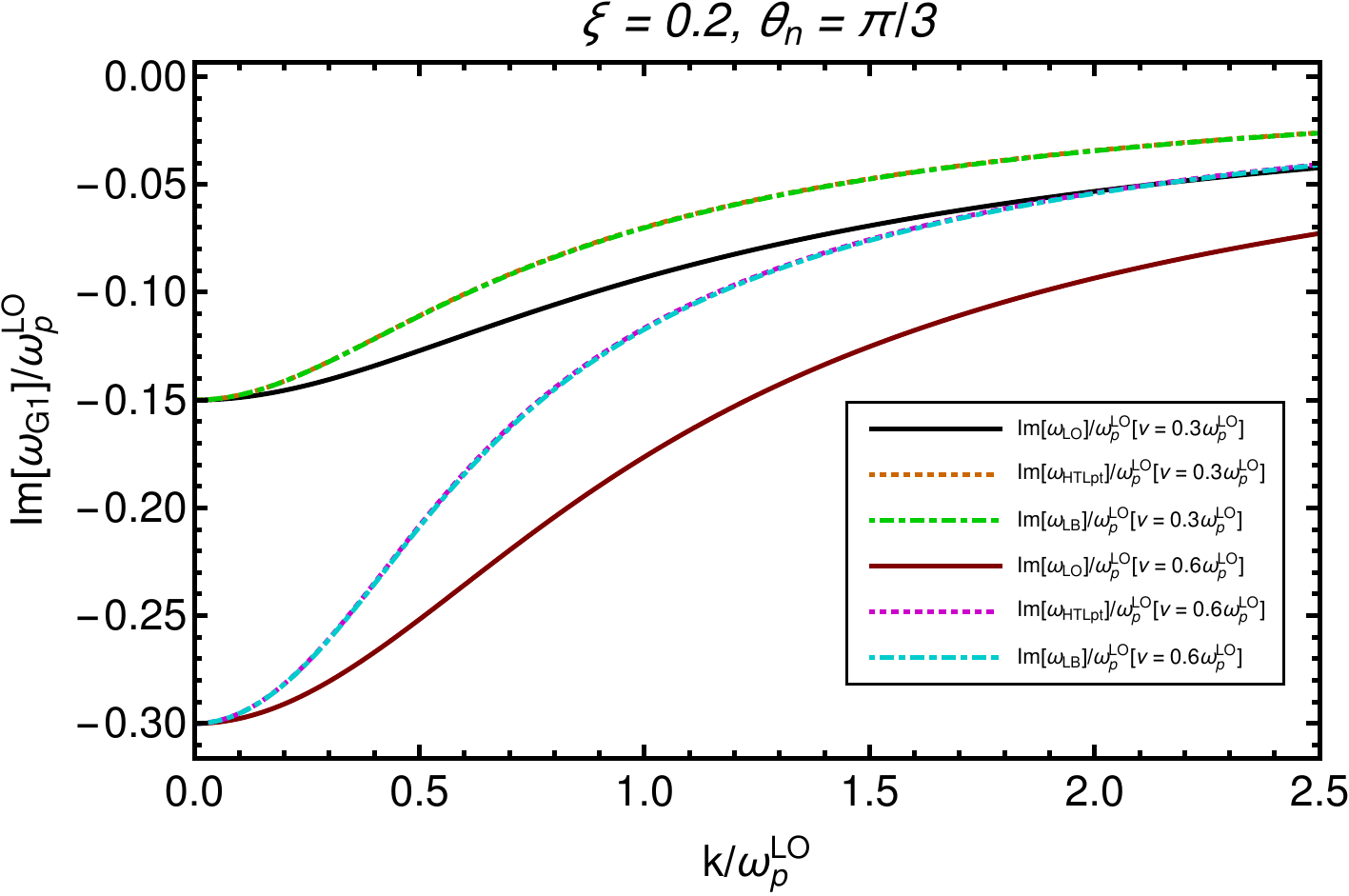}
    \hspace{-1mm}
    \includegraphics[height=5cm,width=5.8cm]{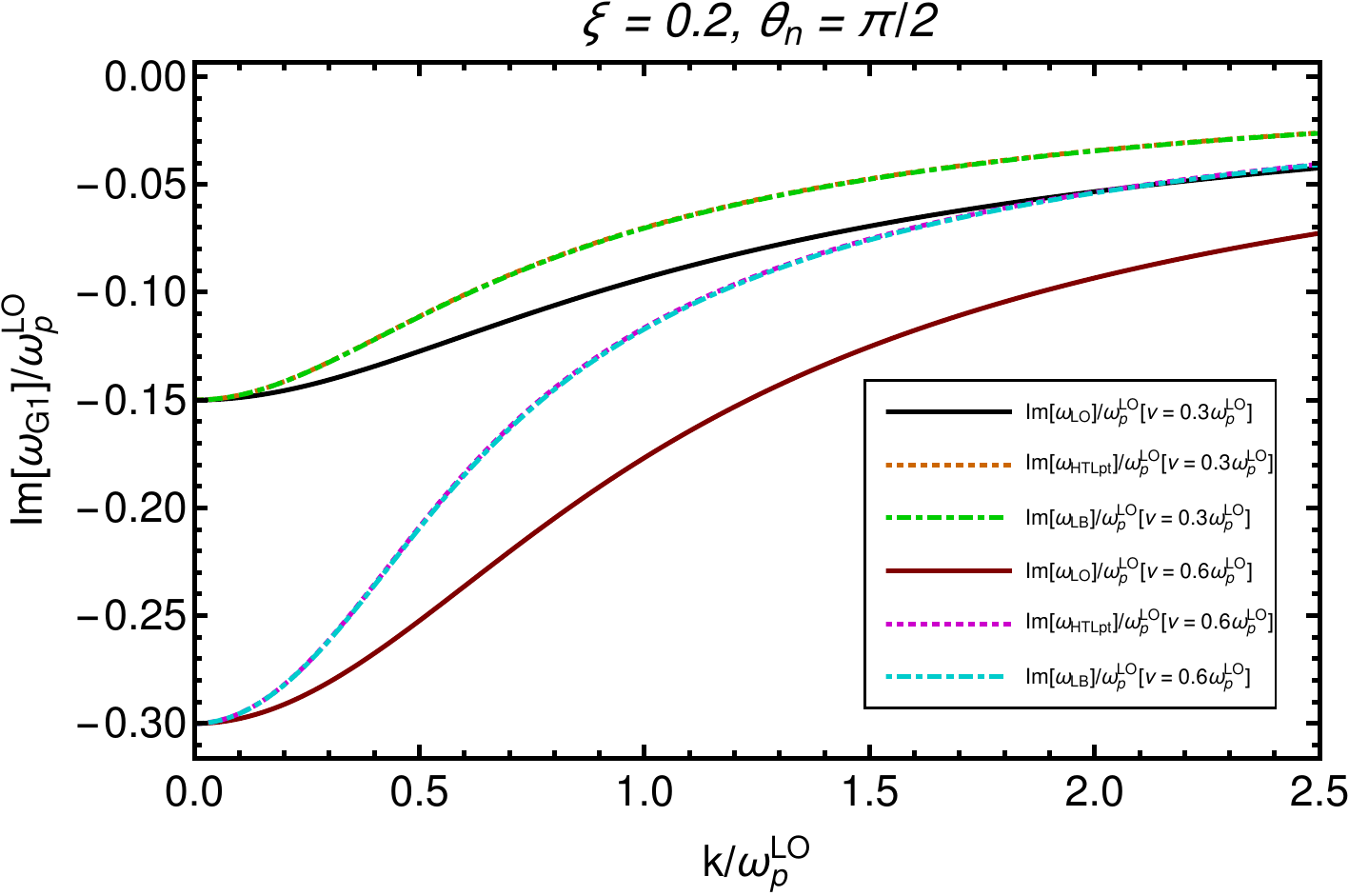}
    \caption{Imaginary G1-mode dispersion curves for various EoSs at $\xi = 0.2$, $T_{c} = 0.17GeV$ and $T = 0.25GeV$ at different $\nu$.}
    \label{fig:Stable_G1_modes_Imaginary}
\end{figure*}
\begin{figure*}    
    \includegraphics[height=5cm,width=5.8cm]{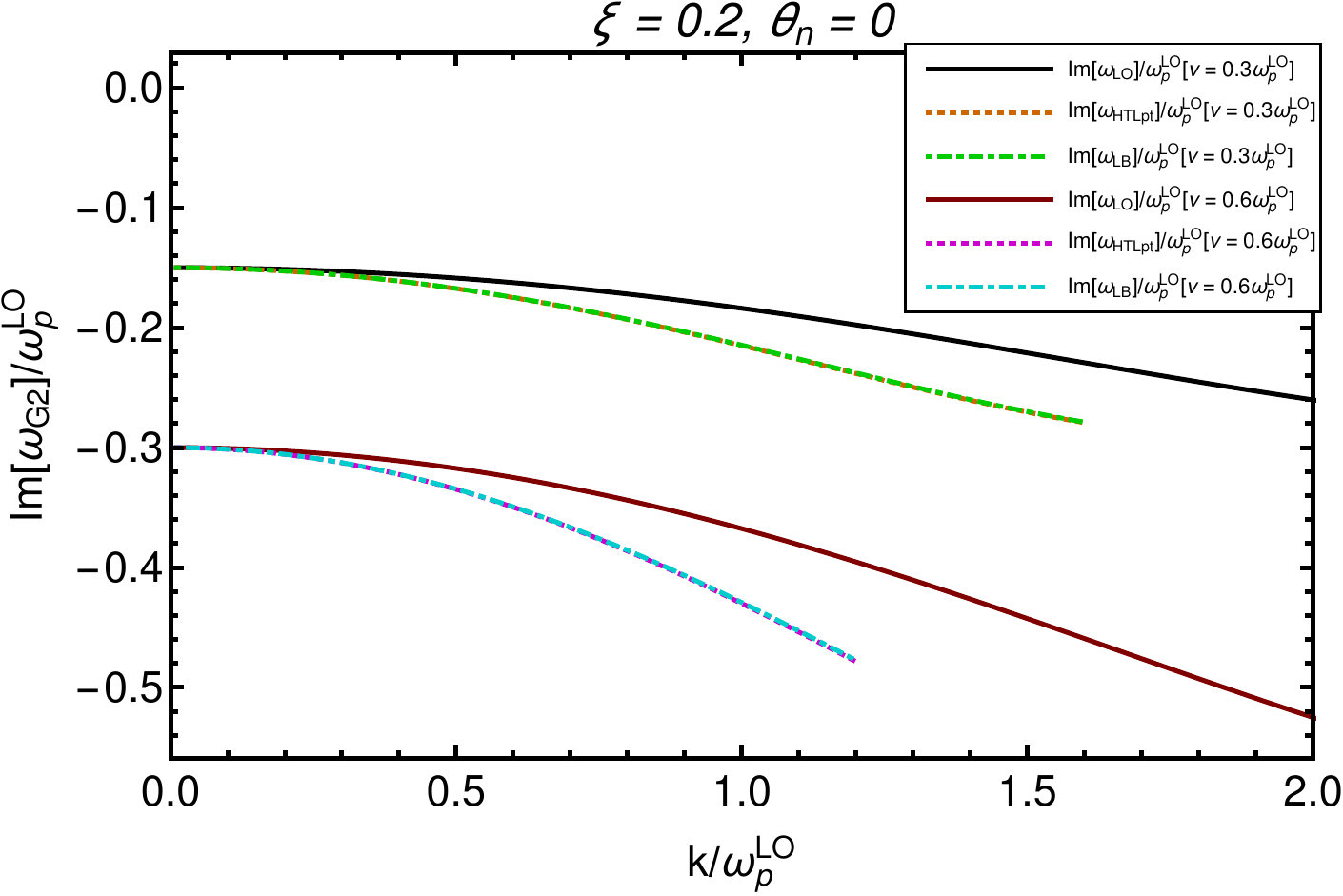}
    \hspace{-1mm}
    \includegraphics[height=5cm,width=5.8cm]{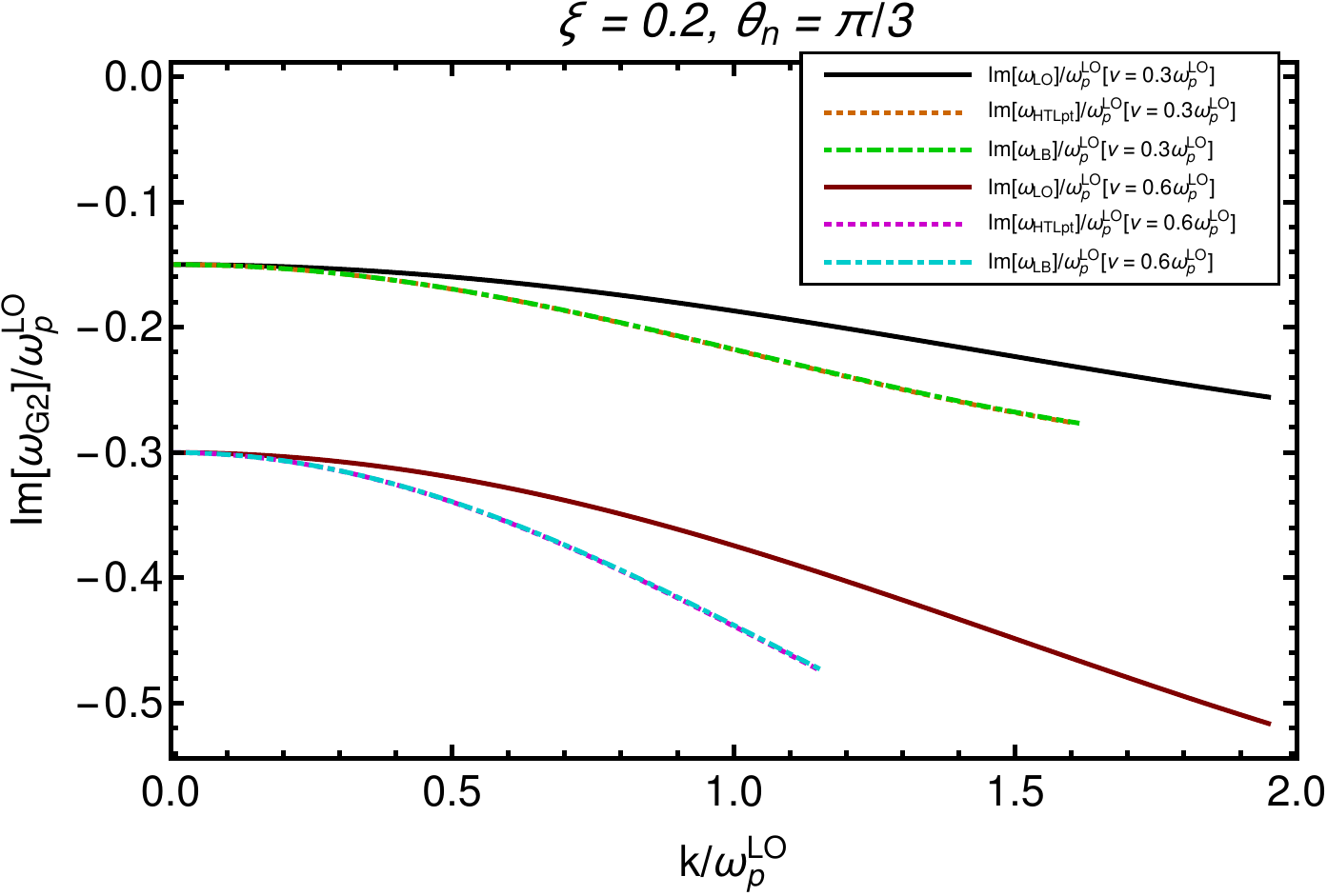}
    \hspace{-1mm}
    \includegraphics[height=5cm,width=5.8cm]{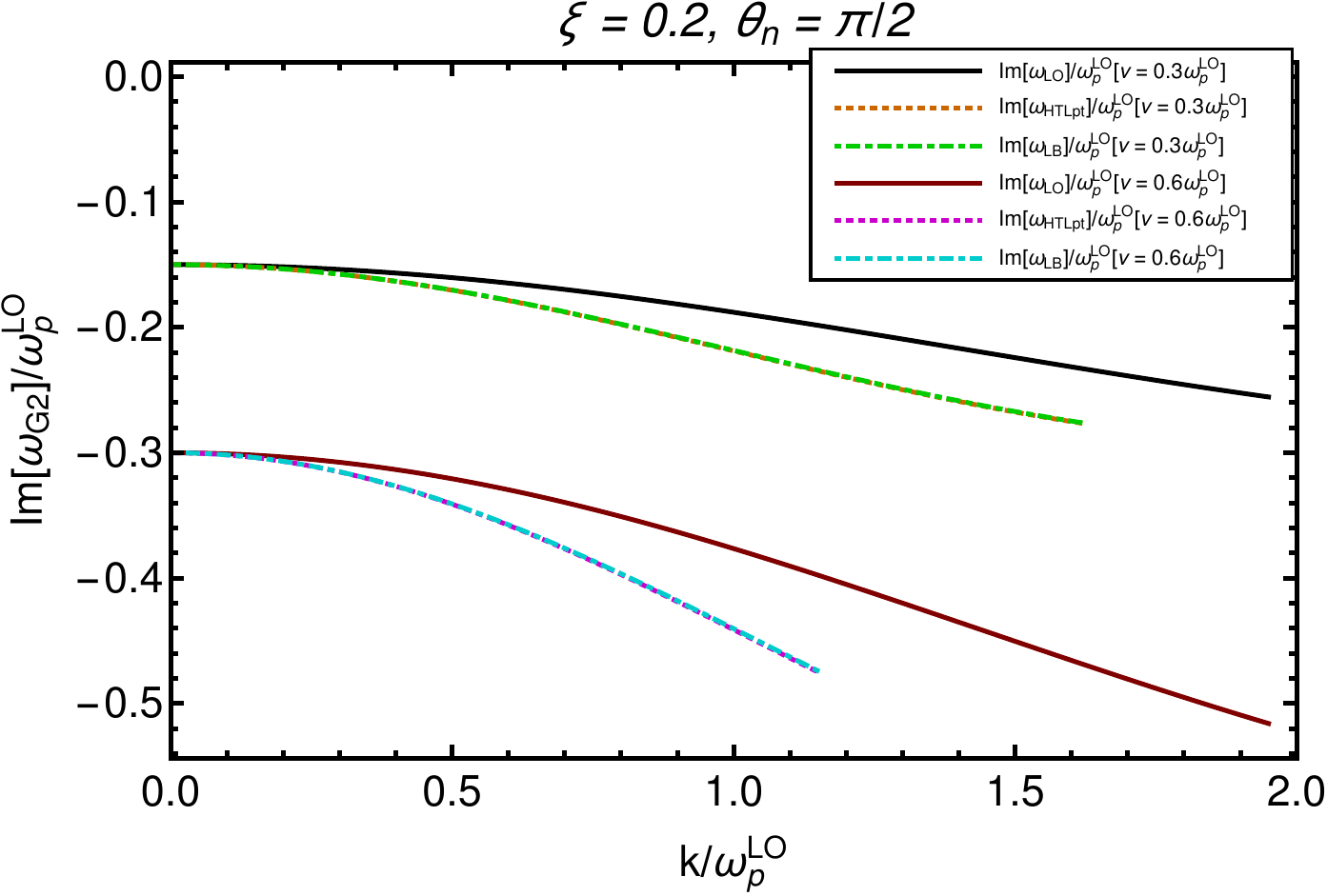}
    \caption{Imaginary G2-mode dispersion curves for various EoSs at $\xi = 0.2$, $T_{c} = 0.17GeV$ and $T = 0.25GeV$ at different $\nu$.}
    \label{fig:Stable_G2_modes_Imaginary}
\end{figure*}
\begin{figure*}    
    \includegraphics[height=5cm,width=5.8cm]{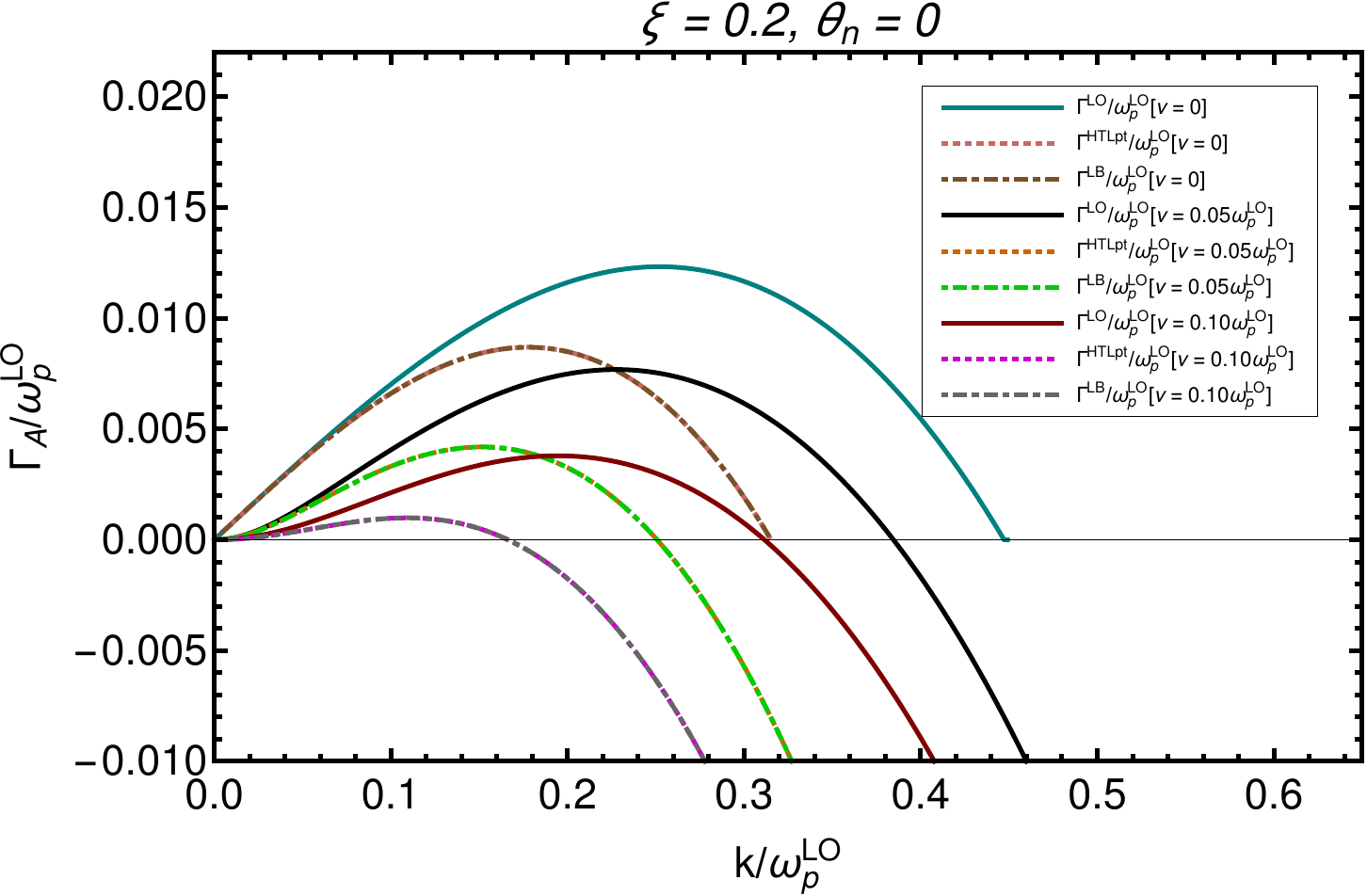}
    \hspace{-1mm}
    \includegraphics[height=5cm,width=5.8cm]{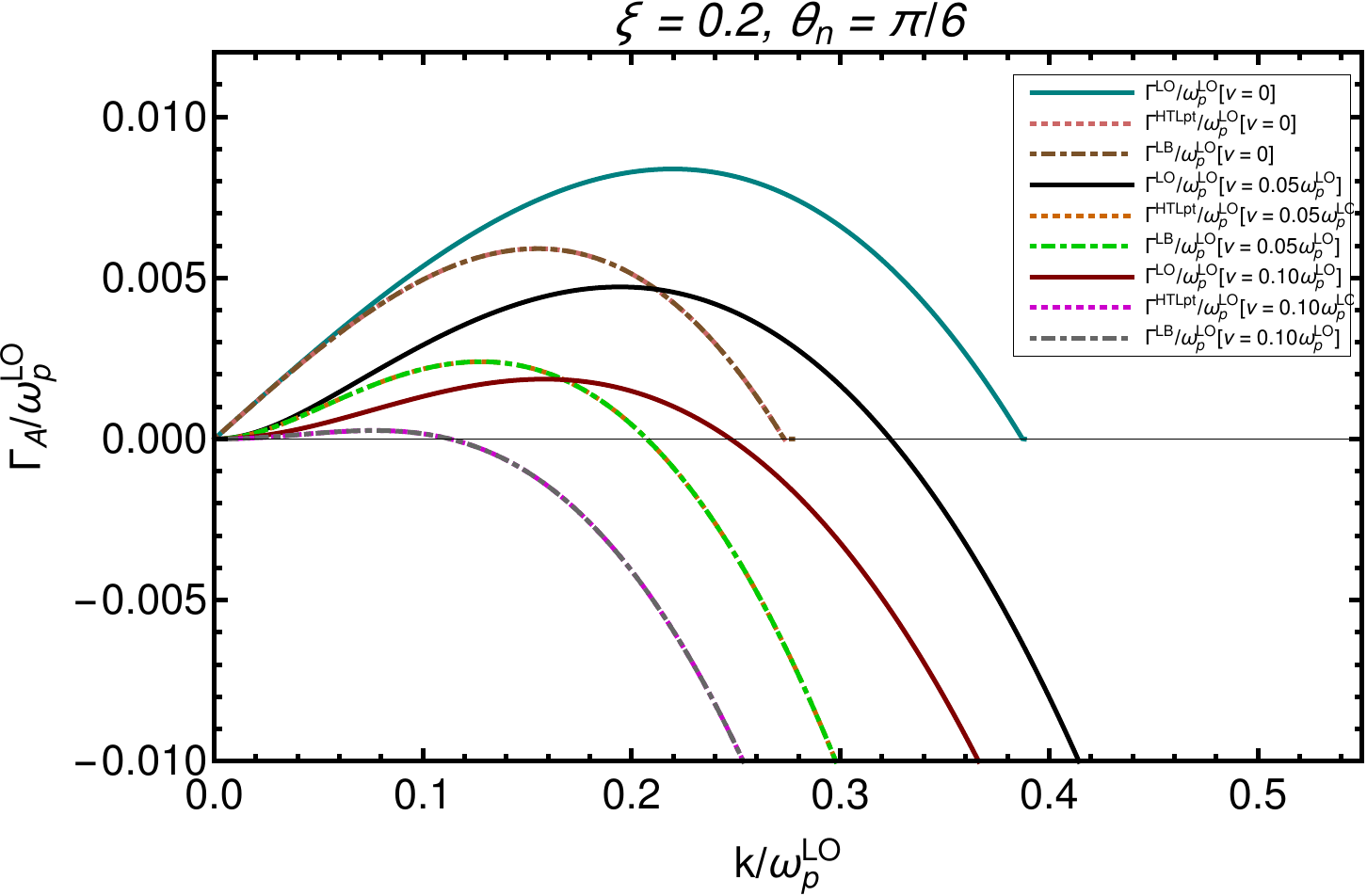}
    \vspace{-1mm}
    \includegraphics[height=5cm,width=5.8cm]{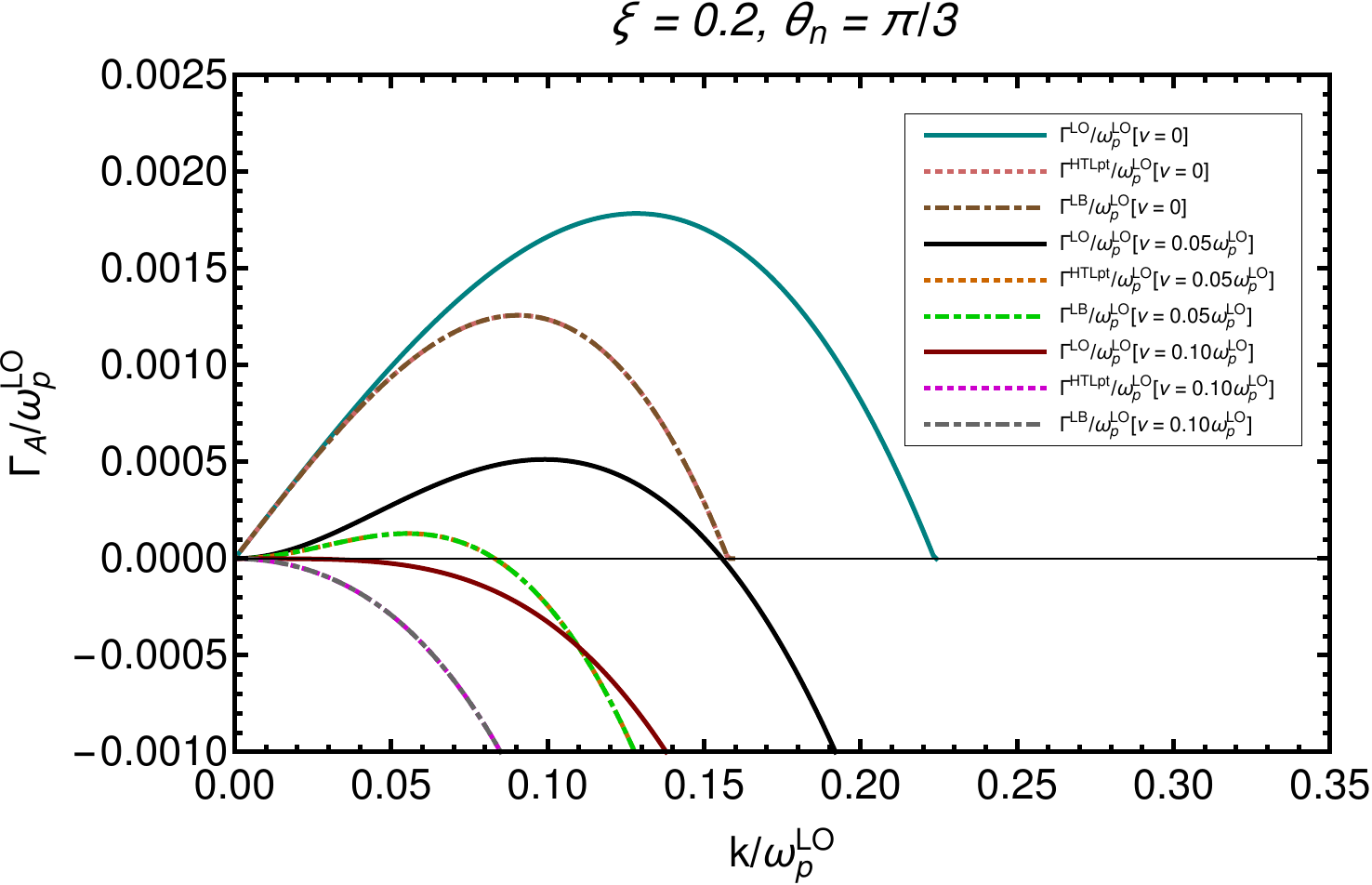}
     \caption{Dispersion curve of unstable A-mode for various EoSs at $\xi = 0.2$, $T_{c} = 0.17GeV$ and $T = 0.25GeV$ with different $\nu$.}
    \label{fig:Unstable_A_mode}
\end{figure*}
 \begin{figure*}    
    \includegraphics[height=5cm,width=8.6cm]{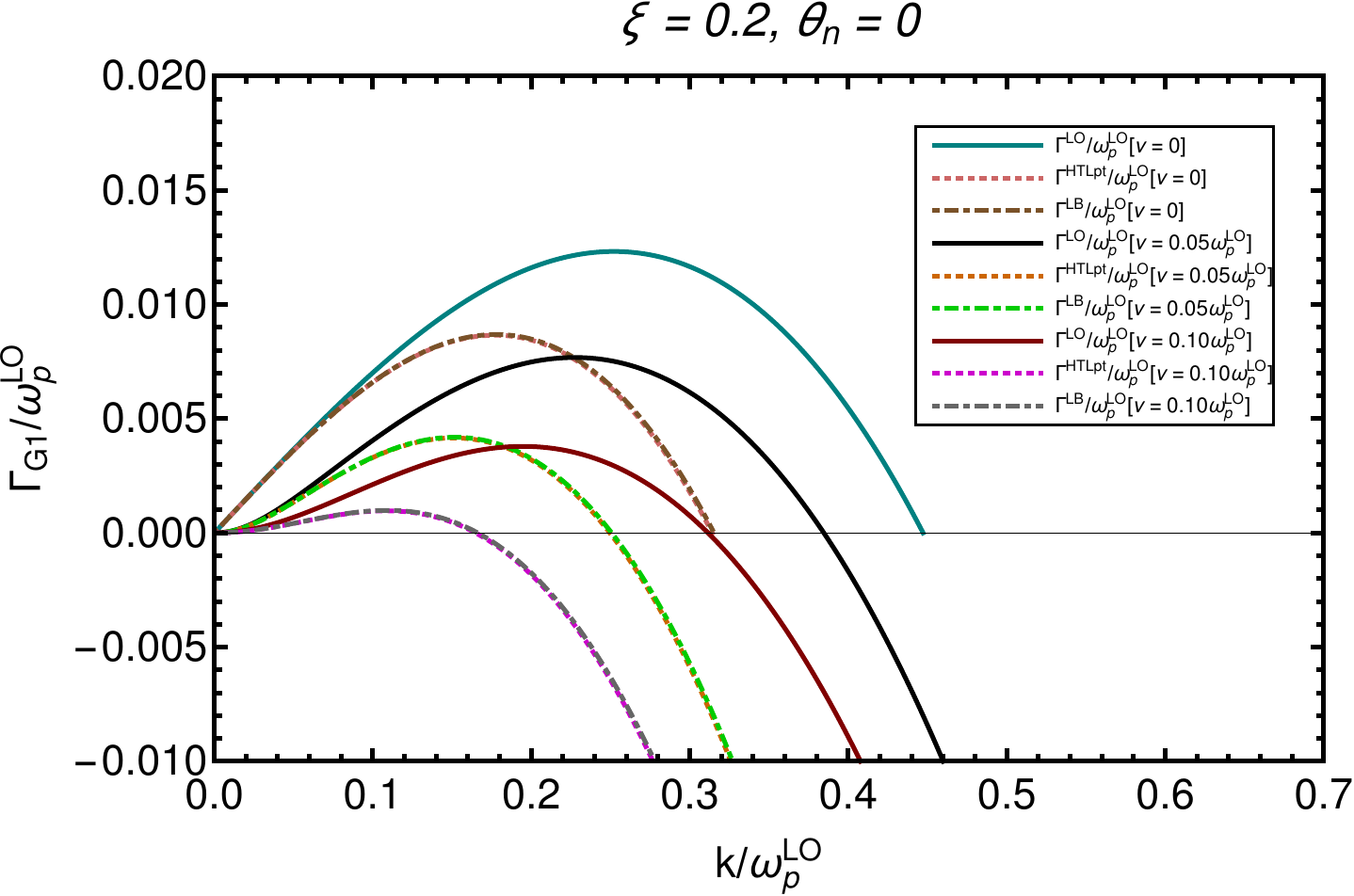}
    \hspace{-1mm}
    \includegraphics[height=5cm,width=8.6cm]{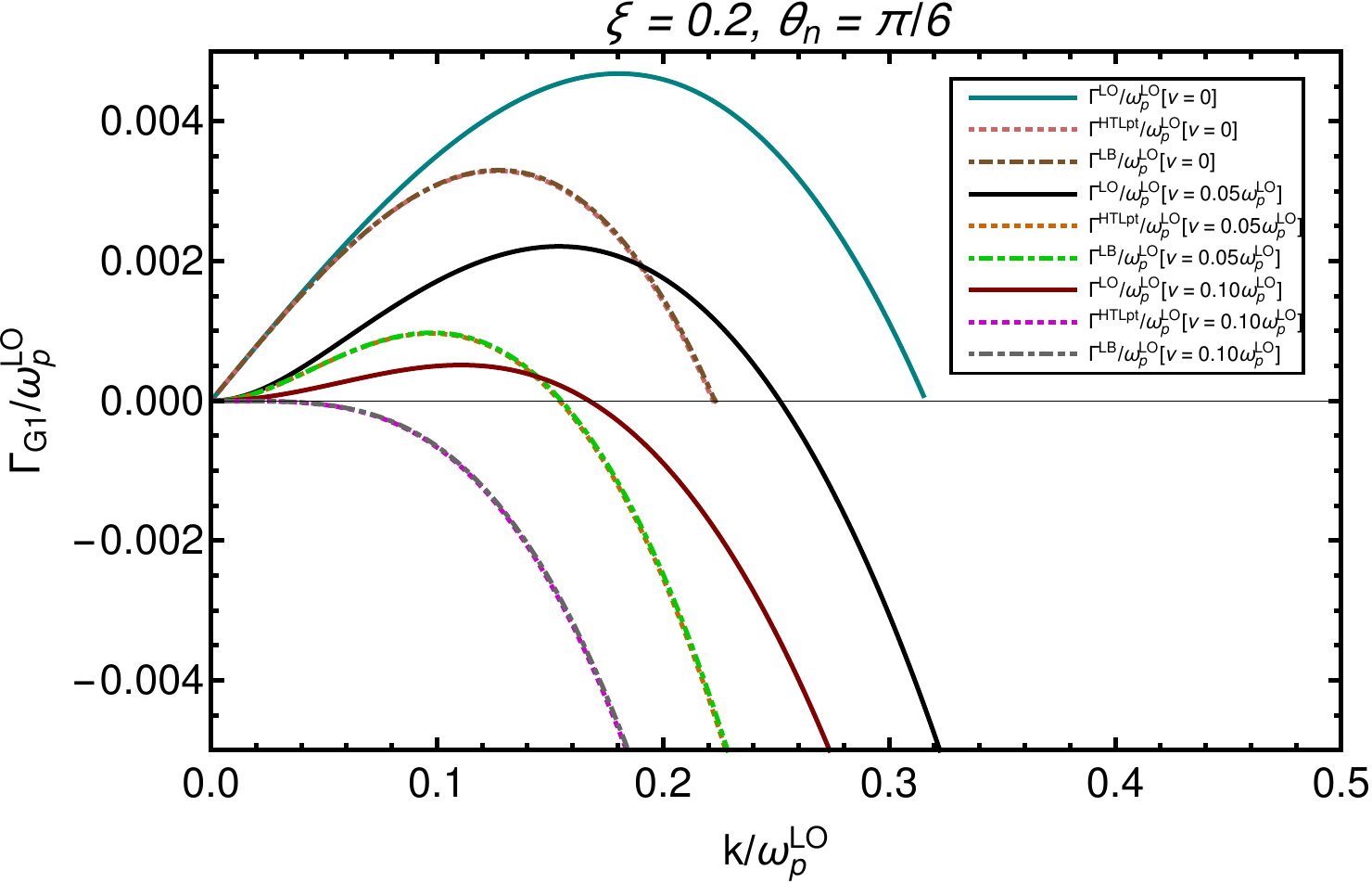}
    \caption{Dispersion curve of G1-mode for various EoSs at $\xi = 0.2$, $T_{c} = 0.17GeV$ and $T = 0.25GeV$ with different $\nu$.}
    \label{fig:Unstable_G1_mode}
\end{figure*}
\begin{figure*}    
    \includegraphics[height=5cm,width=5.8cm]{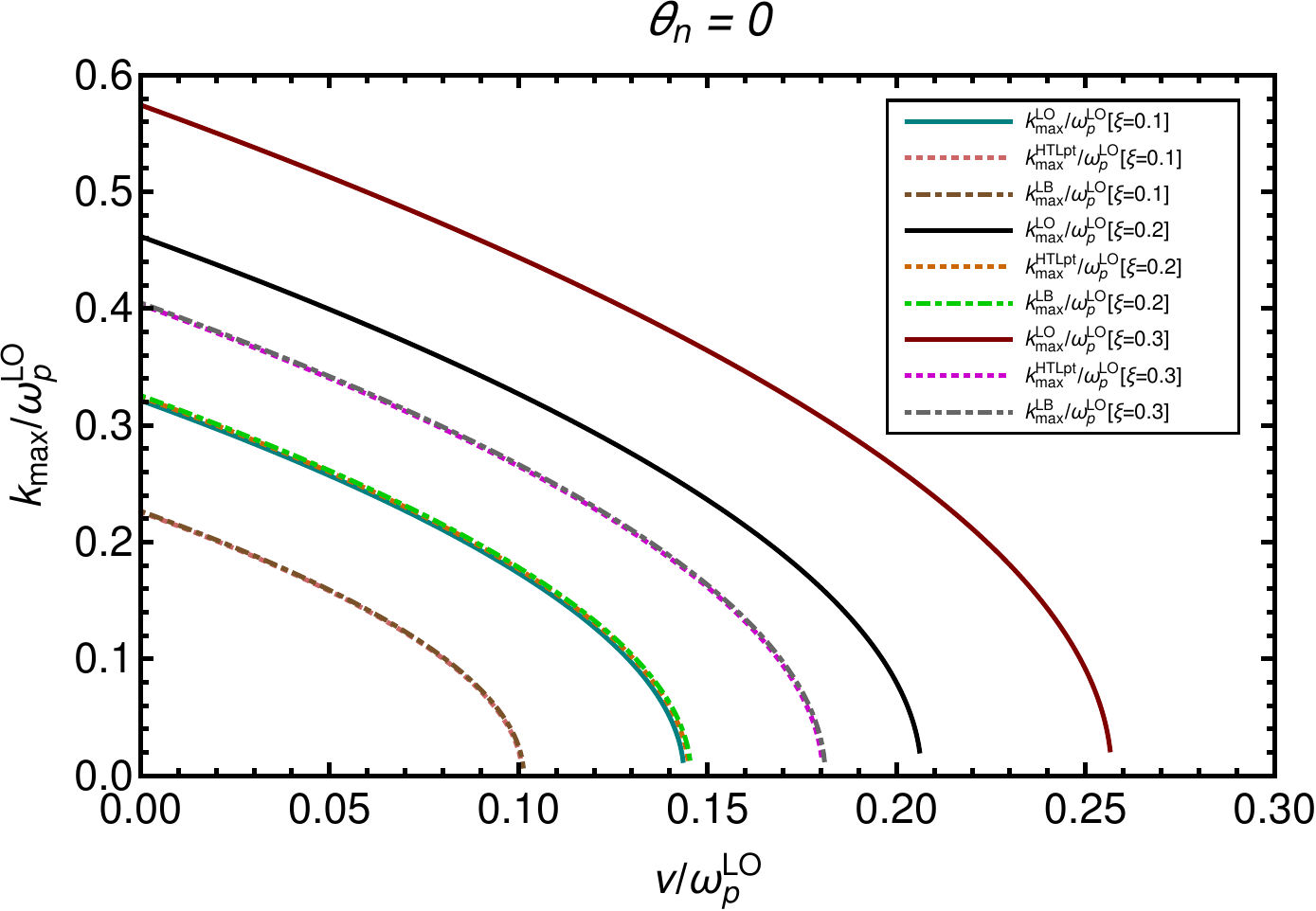}
    \hspace{-1mm}
    \includegraphics[height=5cm,width=5.8cm]{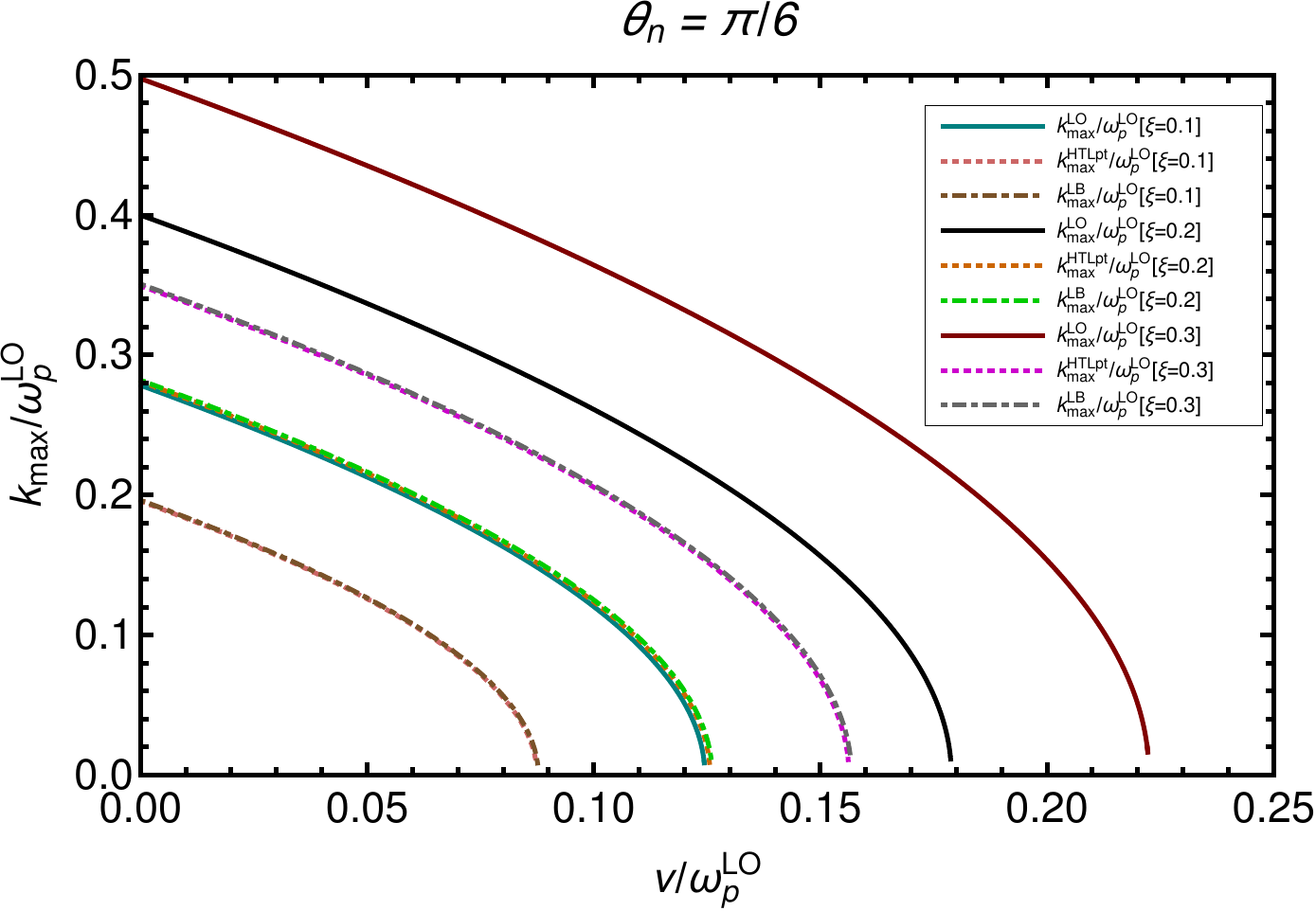}
    \hspace{-1mm}
    \includegraphics[height=5cm,width=5.8cm]{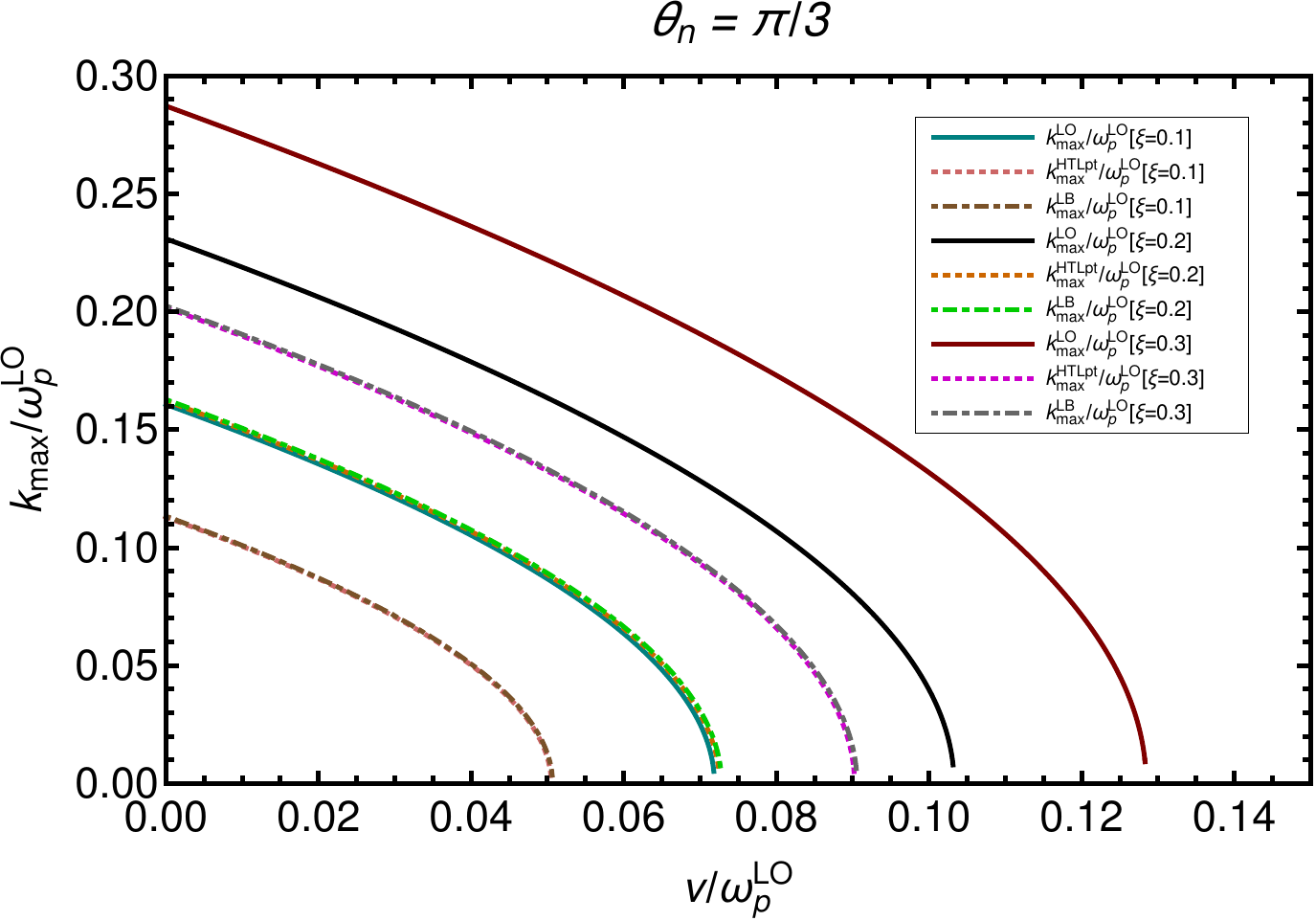}
    \caption{$k_{max}/{\omega^{LO}_{p}}$ vs $\nu/{\omega^{LO}_{p}}$ corresponding to A-mode for various EoSs at $\xi = 0.2$, $T_{c} = 0.17GeV$ and $T = 0.25GeV$ with different  $\theta_n$.}
    \label{fig:kmax_vs_nu_A}
\end{figure*}
\begin{figure*}    
    \includegraphics[height=5cm,width=8.6cm]{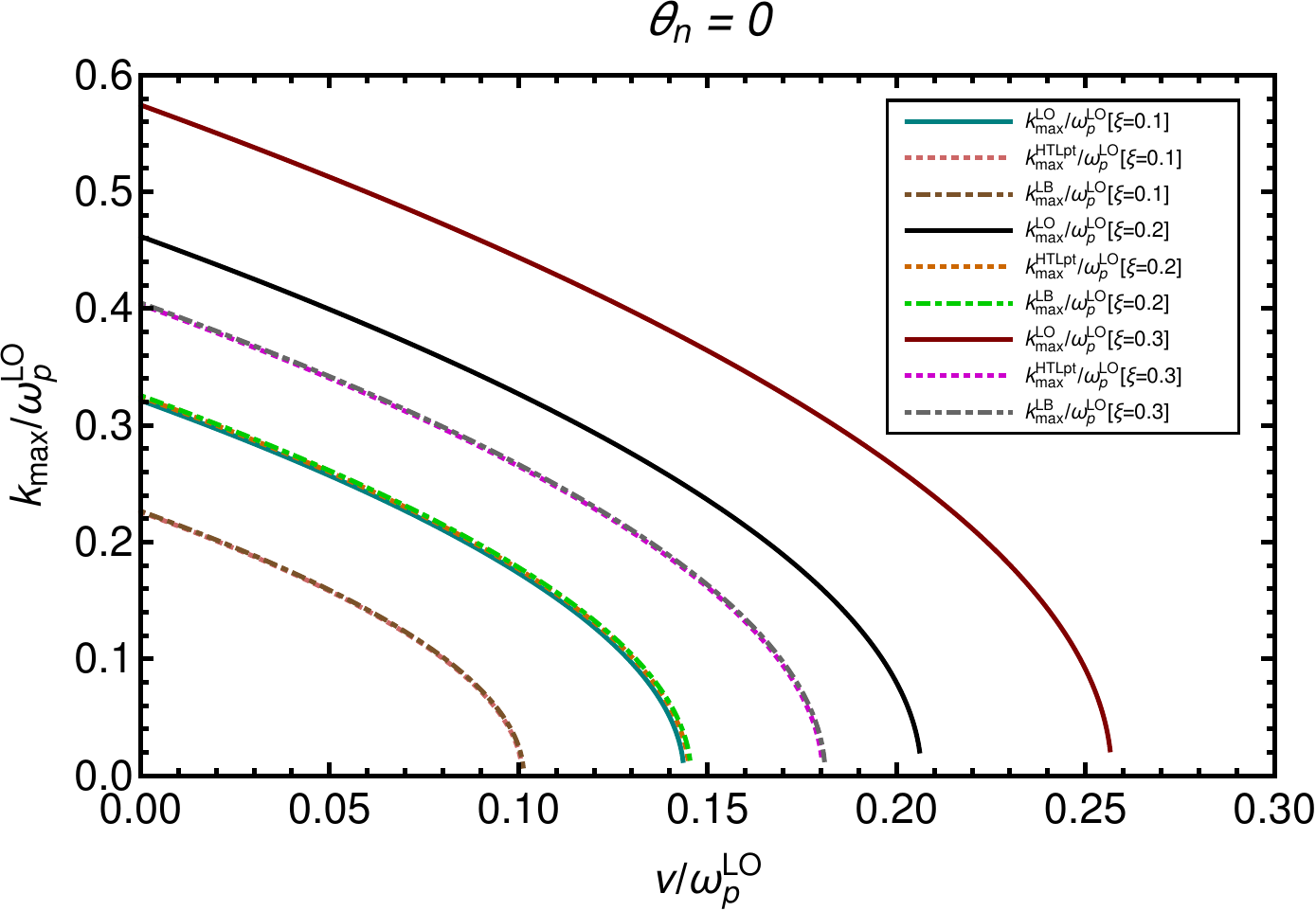}
    \hspace{-1mm}
    \includegraphics[height=5cm,width=8.6cm]{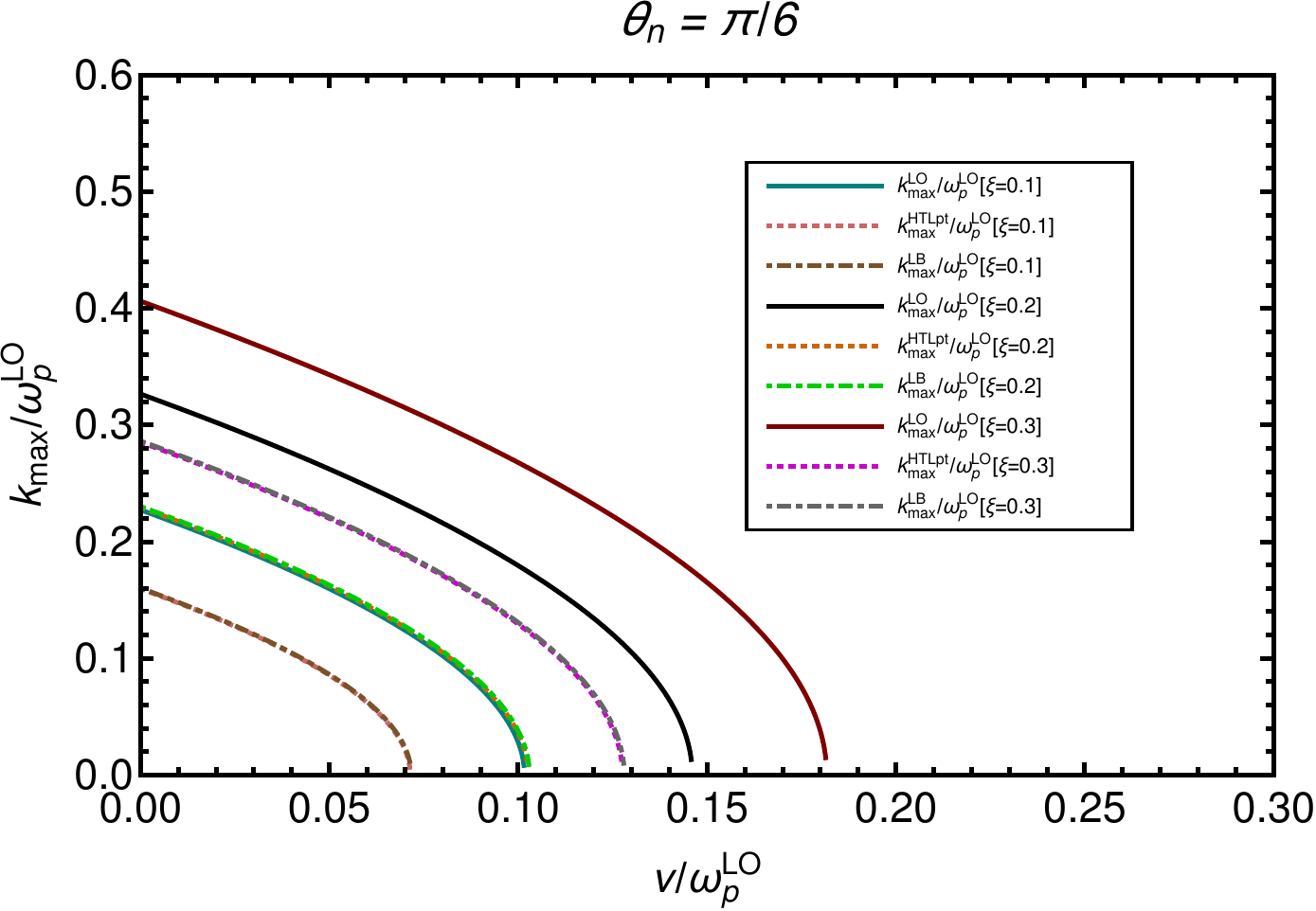}
    \caption{$k_{max}/{\omega^{LO}_{p}}$ vs $\nu/{\omega^{LO}_{p}}$ corresponding to G1-mode for various EoSs at $\xi = 0.2$, $T_{c} = 0.17GeV$ and $T = 0.25GeV$ with different $\theta_n$.}
    \label{fig:kmax_vs_nu_G1}
\end{figure*}
\begin{figure*}    
    \includegraphics[height=5cm,width=5.8cm]{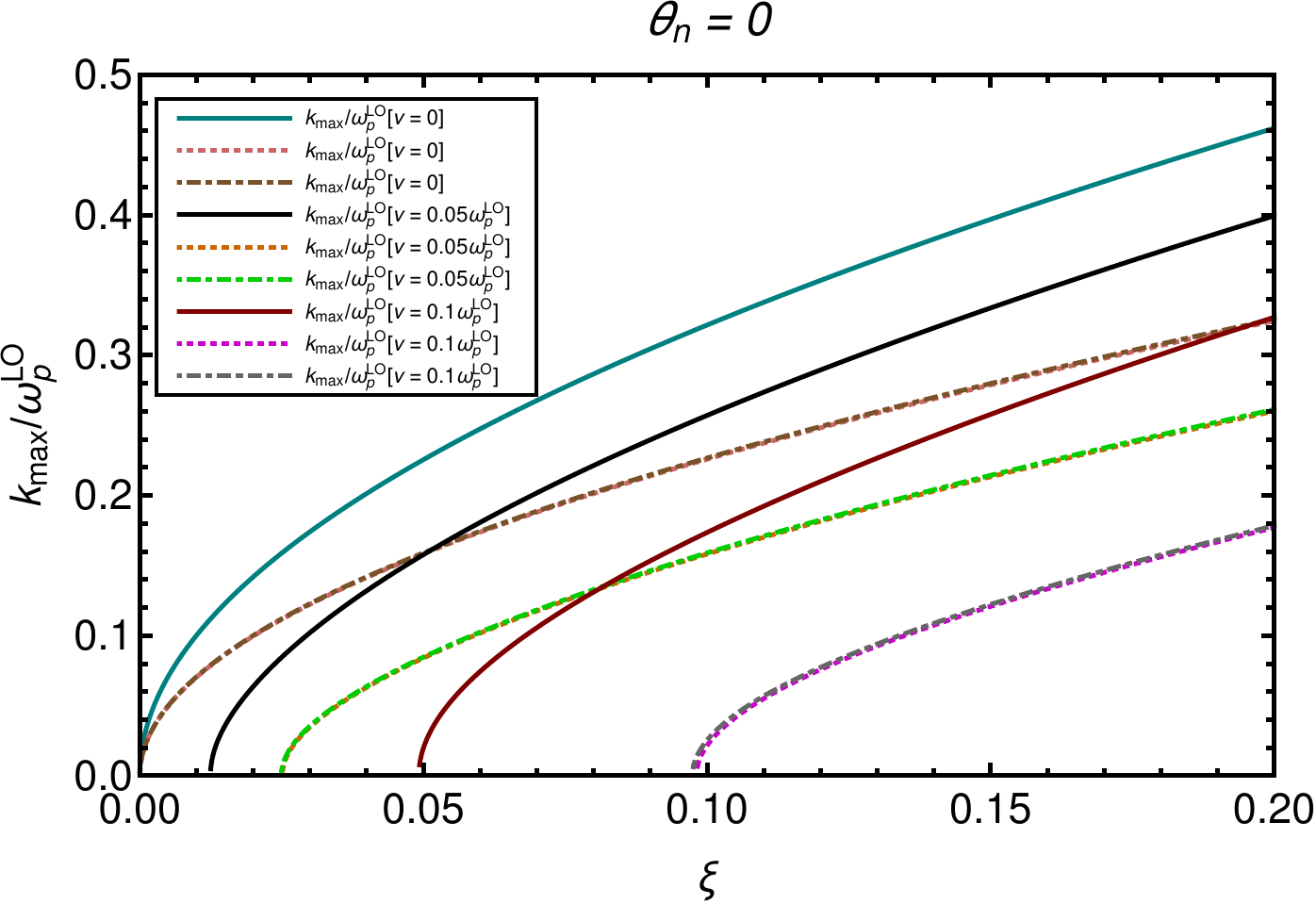}
    \hspace{-1mm}
    \includegraphics[height=5cm,width=5.8cm]{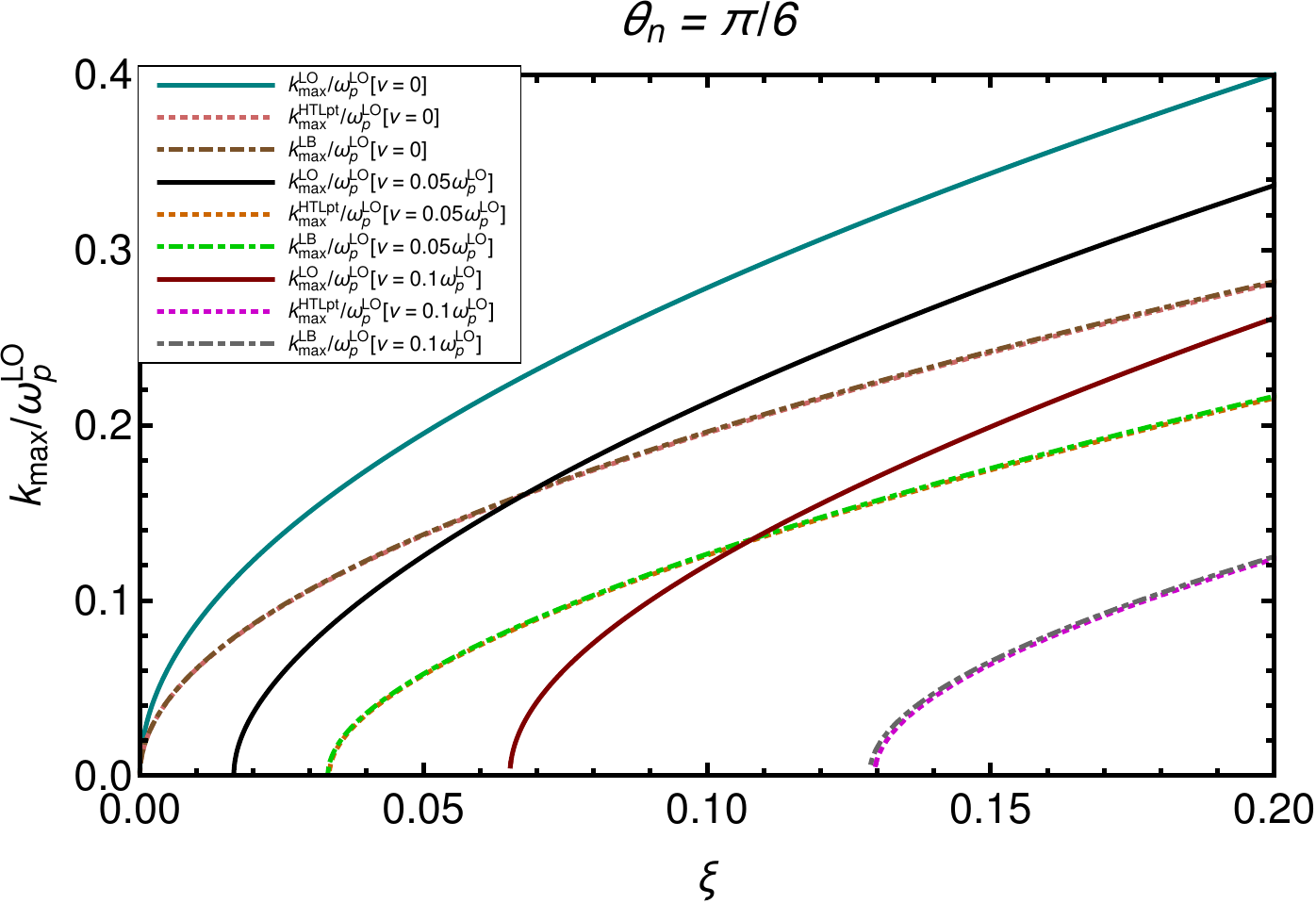}
    \hspace{-1mm}
    \includegraphics[height=5cm,width=5.8cm]{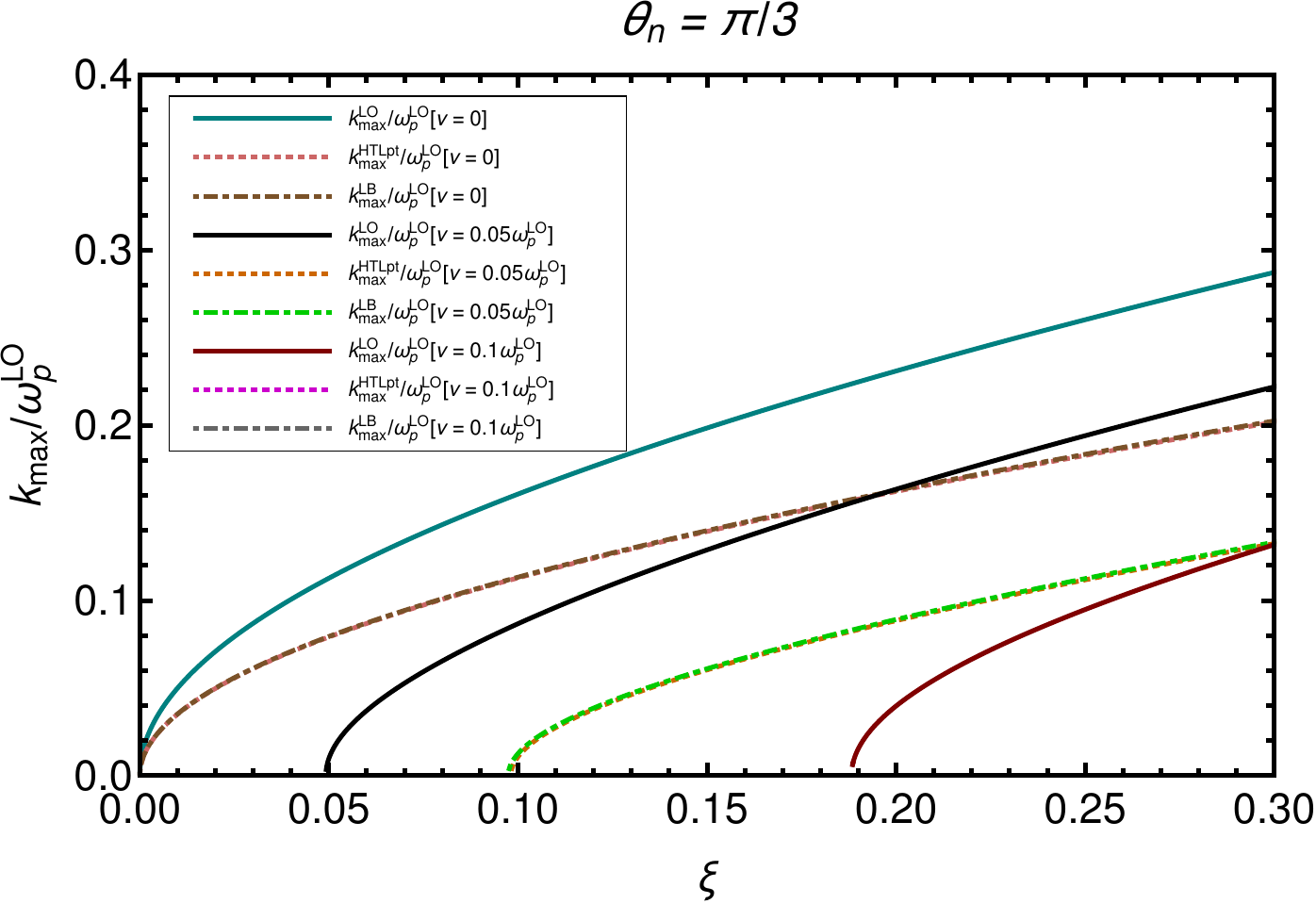}
    \caption{$k_{max}/{\omega^{LO}_{p}}$ vs $\xi$ corresponding to A-mode for various EoSs at $T_{c} = 0.17GeV$ and $T = 0.25GeV$ with different $\nu$ and $\theta_n$.}
    \label{fig:kmax_vs_xi_A}
\end{figure*}
\begin{figure*}    
    \includegraphics[height=5.5cm,width=8.6cm]{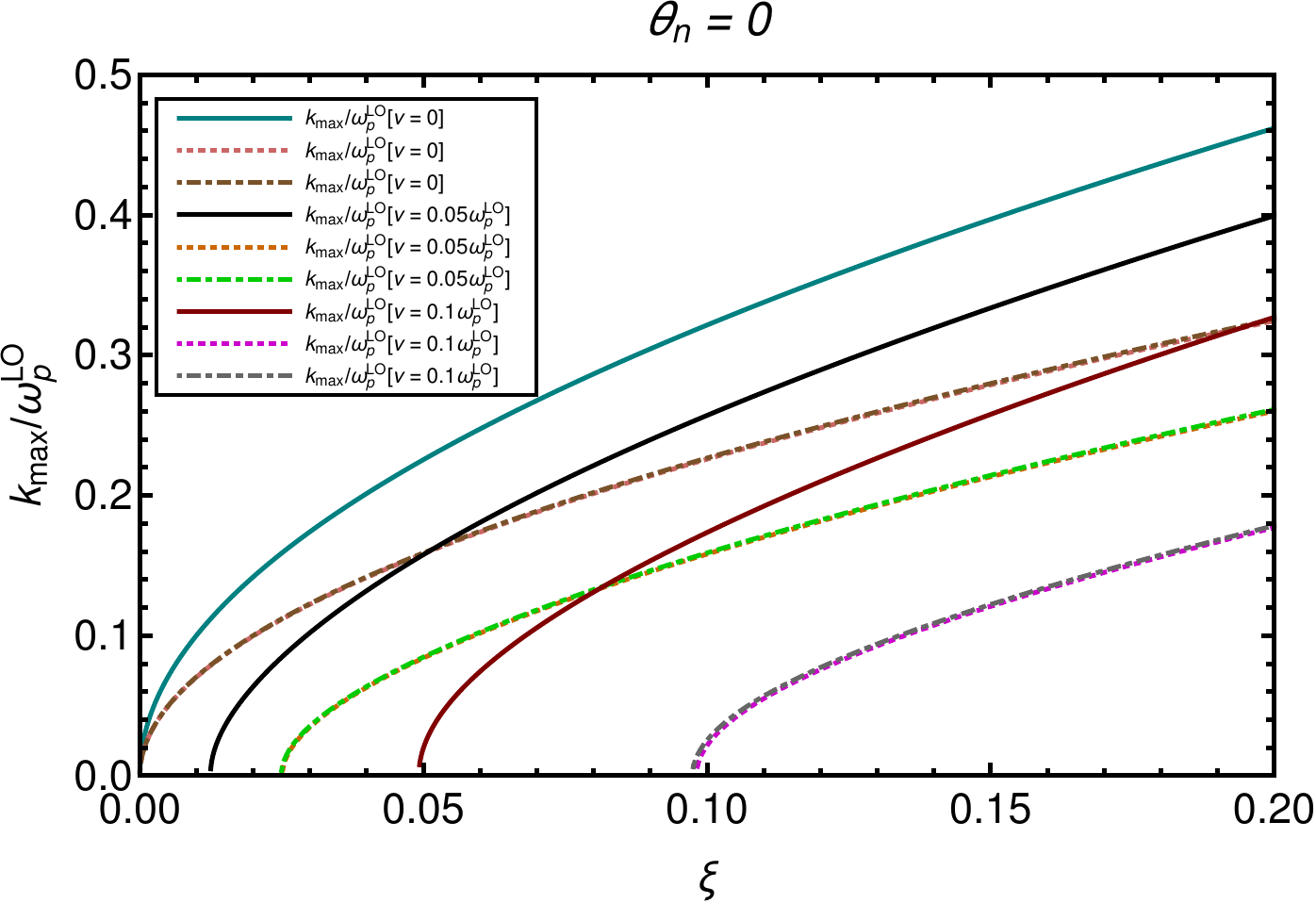}
    \hspace{-1mm}
    \includegraphics[height=5.5cm,width=8.6cm]{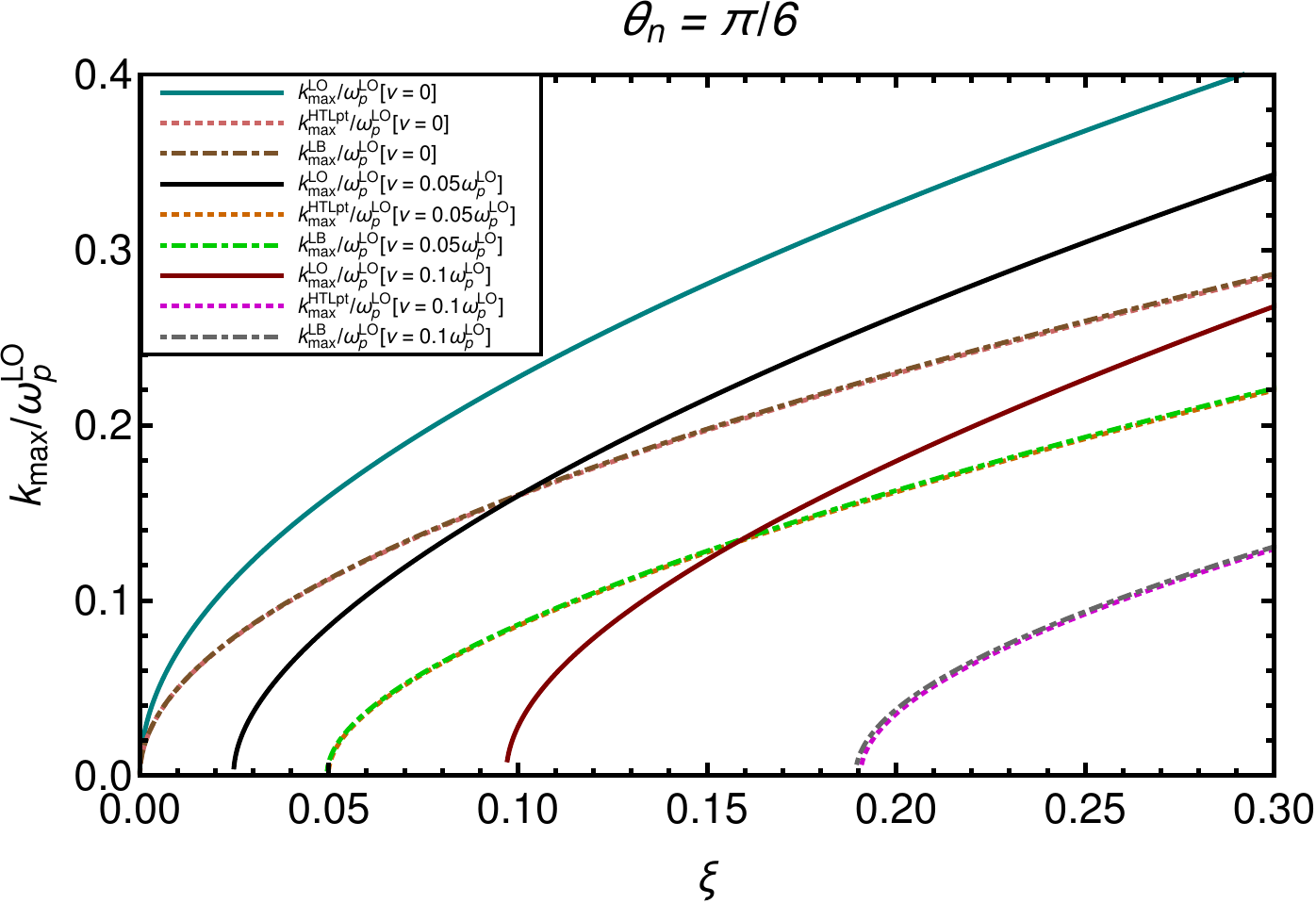}
    \caption{$k_{max}/{\omega^{LO}_{p}}$ vs $\xi$ corresponding to G1-mode for various EoSs at $T_{c} = 0.17GeV$ and $T = 0.25GeV$ with different $\nu$ and $\theta_n$.}
    \label{fig:kmax_vs_xi_G1}
\end{figure*}
\begin{figure*}    
    \includegraphics[height=5cm,width=8.6cm]{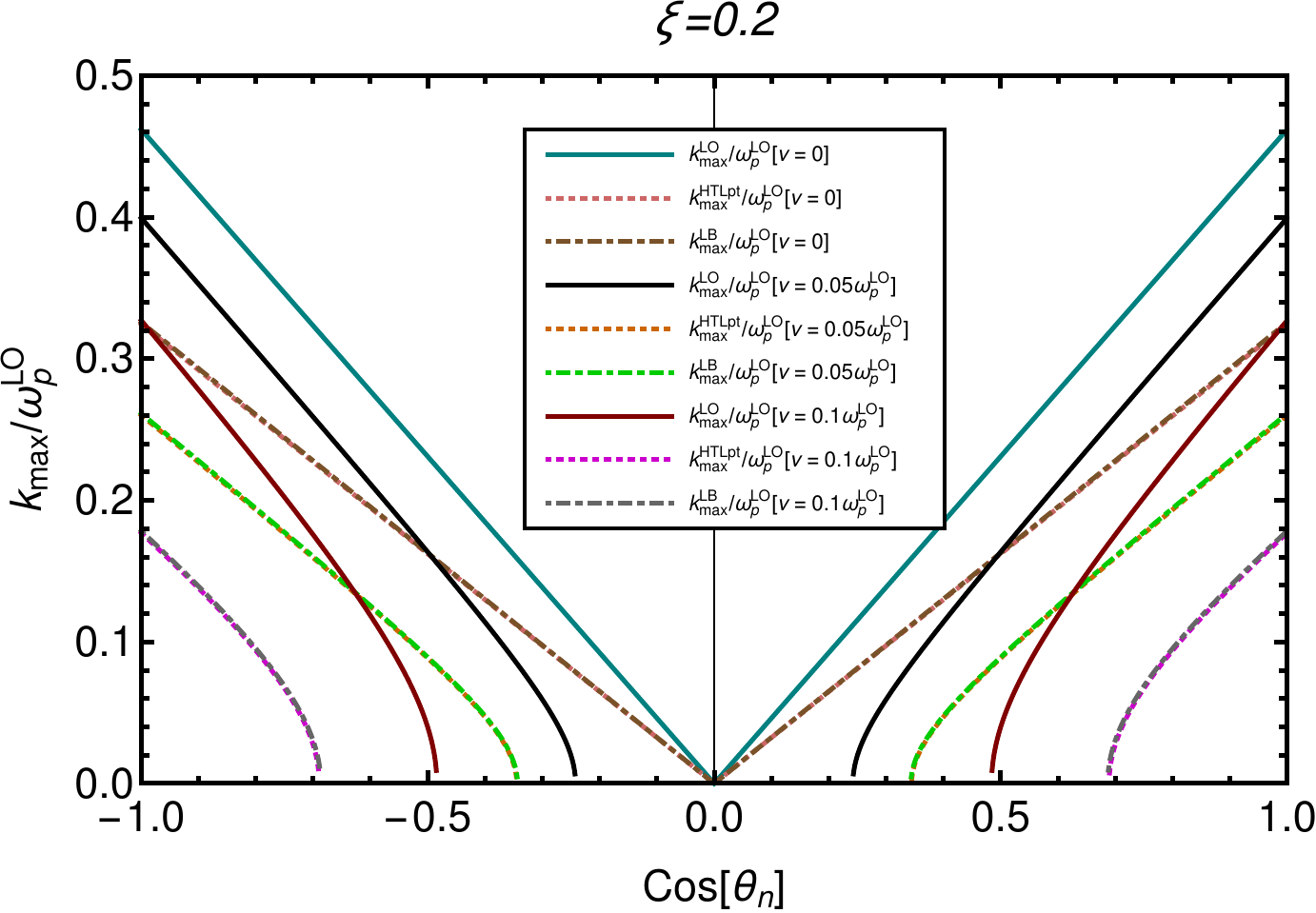}
    \hspace{-1mm}
    \includegraphics[height=5cm,width=8.6cm]{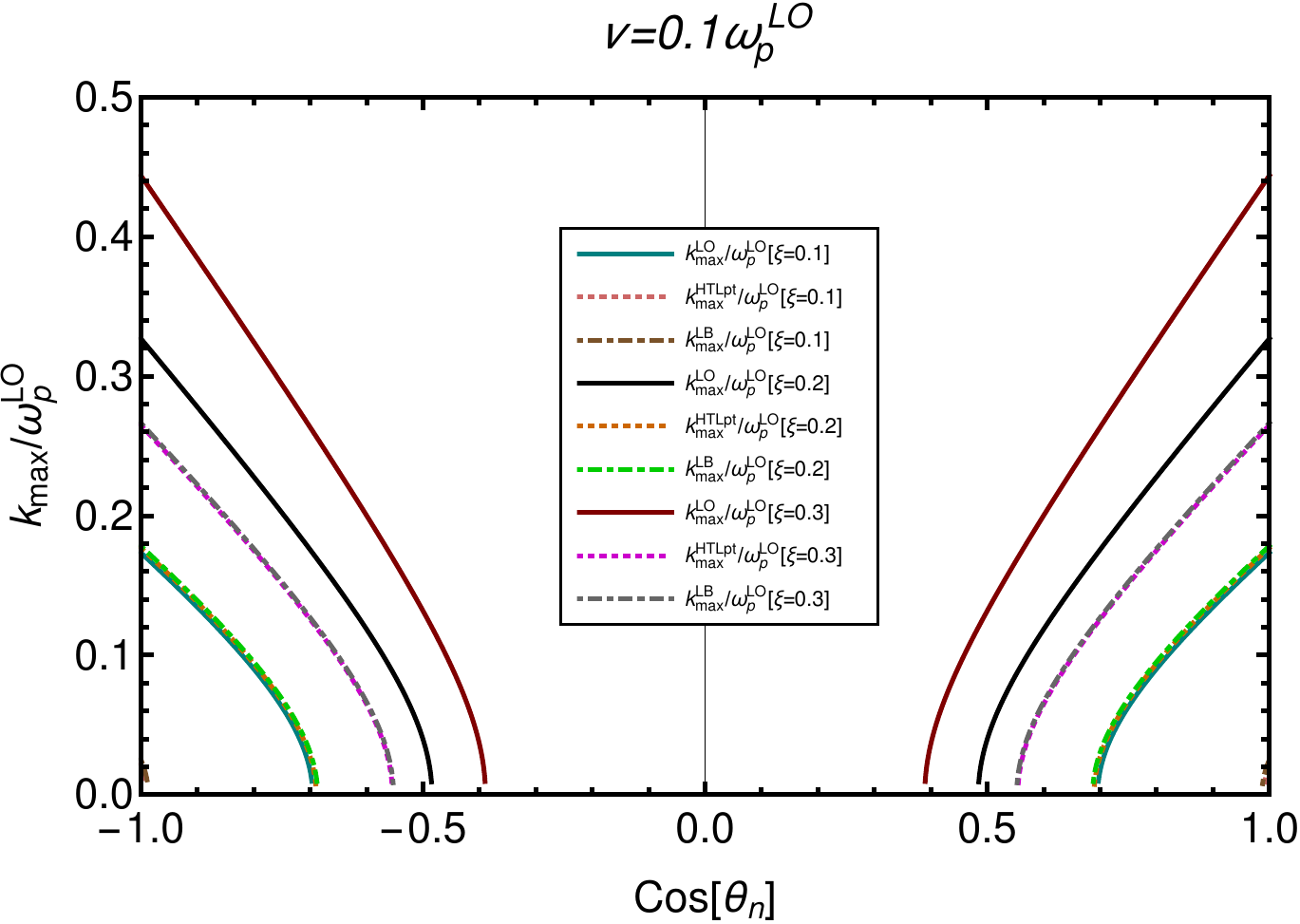}
    \caption{$k_{max}/{\omega^{LO}_{p}}$ vs Cos[$\theta_n$] corresponding to A-mode for various EoSs at $\nu = 0.1\omega^{LO}_{p}$, $T_{c} = 0.17GeV$ and $T = 0.25GeV$.}
    \label{fig:kmax_vs_th_A}
\end{figure*}
\begin{figure*}    
    \includegraphics[height=5cm,width=8.6cm]{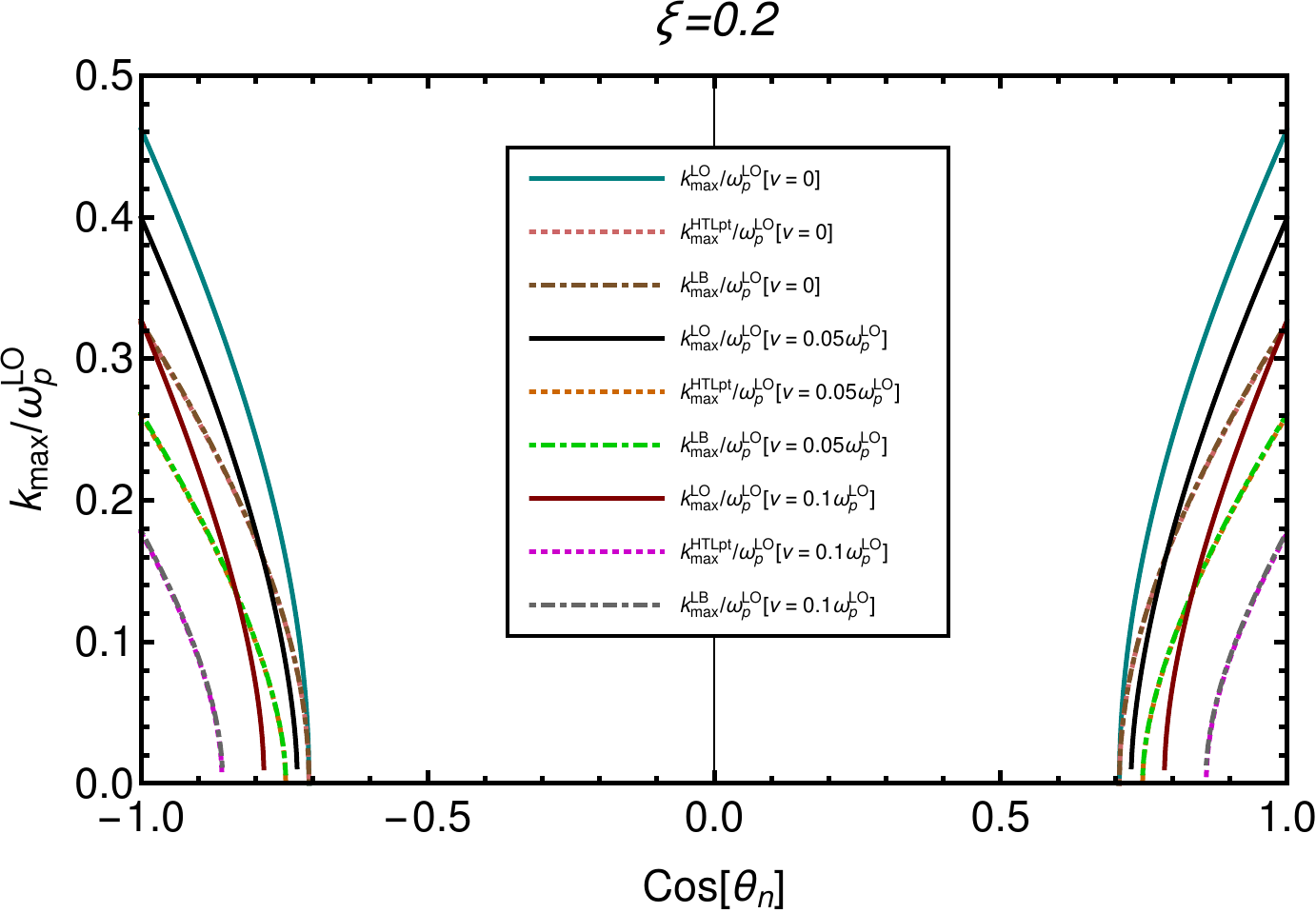}
    \hspace{-1mm}
    \includegraphics[height=5cm,width=8.6cm]{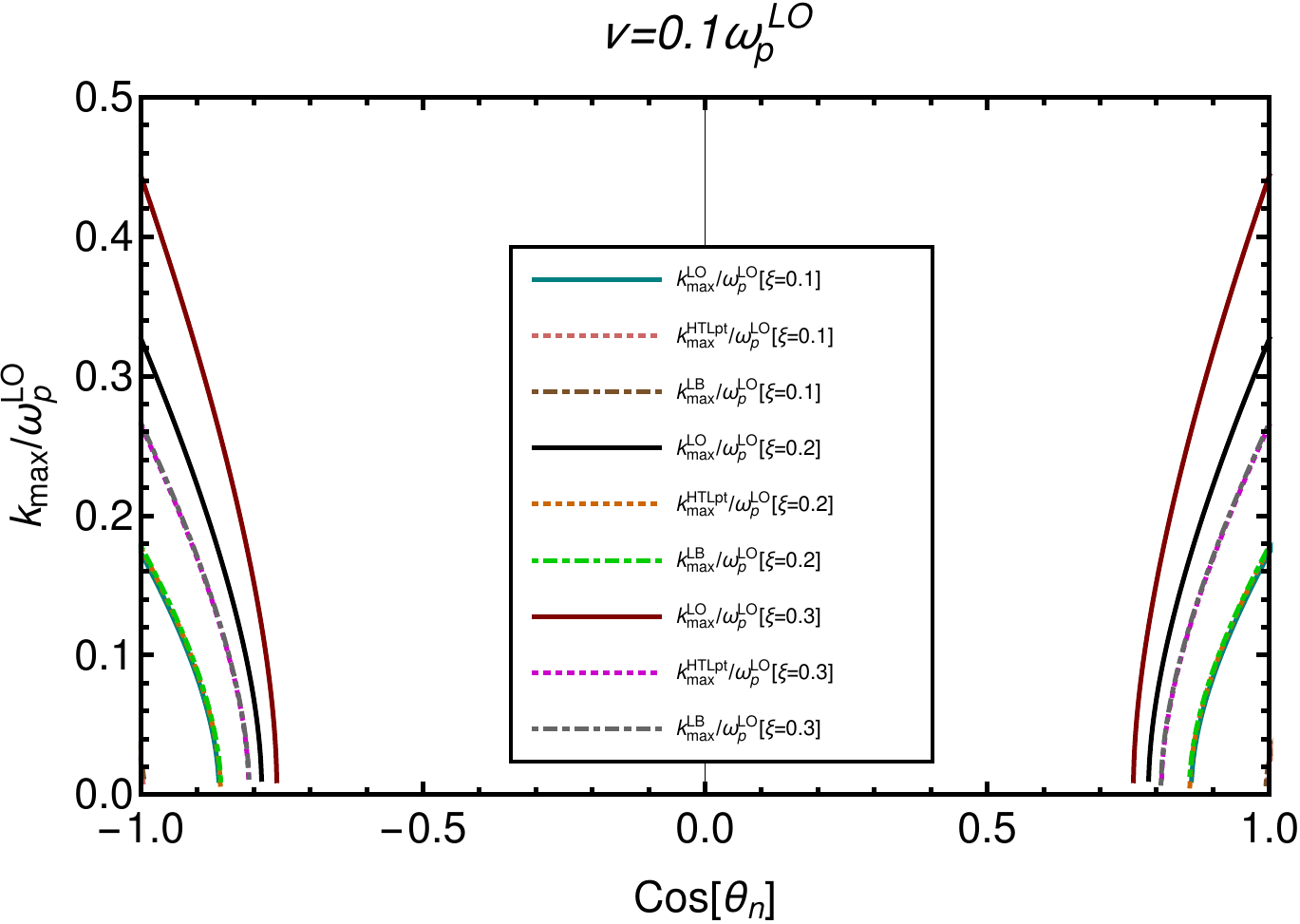}
    \hspace{-1mm}
    \caption{$k_{max}/{\omega^{LO}_{p}}$ vs Cos[$\theta_n$] corresponding to G1-mode for various EoSs at $\nu = 0.1\omega^{LO}_{p}$, $T_{c} = 0.17GeV$ and $T = 0.25GeV$.}
    \label{fig:kmax_vs_th_G1}
\end{figure*}
\begin{figure*}    
    \includegraphics[height=5cm,width=5.8cm]{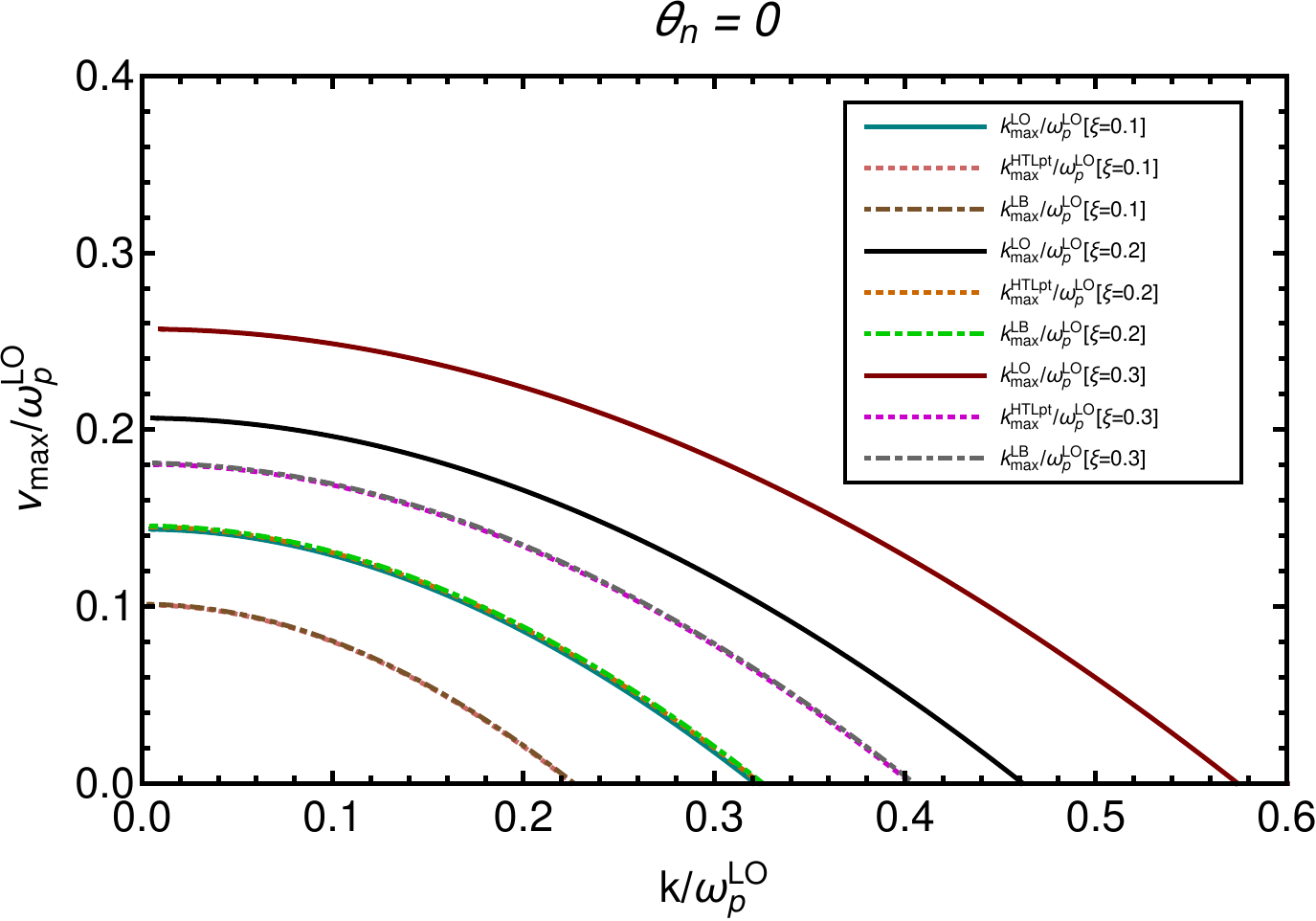}
    \hspace{-1mm}
    \includegraphics[height=5cm,width=5.8cm]{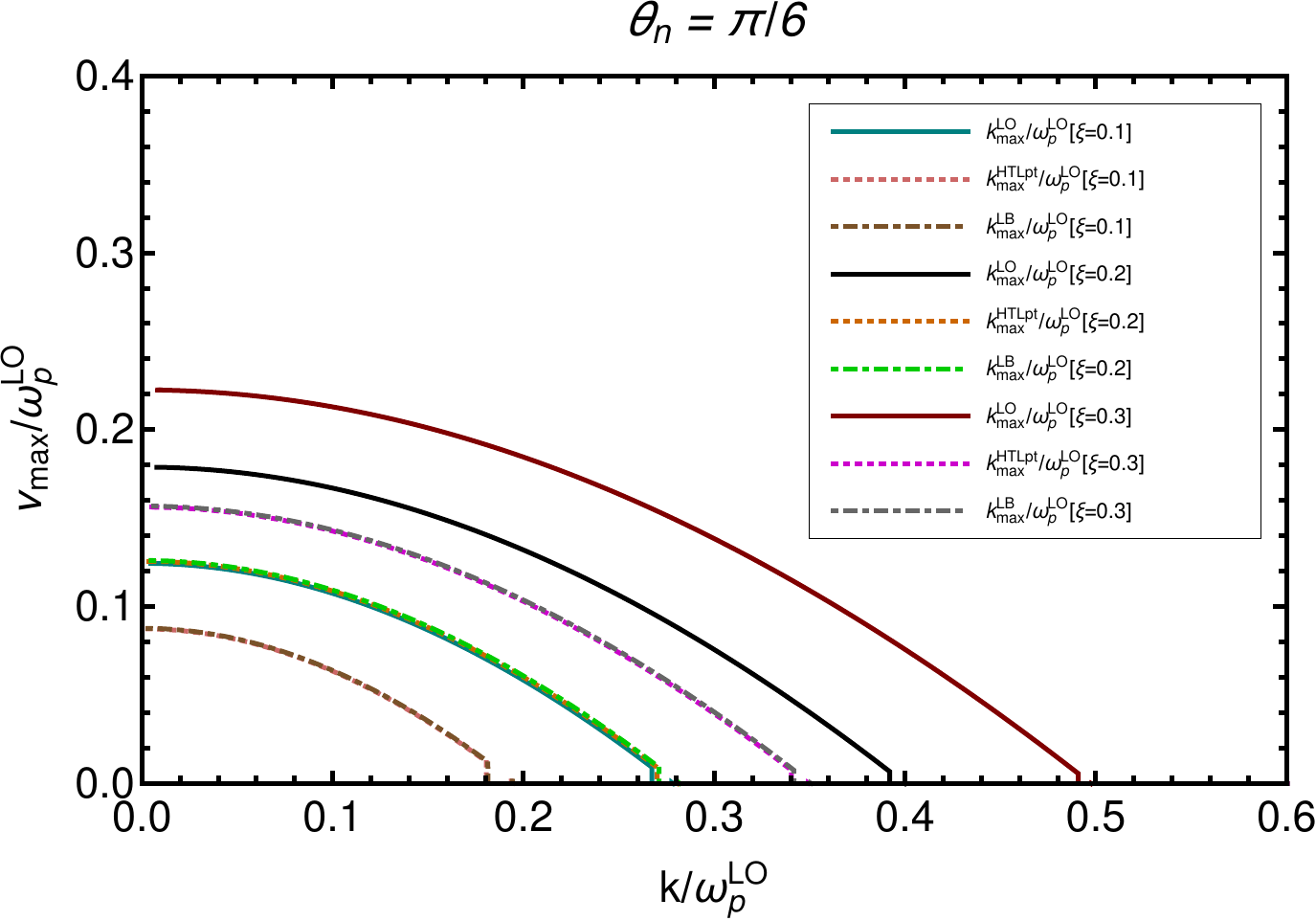}
    \hspace{-1mm}
    \includegraphics[height=5cm,width=5.8cm]{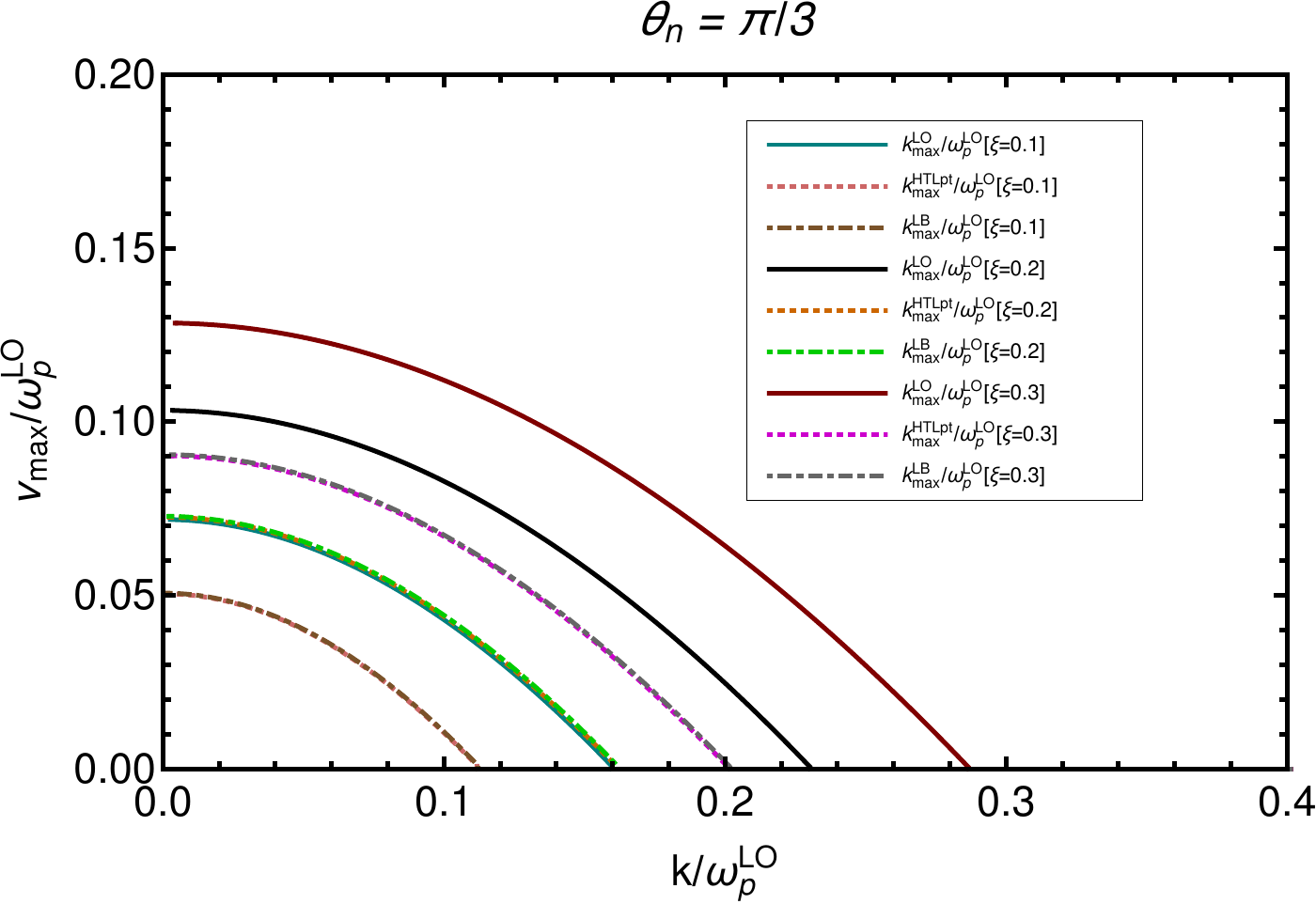}
    \caption{$\nu_{max}/{\omega^{LO}_{p}}$ vs $k/{\omega^{LO}_{p}}$ corresponding to A-mode for various EoSs at $\xi=0.2$, $T_{c} = 0.17GeV$, $T = 0.25GeV$ with different $\xi$ and $\theta_n$.}
    \label{fig:nu_max_vs_k_A}
\end{figure*}
\begin{figure*}    
    \includegraphics[height=5cm,width=8.6cm]{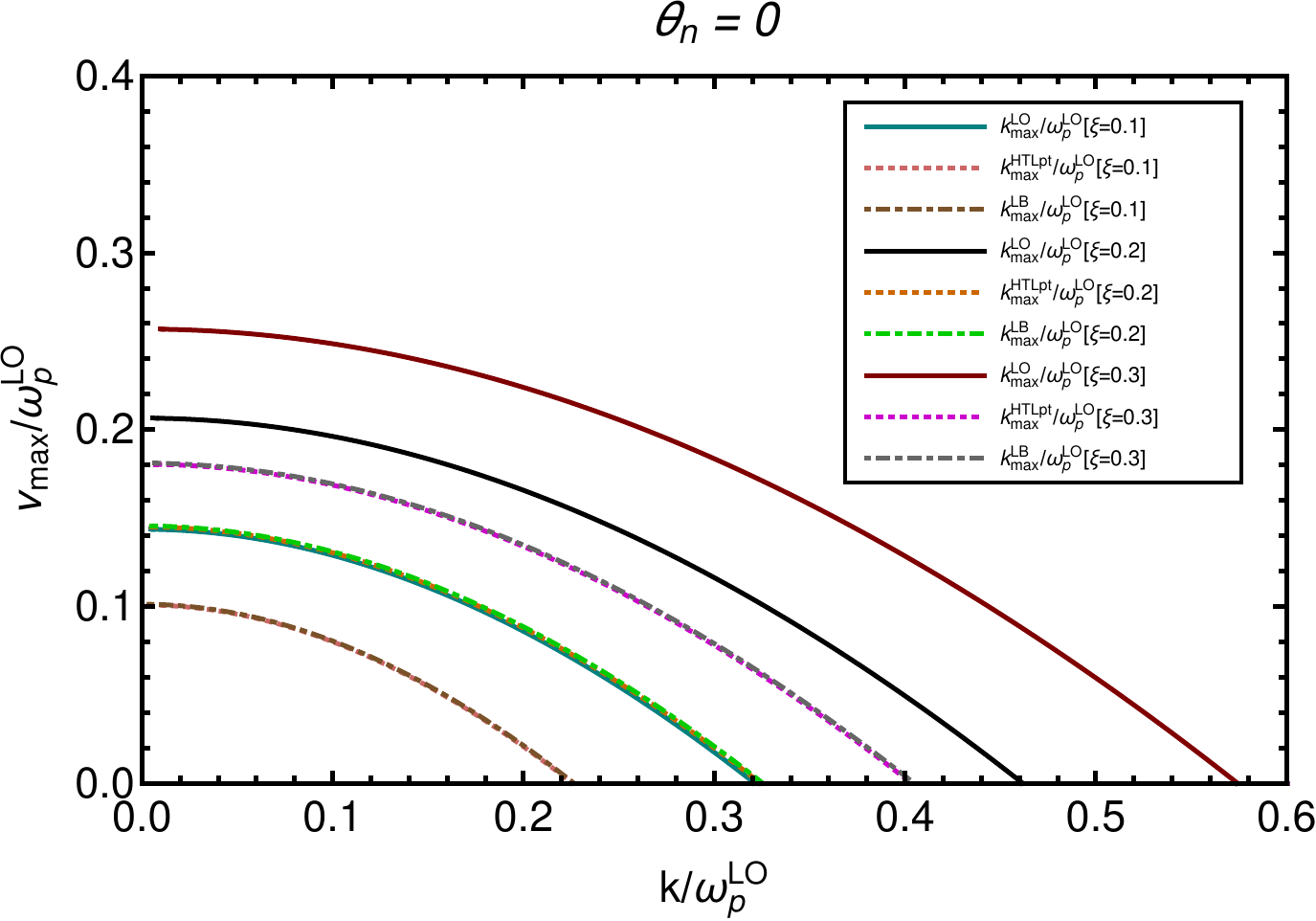}
    \hspace{-1mm}
    \includegraphics[height=5cm,width=8.6cm]{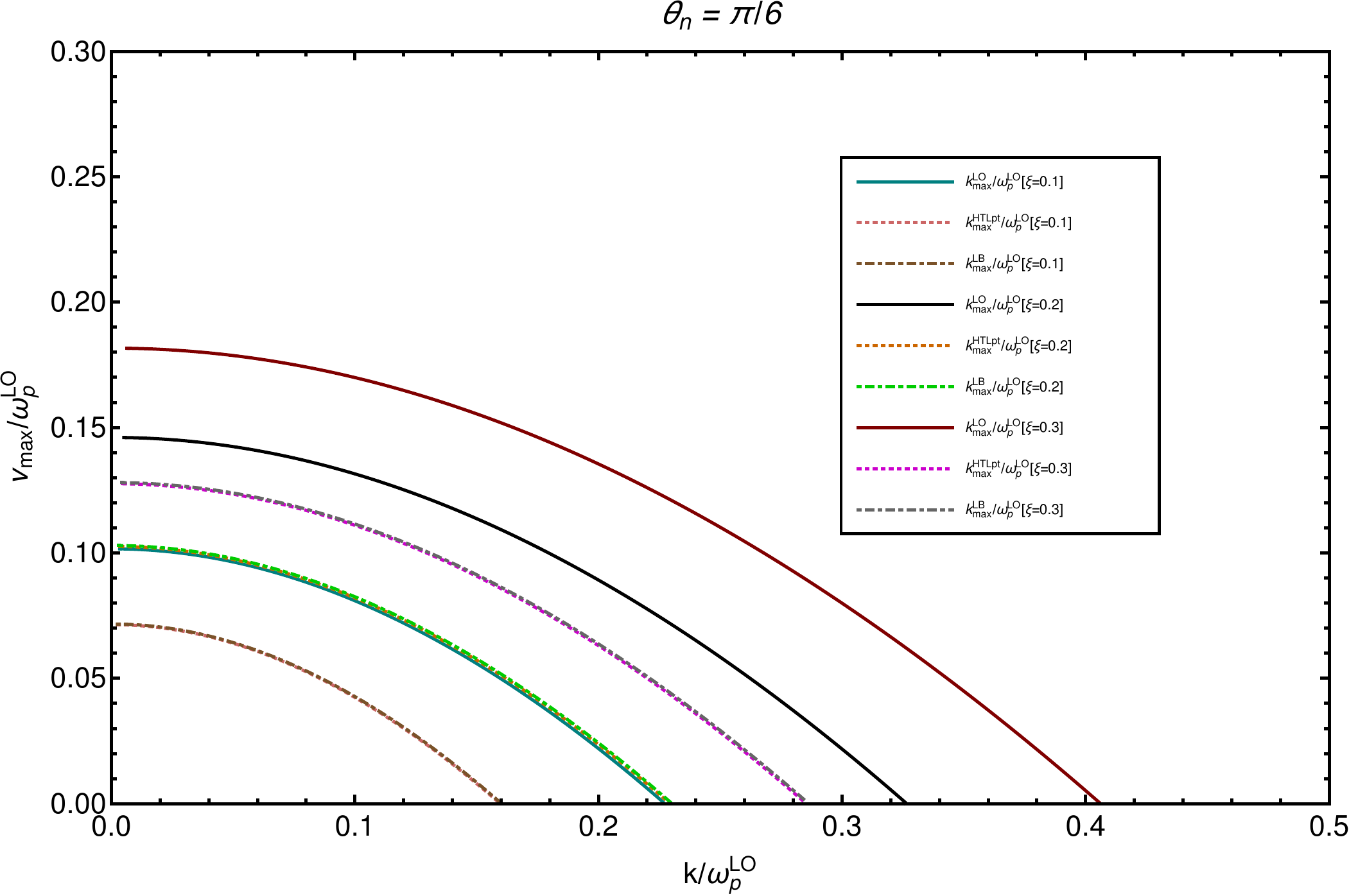}
    \caption{$\nu_{max}/{\omega^{LO}_{p}}$ vs $k/{\omega^{LO}_{p}}$ corresponding to G1-mode for various EoSs at $\xi=0.2$, $T_{c} = 0.17GeV$, $T = 0.25GeV$ with different $\xi$ and $\theta_n$.}
    \label{fig:nu_max_vs_k_G1}
\end{figure*}
\begin{figure*}    
    \includegraphics[height=5cm,width=8.6cm]{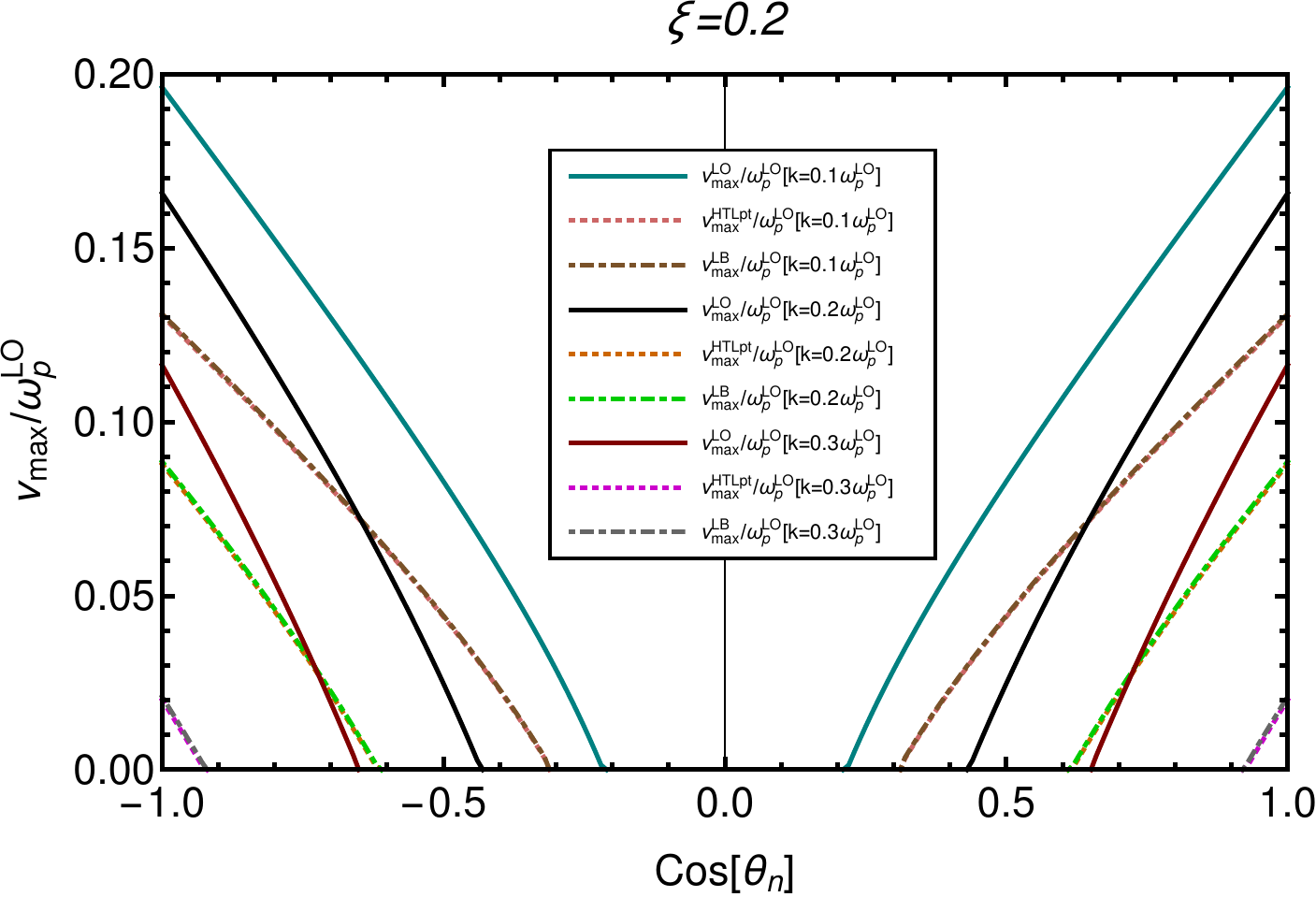}
    \hspace{-1mm}
    \includegraphics[height=5cm,width=8.6cm]{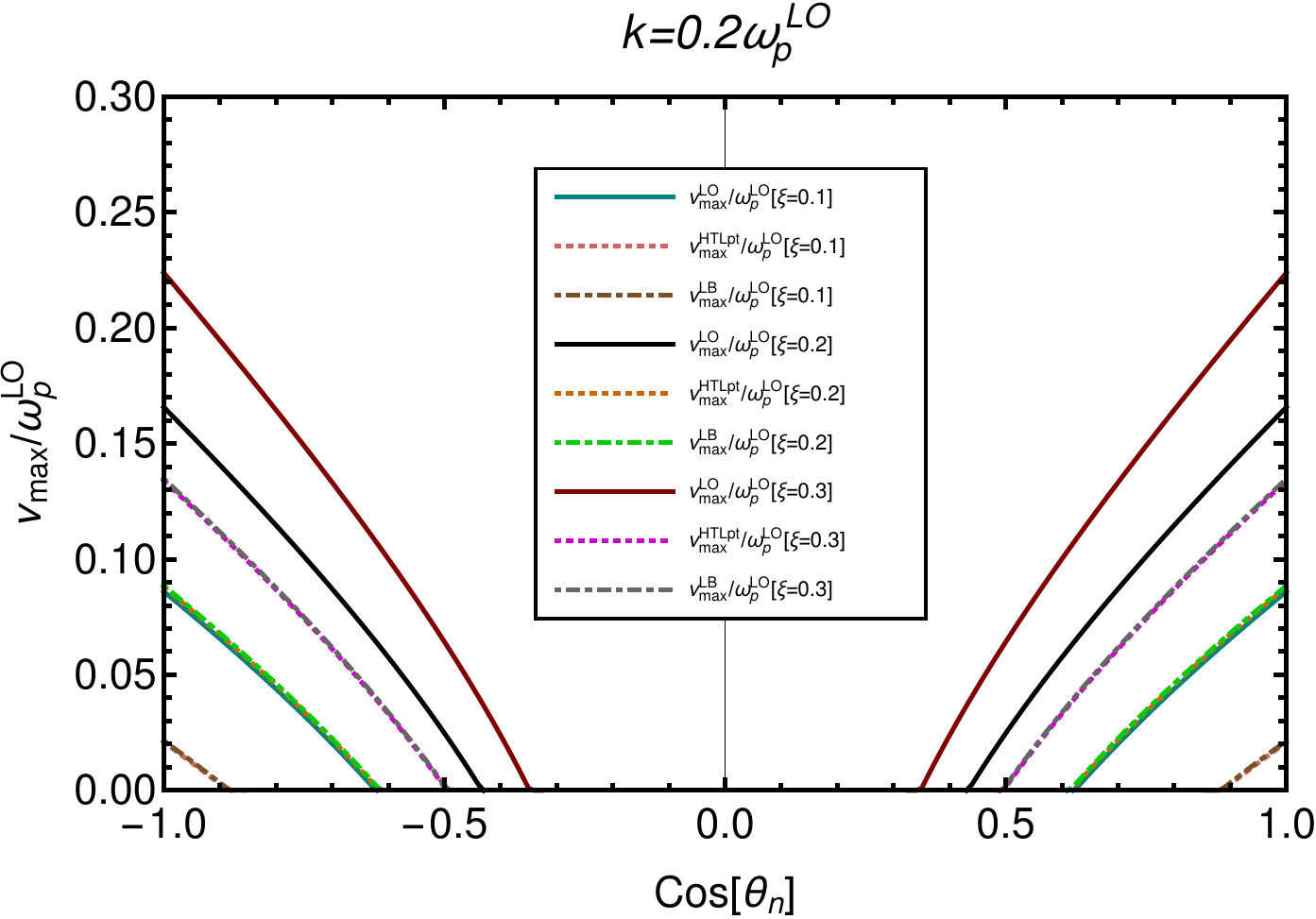}
    \caption{$\nu_{max}/{\omega^{LO}_{p}}$ vs Cos[$\theta_n$] corresponding to A-mode for various EoSs at $T_{c} = 0.17GeV$ and $T = 0.25GeV$.}
    \label{fig:nu_max_vs_th_A}
\end{figure*}
\begin{figure*}    
    \includegraphics[height=5cm,width=8.6cm]{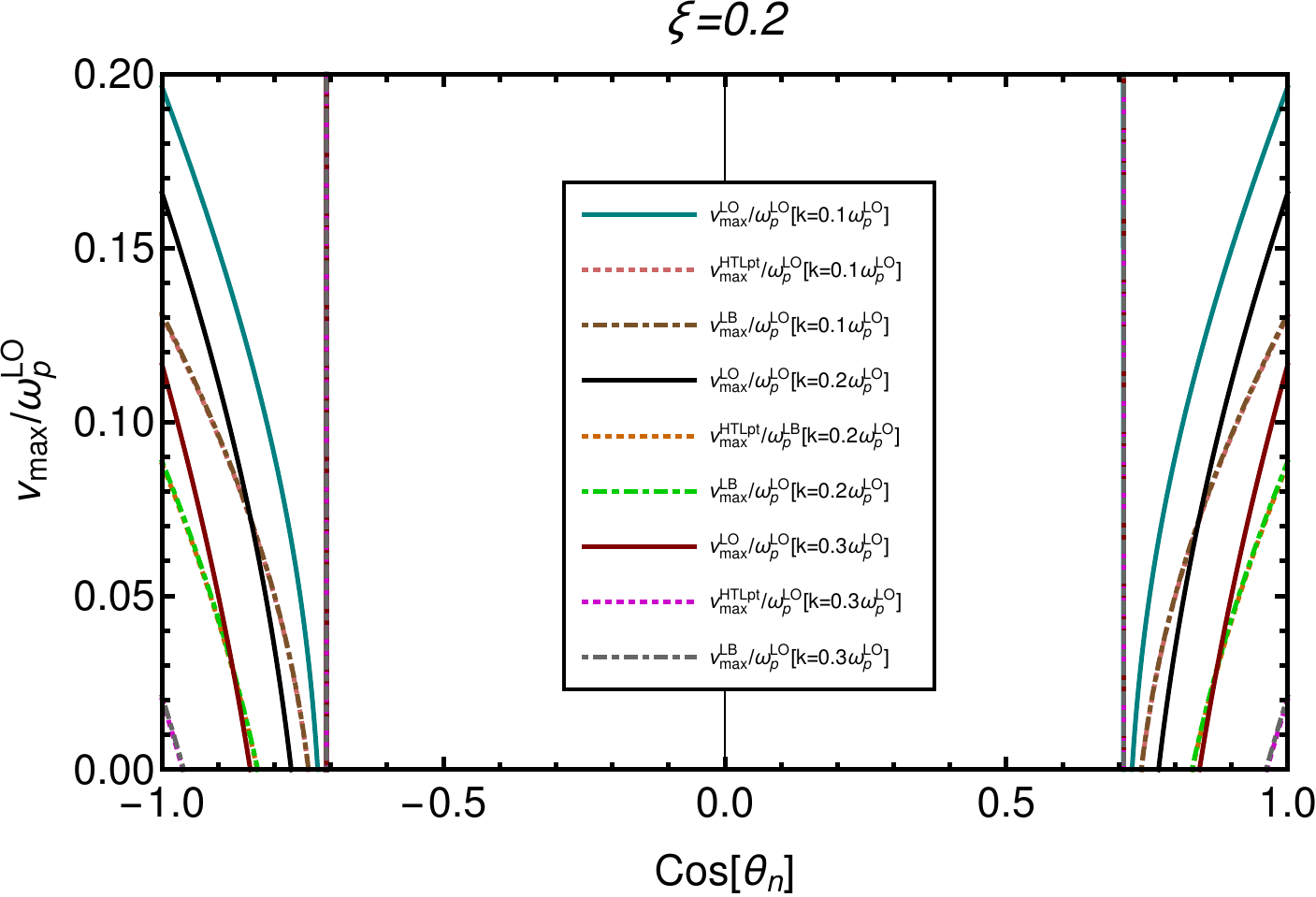}
    \hspace{-1mm}
    \includegraphics[height=5cm,width=8.6cm]{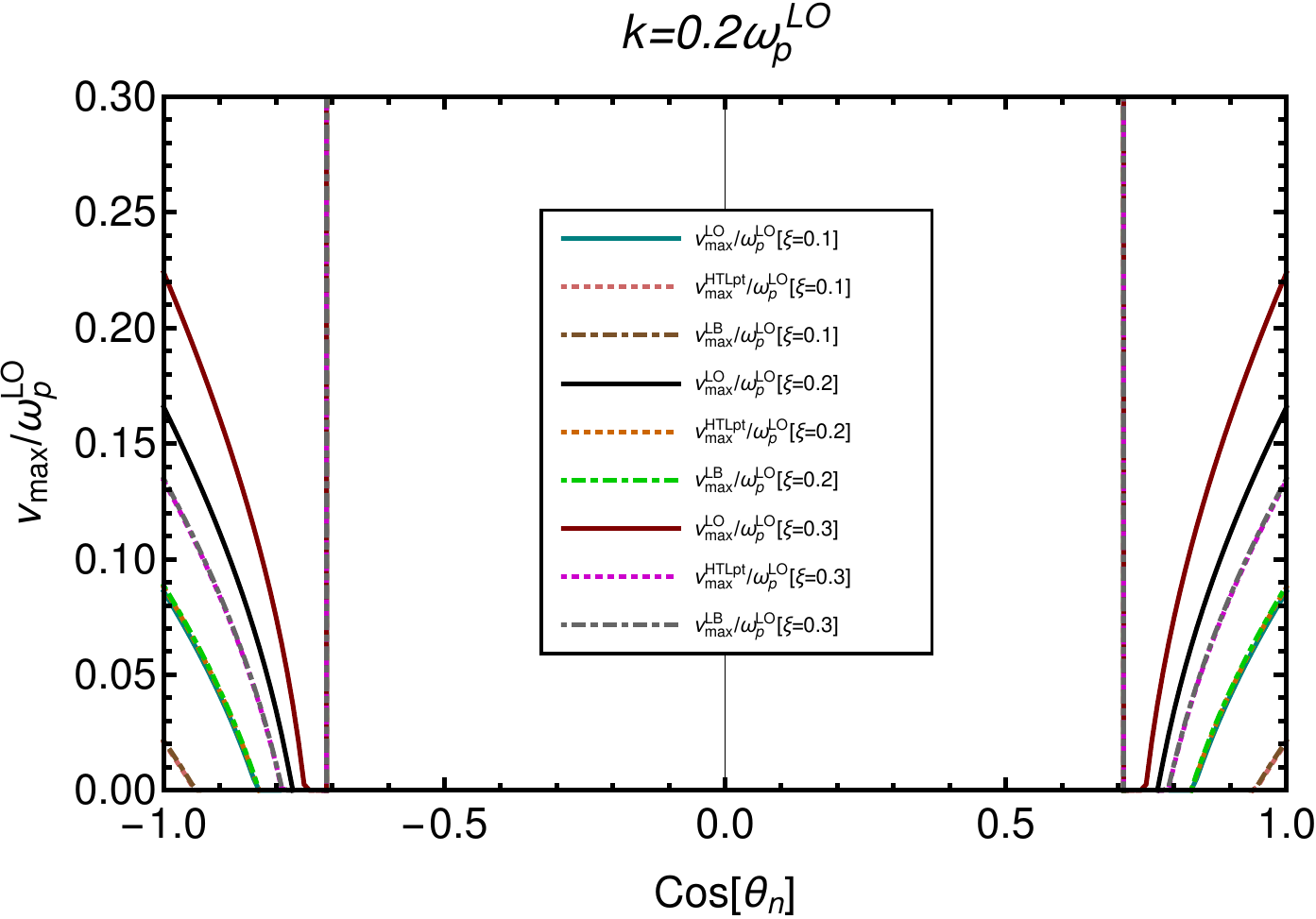}
    \caption{$\nu_{max}/{\omega^{LO}_{p}}$ vs Cos[$\theta_n$] corresponding to G1-mode for various EoSs at $T_{c} = 0.17GeV$ and $T = 0.25GeV$.}
    \label{fig:nu_max_vs_th_G1}
\end{figure*}
\begin{figure*}    
    \includegraphics[height=5cm,width=8.6cm]{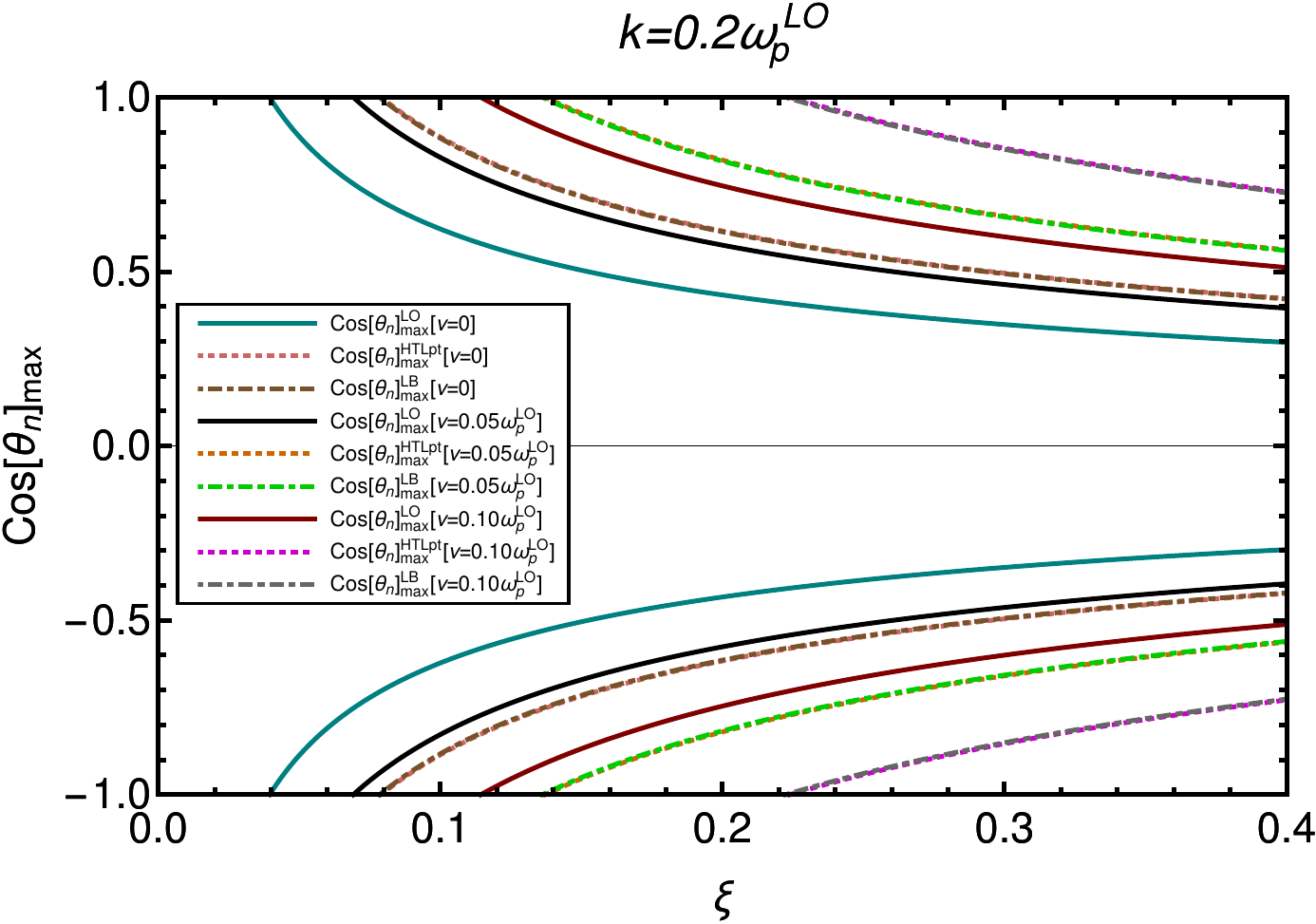}
    \hspace{-1mm}
    \includegraphics[height=5cm,width=8.6cm]{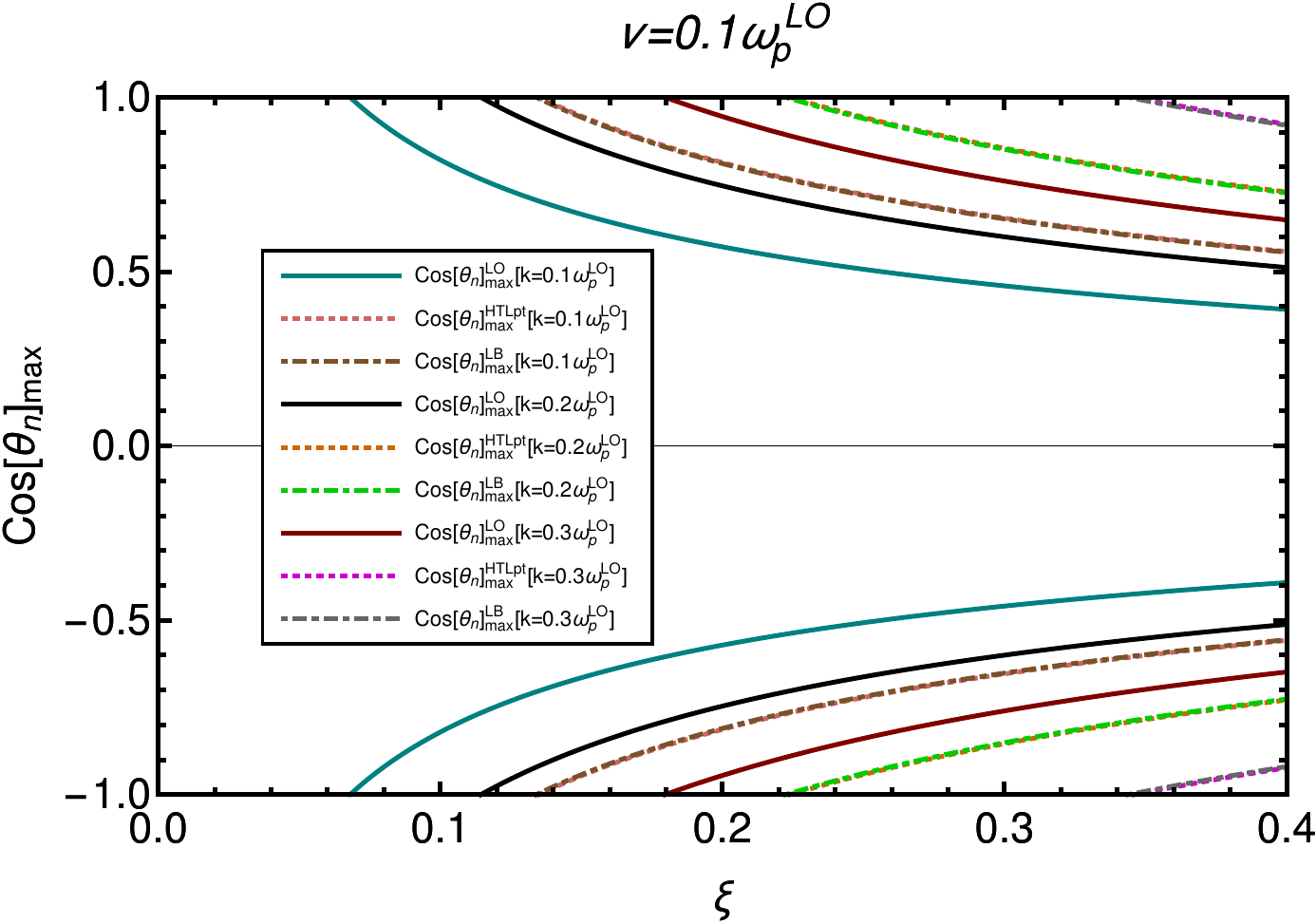}
    \caption{$Cos[\theta_n]_{max}$ vs $\xi$ corresponding to A-mode for various EoSs at $\xi=0.2$, $T_{c} = 0.17GeV$ and $T = 0.25GeV$.}
    \label{fig:th_max_vs_xi_A}
\end{figure*}
\begin{figure*}    
    \includegraphics[height=5cm,width=8.6cm]{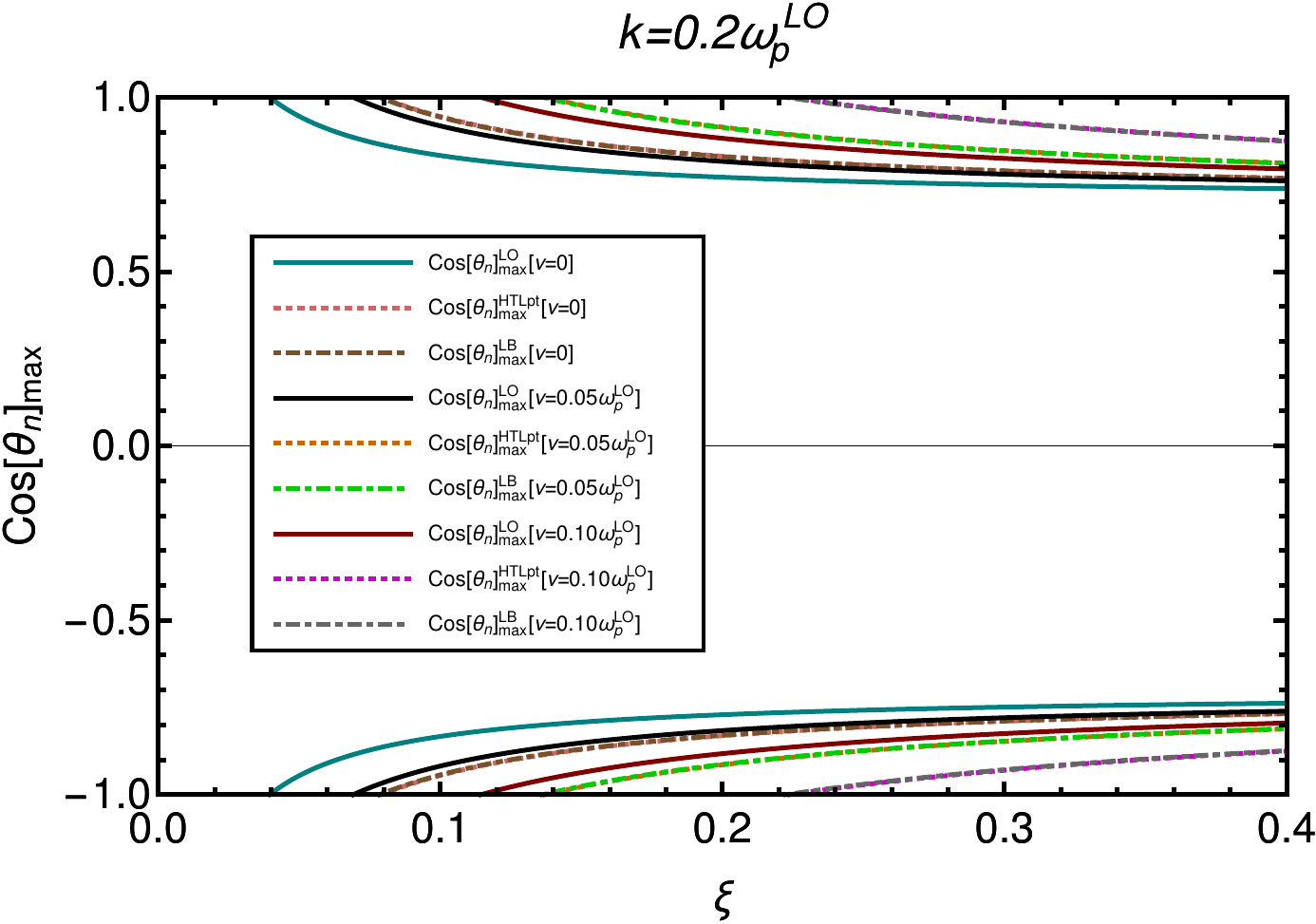}
    \hspace{-1mm}
    \includegraphics[height=5cm,width=8.6cm]{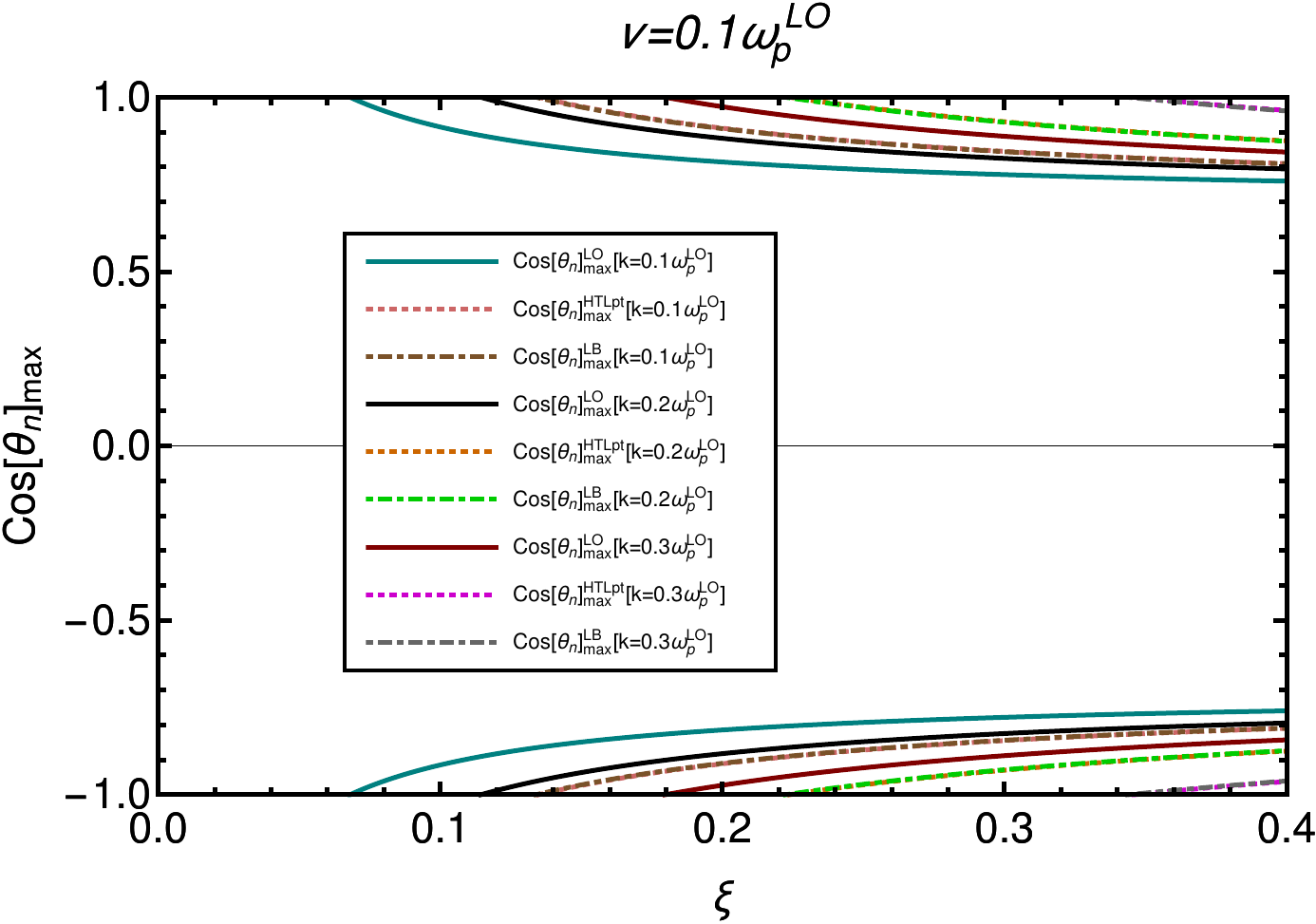}
    \caption{$Cos[\theta_n]_{max}$ vs $\xi$ corresponding to G1-mode for various EoSs at $\xi=0.2$, $T_{c} = 0.17GeV$ and $T = 0.25GeV$.}
    \label{fig:th_max_vs_xi_G1}
\end{figure*}

\subsection{Quasi-particle description of Isotropic hot QCD medium}
\label{QPD}
As mentioned earlier, the isotropic/equilibrium state of the QGP has been 
done within an effective quasi-particle model that describes hot QCD 
regarding temperature dependent effective fugacity parameters for the
gluons and quark/anti-quarks~\cite{chandra_quasi1, chandra_quasi2}. 
This model is denoted as EQPM (effective fugacity quasi-particle model) here.
It is to be noted that there are various quasi-particle models proposed
to describe hot QCD medium effects, {\it viz.} effective mass  models~\cite{effmass1, effmass2},
effective mass models with Polyakov loop~\cite{polya}, 
NJL and PNJL based effective models~\cite{pnjl}, and the EQPM and recent quasi-particle models based on the
Gribov-Zwanziger (GZ)  quantization results were leading to non-trivial IR-improved dispersion~\cite{flor}.

These quasi-particle models have shown their utility while studying transport properties of 
the QGP~\cite{Bluhm,chandra_eta, chandra_etazeta, PJI, Mkap, Mitra:2016zdw}.  Further, thermal conductivity
has also been considered, in addition to the viscosities~\cite{Mkap}, again within the
effective mass model along with electrical conductivity parameter for the QGP~\cite{Greco}, 
within EQPM by  Mitra and  Chandra~\cite{Mitra:2016zdw}
estimated the electrical conductivity and charge diffusion coefficients employing EQPM. 
The EQPM has also been applied to study heavy-quark transport in isotropic~\cite{Das:2012ck} and anisotropic hot QCD medium
~\cite{Chandra:2015gma} along with quarkonia in hot QCD medium~\cite{Chandra:2010xg, Agotiya:2016bqr} and dileptons in the 
QGP medium~\cite{Chandra:2015rdz,Chandra:2016dwy}.  An important point to be noted here is that the above models calculations
were not able to correctly reproduce the $\eta$ and $\zeta$ that are
phenomenologically extracted from the hydrodynamic simulations of the QGP \cite{Ryu, Denicol1}, consistently agreeing with different
experimental observables at RHIC. 
Earlier, we considered the EQPM description of a (2+1)-flavor lattice QCD EoS~\cite{cheng} (LEoS) . In present work, we have 
updated the model with the 3-loop HTL perturbative EoS (NNLO HTLpt EoS) that has recently been computed by 
 N. Haque {\it et,  al.}~\cite{nhaque, Andersen:2015eoa}  and  agrees remarkably
 well with the recent lattice results~\cite{bazabov2014,fodor2014}, and 
 very recent (2+1)-flavor lattice EoS by Bazabov {\it et. al}~\cite{bazabov2014}. 

The basic quantities that we need to set-up the linearized transport equation are,
\begin{itemize}
\item
The 
quasi-particle distribution functions with EQPM,  $ f_{eq}\equiv \lbrace f_{g}, f_{q} \rbrace$ 
(describing the strong interaction effects in terms of effective fugacities $z_{g,q}$):
\be
\label{eq1}
f_{g/q}= \frac{z_{g/q}\exp[-\beta E_p]}{\bigg(1\mp z_{g/q}\exp[-\beta E_p]\bigg)},
\ee
where $E_{p}=|{\bf p}|$ for the gluons and $\sqrt{|{\bf p}|^2+m_q^2}$ for the quark degrees of freedom ($m_q$ denotes the mass of the quarks).
Since the model is valid in the deconfined phase of QCD (beyond $T_c$), therefore, the mass of the light quarks can be neglected as compared to 
the temperature.
\item
The dispersion relation both in the gluonic and quark sectors:
\ba
\omega_{g/q}=E_{p}+T^2\partial_T ln(z_{g/q}).
\label{epp}
\ea
\item The Debye mass parameter  ($m_D$) and the effective coupling are other important quantities that are needed in our analysis throughout.
Following the definition of $m_D$ derived in semi-classical transport theory~\cite{dmass1, dm_rev1, dm_rev2} given below in terms of 
equilibrium gluonic and quark/anti-quark distribution function can be employed here, 
 \ba
 \label{dm}
 m_D^2&=& -4 \pi \alpha_{s}(T) \bigg(2 N_c \int \frac{d^3 p}{(2 \pi)^3} \partial_p f_g ({\bf p})\nn
 &+& 2 N_f  \int \frac{d^3 p}{(2 \pi)^3} \partial_p f_q ({\bf p})\bigg),
 \ea
 where, $\alpha_{s}(T)$ is the QCD running coupling constant at finite temperature~\cite{qcd_coupling}.
 Employing EQPM, we obtain the following expression,
 \ \ba
 \label{dm1}
 {m_D^2}^{l}&=&4 \pi \alpha_{s}(T) T^2  \bigg( \frac{2 N_c}{\pi^2} PolyLog[2,z_g^{l}]\nn&-&\frac{2 N_f}{\pi^2} PolyLog[2,-z_q^{l}]\bigg). 
 \ea
 where $l$ denotes different EoSs employed here with $z_{g/q}\rightarrow1$, corresponds to LO or ideal EoS. The effective coupling constant within EQPM that can be 
 read off from the expression for the Debye mass~\cite{Jamal:2017dqs} as, 
\ba
\label{eff}
\alpha_{eff} \equiv \alpha_{s}(T) g(z_g,z_q),
\ea where, 
 \ba
g(z_g,z_q)&=& \frac{\frac{2 N_c}{\pi^2} PolyLog[2,z_g]-\frac{2 N_f}{\pi^2} PolyLog[2,-z_q]}{\frac{N_c}{3}+\frac{N_f}{6}}.\nn
\ea
\end{itemize}
In the next sub-section, we will discuss the calculation of gluon self-energy while incorporating the BGK-colliosional kernel.

\subsection{Calculation of gluon self-energy}
\label{CGSE}
In order to describe the anisotropic hot QCD medium we
follow the arguments of Ref. \cite{Romatschke:2003ms}
where, the anisotropic distribution function was
obtained from a isotropic distribution function by
rescaling (stretching and squeezing) of one direction in
the momentum space as follows. 
\ba
f({\mathbf{p}})\equiv f_{\xi}({\mathbf{p}}) = C_{\xi}f(\sqrt{{\bf p}^{2} + \xi({\bf p}\cdot{\bf \hat{n}})^{2}}).
\label{aniso_distr}
\ea
where, ${\mathbf{\hat{n}}}$ is an unit vector (${\bf \hat{n}}^{2} = 1$) showing
the direction of momentum anisotropy. $C_{\xi}$ is the normalization constant.
$\xi$ is the anisotropy parameter which describes the amount of squeezing($\xi > 0$) or stretching($-1<\xi<0$)
of the distribution function in the ${\bf \hat{n}}$ direction. 

Some authors have considered $C_{\xi}$ to be unity ~\cite{Romatschke:2003ms} and later on they normalize
anisotropic number density to the isotropic one ~\cite{Romatschke:2004jh}.
We want Debye mass to remain undisturbed with the effects of anisotropy so that the effects of various EoSs can be seen clearly.
Hence we normalize the Debye mass, also done in Ref.~\cite{Carrington:2014bla}, we get,
\ba
C_{\xi} = \frac{\sqrt{|\xi|}}{arctan\sqrt{|\xi|}}.
\label{aniso_const}
\ea
Thus, the Debye mass is not going to get affected by anisotropy but only contains the effects of 
various EoSs. In small-$\xi$ limit $C_{\xi}$ can be written as,
\ba
C_{\xi} = 1 +\frac{\xi}{3} +O({\xi^2}).
\label{small_aniso_const}
\ea
Now writing down the equation for the self-energy
(\ref{selfenergy}) in temporal gauge for anisotropic hot QCD medium with
quasi particle description and making a change of
variable as ${\mathbf{\tilde p}}\equiv \sqrt{{\bf p}^{2}
+ \xi({\bf p}\cdot{\bf \hat{n}})^{2}}$,  we can obtain
the following expression,
\begin{widetext}
\ba
\Pi^{ij}(K)&=&m_D^2~ C_{\xi}\int\frac{d\Omega}{4\pi}v^i\frac{v^l+\xi(\mathbf{v}\cdot\mathbf{\hat{n}})n^l}{(1+\xi(\mathbf{v}\cdot\mathbf{\hat{n}})^2)^2} \left[\delta^{jl}(\omega-\mathbf{k}\cdot\mathbf{v})+v^jk^l\right]D^{-1}(K,\mathbf{v},\nu)+(i\nu) m_D^2~ C_{\xi}\int\frac{d\Omega^{\prime}}{4\pi}(v^{\prime})^i\nn
&&\times D^{-1}(K,\mathbf{v}^{\prime},\nu) \int\frac{d\Omega}{4\pi}\frac{v^l+\xi(\mathbf{v}\cdot\mathbf{\hat{n}})n^l}{(1+\xi(\mathbf{v}\cdot\mathbf{\hat{n}})^2)^2}\big[\delta^{jl}(\omega-\mathbf{k}\cdot\mathbf{v})+v^jk^l\big] D^{-1}(K,\mathbf{v},\nu)\mathcal{W}^{-1}(K,\nu),
\label{pimunu}
\ea
\end{widetext}
where the Debye mass,
\ba
m_D^2=-\frac{g^2}{2\pi^2}\int_0^{\infty}d\tilde{p}\,\tilde{p}^2\frac{d f_{\text{iso}}(\tilde{p})}{d\tilde{p}}\text{\,.}
\ea
is same as given in Eq.(\ref{dm}) and (\ref{dm1}). Now as we can see,
Eq.(\ref{pimunu}) is a tensorial equation and hence one can not simply integrate it.
We need to construct an analytical form of the gluon self-energy
using the available symmetric tensors. In the next sub-section~\ref{DSS},
we shall construct $\Pi^{ij}$, analytically and then solve it.  

\subsubsection{Decomposition of self-energy in terms of structure functions}
\label{DSS}
For isotropic hot QCD plasma we need only the transverse
$P^{ij}_{T}=\delta^{ij}-k^{i}k^{j}/{k^{2}}$ and the longitudinal
$P^{ij}_{L}=k^{i}k^{j}/{k^{2}}$ tensor
projectors to decompose $\Pi^{ij}$. Due to presence of
anisotropy vector 
${\bf \hat{n}}$, we have to take into account two more
projectors $P^{ij}_{n}={\tilde{n}}^i{\tilde{n}}^j/{\tilde{n}}^2$ 
and $P^{ij}_{kn}=k^{i}{\tilde{n}}^{j}+k^{j}{\tilde{n}}^{i}$
\cite{Kobes:1990dc,Romatschke:2003ms, Romatschke:2004jh} where, $\tilde{n}^i=(\delta^{ij}-\frac{k^i k^j}{k^2})\hat{n}^j$
 is a vector orthogonal to $k^i$ {\it i.e.}   $\mathbf{\tilde{n}}\cdot{\mathbf{k}}=0$.
Thus we can decompose the self-energy given in Eq.(\ref{selfenergy})
into following four basis as follows,
\ba
\Pi^{ij}=\alpha{P}^{ij}_{T}+\beta{P}^{ij}_{L}+\gamma{P}^{ij}_{n}+\delta{P}^{ij}_{kn},
\label{seexpan}
\ea
where 
$\alpha$, $\beta$, $\gamma$ and $\delta$ are 
some scalar functions which are called structure functions. They can be
determined by taking the appropriate projections of the
Eq.(\ref{selfenergy}) as follows,  
\ba
\alpha&=&({P}^{ij}_{T}-{P}^{ij}_{n})\Pi^{ij}, ~~~\beta={P}^{ij}_{L}\Pi^{ij} \nonumber,\\
\gamma&=&(2{P}^{ij}_{n}-{P}^{ij}_{T})\Pi^{ij},~~~\delta=\frac{1}{2 k^{2}{\tilde{n}}^2}{P}^{ij}_{kn}\Pi^{ij}.
\label{structurefunctions}
\ea 
The structure functions mainly depend on $k$, 
$\omega$, $\xi$ $\nu$ and $\mathbf{k\cdot{\hat{n}}}=\cos\theta_n$. 
In the limit $\xi\rightarrow0$ it can shown that,  
$\alpha_{\arrowvert_{\xi,\nu=0}}=\Pi_{T}$, 
$\beta_{\arrowvert_{\xi,\nu=0}}=\frac{\omega^2}{k^2}\Pi_{L}$, $\gamma_{\arrowvert_{\xi,\nu=0}}=0$, $\delta_{\arrowvert_{\xi,\nu=0}}=0$, 
where,
\ba
\Pi_{T}&=&m^{2}_{D}\frac{\omega^{2}}{2 k^{2}}\left[1+\frac{k^{2}-{\omega}^{2}}{2\omega k}\ln\frac{\omega+k}{\omega-k}\right],\nn
\Pi_{L}&=&m^{2}_{D}\left[\frac{\omega}{2 k}\ln\frac{\omega+k}{\omega-k}-1\right].
\label{isotropic:eq}
\ea
Functions $\Pi_T$ and $\Pi_{L}$ respectively represent
the transverse and longitudinal part of the
self-energy for isotropic($\xi=0$) collisionless($\nu = 0$)case.

\subsubsection{Structure functions in weak anisotropy limit}

In the context of heavy ion collisions the anisotropy
parameter is defined as,
$\xi=\frac{1}{2}\frac{\langle {P^2_{T}}\rangle}{\langle{P^2_{L}}\rangle}-1$. 
It is essential to note that the system we are studying is in near-equilibrium,
and therefore, we are only considering small values of $\xi$.

In the small anisotropy ($\xi<1$) limit all the structure functions can be calculated analytically.
The following expressions for the structure
functions can be obtained by expanding the Eq.(\ref{selfenergy}) upto linear order in $\xi$, 
\begin{widetext}
\ba
\alpha \left(\omega ,k,\xi ,\nu ,\theta _n\right)&=&\frac{m_D^2}{48 k}\Bigg(24 k z^2-2 k \xi  \left(9 z^4-13 z^2+4\right)+2 i \nu  z \left(\xi  \left(9 z^2-7\right)-12\right)-
2 \xi \cos{2\theta _n}\Big(k \big(15 z^4-19 z^2+4\big)\nn
&&+i \nu z\big(13-15 z^2\big)\Big)+\left(z^2-1\right)\big(3 \xi \left(k z \left(5 z^2-3\right)+i \nu  \left(1-5 z^2\right)\right)\cos{2\theta _n} + k z \left(-7 \xi +9 \xi  z^2-12\right)\nn
&&+i \nu  \left(\xi -9 \xi  z^2+12\right)\big)\ln\frac{z+1}{z-1}
\Bigg),
\ea
\ba
\beta\left(\omega ,k,\xi ,\nu ,\theta _n\right)&=&-\frac{(2 k (k z-i \nu )^2) \left(m_D^2\right)}{k^2 \left(\nu  \ln \frac{z+1}{z-1}+2 i k\right)} 
\Bigg(1-\frac{1}{2} z \ln\frac{z+1}{z-1}+\frac{1}{12}\xi\left(1+3 \cos{2\theta _n}\right)\left(2-6 z^2+\left(3 z^2-2\right)z\ln {\frac{z+1}{z-1}}\right)\Bigg),\nn\label{beta}
\ea
\ba
\gamma \left(k,\omega ,\xi ,\theta _n,\nu \right)=-\frac{m_D^2}{12 k}{\xi\left(k \left(z^2-1\right)-i \nu  z\right)\left(4-6 z^2+3 \left(z^2-1\right) z \ln\frac{z+1}{z-1}\right) \sin ^2{\theta _n}},
\label{eq:gamma}
\ea
\ba
\delta\left(\omega ,k,\xi ,\nu ,\theta _n\right)&=&\frac{\xi m_D^2 (k z-i \nu ) \cos{\theta _n}}{24 k^2 \left(2 k-i \nu  \ln\frac{z+1}{z-1}\right)}\Bigg(k \left(88 z-96 z^3\right)+8 i \nu  \left(6 z^2-1\right)+\ln\frac{z+1}{z-1}\nn
&&\times\left(12 k \left(4 z^4-5 z^2+1\right)-10 i \nu  z-3~ i \nu  \left(4 z^4-5 z^2+1\right) \ln\frac{z+1}{z-1}\right)\Bigg), 
\ea
\end{widetext}
where $z=\frac{\omega+i\nu}{k}$, and\be
\ln\frac{z+1}{z-1} =  \ln\frac{|z+1|}{|z-1|}+ i \bigg[arg\bigg(\frac{z+1}{z-1}\bigg) +2\pi N \bigg].
\ee
Here $N$- corresponds to the number of Riemannian sheets.
In the next section, we shall discuss the formalism to find the gluonic collective modes.

\section{Finding the poles of the propagator (collective modes)}
\label{PP}
We can also decompose $[\Delta^{-1}(K)]^{ij}$ appearing in Eq.(\ref{invprop}) as,
\ba
[\Delta^{-1}(K)]^{ij}&=&(k^{2}-{\omega}^{2}+\alpha){P}^{ij}_{T}+(-{\omega}^{2}+\beta){P}^{ij}_{L}\nn
&&+\gamma{P}^{ij}_{n}+\delta{P}^{ij}_{kn}.
\label{invpropexpan}
\ea
In order to find out the poles of the propagator $[\Delta(K)]^{ij}$, we first need to know the 
exact form of $[\Delta(K)]^{ij}$. To achieve that we shall first obtain the inverse of $[\Delta(K)]^{ij}$. 
We know that if a tensor is exist in a space spanned
by some basis vector (projection operators) then its inverse
should also exist in the same space, therefore we can also
expand $[\Delta(K)]^{ij}$ in the tensor projector basis as of
$[\Delta^{-1}(K)]^{ij}$ as follows,
\ba
[\Delta(K)]^{ij}=a{P}^{ij}_{L}+b{P}^{ij}_{T}+c{P}^{ij}_{n}+d{P}^{ij}_{kn}. \label{propagator}
\ea
Now, using the relation 
$[\Delta^{-1}(K)]^{ij} [\Delta(K)]^{jl}=\delta^{il}$ one can obtain the expression for the
coefficients $a$, $b$, $c$, $d$ which will yield the following result for the propagator 
\ba
[\Delta(K)]^{ij}&=&\Delta_A({P}^{ij}_{T}-{P}^{ij}_{n})+\Delta_G\big[(k^2-\omega^2+\alpha+\gamma){P}^{ij}_{L}\nn
&&+(\beta-\omega^2){P}^{ij}_{n}-\delta {P}^{ij}_{kn}\big]\,\text{,}
\label{propagator1}
\ea
with the poles given by
\ba
\Delta^{-1}_{A}(K)&=&k^2-\omega^2+\alpha=0,\label{mode_a}\\
\Delta^{-1}_{G}(K)&=&(k^2-\omega^2+\alpha + \gamma)(\beta-\omega^2)-k^2 \tilde n^2 \delta^2=0.\nn
\label{mode_g}
\ea
In the linear $\xi$ approximation we can neglect the term containing 
$\delta^2$ as it will be of order $\xi^2$, thus we will have 
\ba
\Delta_G^{-1}(K) = (k^2 - \omega^2 + \alpha + \gamma)(\beta-\omega^2)=0.
\ea
$\Delta_G^{-1}(K)$ can further be written as\\
\ba
\Delta_G^{-1}(K) = \Delta_{G1}^{-1}(k) ~\Delta_{G2}^{-1}(k)=0,
\ea
Thus, we have two more dispersion equations,
\ba
\Delta_{G1}^{-1}(K) &=& k^2 - \omega^2 + \alpha + \gamma=0,\label{mode_g1}\\
\Delta_{G2}^{-1}(K) &=& \beta-\omega^2=0.\label{mode_g2}
\ea
Note that here we have got three dispersion equations \ref{mode_a}, \ref{mode_g1} and \ref{mode_g2}. 
We call these as A-, G1- and G2-mode dispersion equations respectively. 
In the next section, we analyze the obtained dispersion equation and present our results.
In particular, we explore the instabilities in collisional
QGP in small anisotropy($\xi$) and cover the whole range of $\theta_n$ ({\it i.e.,} the angle between
the propagation vector($k$) and the direction of anisotropy($\hat{n}$)).

\section{Results and Discussions}
\label{RD}

We solve the dispersion equations (\ref{mode_a}), (\ref{mode_g1}) and
(\ref{mode_g2}) numerically and discuss the results for the
stable and unstable modes in the  subsections~\ref{SM} and~\ref{UM} respectively. 
To distinguish the effects of various EoSs (3-loop HTLpt and Lattice Bazabov {\it et. al},  2014) from ideal EoS (LO), 
we normalize the frequency $\omega$ and wave-number $k$ by $\omega^{LO}_{p}~(=m_D/\sqrt{3})$ {\it i.e.,} the leading 
order plasma frequency.

\subsection{Stable modes}
\label{SM}
 
The results for real part of the stable A-, G1-, and G2-modes are shown in Fig.\ref{fig:Stable_A_modes_Real}, 
 \ref{fig:Stable_G1_modes_Real} and \ref{fig:Stable_G2_modes_Real}, while their imaginary parts are shown in
Fig.\ref{fig:Stable_A_modes_Imaginary}, \ref{fig:Stable_G1_modes_Imaginary} and \ref{fig:Stable_G2_modes_Imaginary},
 respectively. These imaginary parts are coming only because of the collisional effects. If we consider the
 collision-less ($\nu=0$) plasma, these effects vanish. This has been already shown in the studies by different 
 groups~\cite{Romatschke:2003ms,Carrington:2014bla,Jamal:2017dqs}. We also did not get the imaginary part of the
 stable modes for the collision-less case and hence plotted only for non-zero $\nu$.
  
In Fig.\ref{fig:Stable_A_modes_Real} we have plotted the real A-modes 
at fixed anisotropy parameter, $\xi=0.2$ for the cases when $\theta_n$ is equal to $0,~\pi/3$ and $\pi/2$, respectively. 
For each case we have shown the variation of Re$({\omega_{A}})/\omega^{Lo}_{p}$ with respect to $k/\omega^{Lo}_{p}$ at 
different values of $\nu$ for all three 
EoSs. It can be noticed that for each case (mentioned above), the three curves which starts from the value of Re$({\omega_{A}})/\omega^{Lo}_{p}$ nearly 
equal to unity 
corresponds to the ideal EoS where with an increase of the collision frequency ($\nu = 0.0, 0.3\omega^{Lo}_{p}, 0.6\omega^{Lo}_{p}$),
the modes are marginally suppressed. When one consider the non-ideal EoS the similar pattern repeats but the
A-modes get more suppressed due to the non-ideal effect in comparison with the ideal or leading order results.
We note here that HTLpt and the lattice EoS results overlap with each other for a given $\nu$ and it is expected.

In a similar way, we have plotted the real parts of stable G1-and G2-modes in
Fig-\ref{fig:Stable_G1_modes_Real} and Fig.\ref{fig:Stable_G2_modes_Real}, respectively. 
One can notice by observing  Fig-\ref{fig:Stable_A_modes_Real}~and
\ref{fig:Stable_G1_modes_Real}, that they do not differ significantly.
This can be understood from corresponding dispersions equations 
(Eq.(\ref{mode_a}) and (\ref{mode_g1})). The difference is only
because of an additional contribution of the structure
constant ($\gamma$) that has negligible effect on the results even for different 
$\theta_n$. This is mainly because of the small dependence of $\xi$.
One can also notice from Eq.(\ref{eq:gamma}) that the
structure constant($\gamma$) vanishes at $\theta_n =0$. 
Hence both the modes overlap which can be clearly seen
in Fig. \ref{fig:Stable_G1_modes_Real}~and \ref{fig:Stable_G2_modes_Real} for $\theta_n=0$.
We have also got almost similar pattern for the stable 
G2-mode as shown in Fig.\ref{fig:Stable_G2_modes_Real}. 
However, in case of G2-mode, 
the behavior is slightly different after a certain value of $k/\omega^{Lo}_{p}$ as the dispersion 
curve becomes space like (Re$(\omega)<k$).

Fig.\ref{fig:Stable_A_modes_Imaginary} is plotted for imaginary A-mode
for the same values of the parameters $\nu$, $\xi$ and $\theta_n$ 
as discussed in  Fig.\ref{fig:Stable_A_modes_Real}. In this case, 
we did not get the negative imaginary modes for $\nu = 0.0$. Thus, we have    
plotted the dispersion curves only for $\nu = 0.3\omega^{Lo}_{p},
0.6\omega^{Lo}_{p}$.
For the imaginary part of stable A-mode, one can observe from 
the Fig.\ref{fig:Stable_A_modes_Imaginary} that as we increase 
the value 
of $\nu$, we get Im$[\omega]/\omega^{LO}_p$ to be more negative. The non-ideal effects (effect of EoSs) 
are causing the dispersion curve to be less negative as the $k/\omega^{LO}_p$ increases 
though the curves start from the same point. Here,  
it can also be noted that the observations at fixed $\xi=0.2$ and 
different $\theta_n$ ($0,~\pi/3$ and $\pi/2$) are 
quite similar. This is due to the fact that negative imaginary modes do not
depend on $\xi$ and $\theta$ and is a kind of check of our result that there 
is no imaginary mode at $\nu = 0.0$.

Similarly, we have plotted the imaginary parts
of stable G1- and G2-modes in
Fig-\ref{fig:Stable_G1_modes_Imaginary} and
\ref{fig:Stable_G2_modes_Imaginary}, respectively. For
G1-mode we can see that the dispersion curves follow the
same pattern as in the case of A-mode. Here
also one can notice that results for A- and
G1-modes does not seem to differ for the same reason as discussed for the stable G1 mode. 
For the G2-mode, unlike the case of A- and G1-modes one
can see that dispersion curves goes down as we increase
$k/\omega^{LO}_p$. This is because of the difference in
the behavior of the structure functions.    

\subsection{Unstable modes}
\label{UM}

Unstable modes are the positive imaginary solutions of 
$\omega$ in the dispersion equations (\ref{mode_a}),
(\ref{mode_g1}) and (\ref{mode_g2}).
If we substitute $\omega$ to be purely imaginary
{\it{i.e.}} $\omega=i\Gamma$,   
it can be easily seen from Eq.(\ref{beta}) that $\beta>0$. Thus for G2-mode the
dispersion equation (Eq.(\ref{mode_g2})) that transform to $\Gamma^2+\beta=0$, will never
be satisfied. This is the similar case as shown for collision-less ($\nu = 0$)
case in earlier studies~\cite{Romatschke:2003ms,Carrington:2014bla,Jamal:2017dqs}. 
Thus out of all three modes there can
be only two unstable modes (A and G1). Here we note that G1-mode was not reported in
Ref.\cite{Schenke:2006xu}. This was due to fact that in Ref.\cite{Schenke:2006xu} only the case $\theta_n=0$ was 
considered (In this situation  
the structure function $\gamma$ vanishes and A- and G1-modes gets merged). To study unstable
A- and G1-modes we have solved the corresponding dispersion
equations (Eqs.(\ref{mode_a}) and (\ref{mode_g1}))
numerically and shown our results in Fig.\ref{fig:Unstable_A_mode}
and \ref{fig:Unstable_G1_mode} for the case of weakly
squeezed plasma for non-zero collisional rate ($\nu \neq 0$) at
different angles($\theta_n$).

 In Fig.\ref{fig:Unstable_A_mode}, we have plotted  unstable A-mode 
 at fixed anisotropy, $\xi = 0.2$ but different angles 
 ($\theta_n = 0, \pi/6, \pi/3~ \text{and}~ \pi/2$).
 In order to see the effect of collisions we have taken the different collisional
 rate ($\nu$) as shown in the figure. 
 In a similar way, we have
 plotted unstable G1-mode in Fig.\ref{fig:Unstable_G1_mode}.
 In both the cases, we find out that with the increase in $\nu$,
 the unstable modes suppress. The results from both the EoSs are 
 overlapping but suppressing the instability quicker than
 LO case. For the reason discussed earlier, here also the results of A- and G1-modes are same
 at $\theta_n = 0$. As we increase the value of
 $\theta_n$ unstable G1-mode suppresses faster than
 A-mode. This can be seen from Fig.\ref{fig:Unstable_G1_mode} and 
 Fig.\ref{fig:Unstable_A_mode} (see the case $xi=0.2$ and $\theta_n=\pi/6$) 
 by comparing the value of 
 $\Gamma_{G1}/\omega^{LO}_{p}$ with that of $\Gamma_{A}/\omega^{LO}_{p}$ at 
 a particular $k/\omega^{LO}_{p}$.  
   The reason is pretty clear here that
  G1-mode have additional contribution of structure
  function $\gamma$ which tries to stabilize the modes as 
  we increase $\theta_n$ (because of $\sin{\theta_n}$ term). 
  However, it is important to note that both the unstable
  modes decreases as we increase the value of $\theta_n$. 
   
 As mentioned earlier the unstable modes critically depend on
four the parameters $k, \theta_n, \nu ~\text{and}~ \xi$.
Therefore, for a given set of any three parameters there
must exists a maximum value of fourth parameter (which
will be a function of the remaining three) at which
instability will completely suppress.  
 In the next subsections~\ref{KM},~~\ref{NM}, and ~\ref{TM}, we shall discuss the suppression
 of instability at the maximum value of the
 parameters $k, ~\nu ~\text{and} ~\theta_n $, respectively.

\subsubsection{Maximum values of the k at which instability completely suppresses}
\label{KM}

The maximum values of $k$ ($k_{max}$) at which instability
for modes (A and G1) completely suppressed can be obtained
by substituting $\omega=0$ in their dispersion equations.
In Fig. \ref{fig:kmax_vs_nu_A} we have shown the behavior of 
$k_{max}$ corresponding to A-mode with respect to $\nu$ scaled with
$\omega_{p}^{LO}$ at different
$\xi$ ($\xi = 0.1, 0.2 ~\text{and}~ 0.3$) 
for $\theta_n = 0, \pi/6 ~\text{and}~\pi/3$. 
In a similar
way the behavior of $k_{max}$ vs $\nu$ scaled with
$\omega_{p}^{LO}$ for G1-mode for $\xi = 0.1, 0.2 ~\text{and}~ 0.3$ at 
$\theta_n = 0 ~\text{and}~ \pi/6$ is shown in Fig.\ref{fig:kmax_vs_nu_G1}. 
In both the cases we have found that with the increase in $\nu$,
$k_{max}$ decreases. The same is the case when we increase $\theta_n$.
Note that unlike for A-mode, we have not shown the plot of $k_{max}$ vs $\nu$ scaled with
$\omega_{p}^{LO}$ at $\theta_n = \pi/3$. This is due to
fact that unstable G1-mode completely suppresses at
$\theta_n = \pi/3$ irrespective of the value of $k$. 
One can also note that with the increase in anisotropy,   
value of $k_{max}$ increases
and hence for higher anisotropy, instability can sustain
for larger $k$ values. At $\theta_n= 0$, $k_{max}$ for 
both the modes is same but at higher value of
$\theta_n$ ($\theta_n = \pi/6$), it suppresses more for
G1-mode. 
To cross check the above facts, we have plotted,
respectively, Fig. \ref{fig:kmax_vs_xi_A} and
\ref{fig:kmax_vs_xi_G1} for unstable A- and G1-mode
for $k_{max}$ with respect to $\xi$ at different values
of $\nu$ and got the similar results. 
Similarly, we have plotted Fig.\ref{fig:kmax_vs_th_A} and
\ref{fig:kmax_vs_th_G1} for $k_{max}$ with respect to 
$Cos[\theta_n]$ for A- and G1-mode, respectively. We can  
observe that as we move from $\theta_n = \pi/2$ to $\theta_n = 0$ or from $\theta_n = \pi/2$ to 
$\theta_n = \pi$, there is symmetry in values of
$k_{max}$. This shows that there is a symmetry in the system for the values of $k$, where the instability
completely suppress. Also at $\theta_n = 0 ~\text{and}~
\pi$, the values of $k_{max}$ for A- and G1- mode are overlapping. This
was also expected, as discussed earlier, at $\theta_n=0$, $\gamma$ is
zero which lead to the same dispersion equations of both the modes.
For A-mode, one can also notice that at $\nu =0$, $k_{max}$ is
going upto $\theta_n = \pm \pi/2$ while for
 G1-mode it is only going upto $\theta_n =\pm \pi/3$.
 Further more, in all of the above cases we found that
 the results of other EoSs 
follows the similar pattern as LO with slightly different
numbers.

\subsubsection{Maximum values of the $\nu$ at which instability completely suppresses}
\label{NM}

In the similar way as we did for $k_{max}$ the value of
$\nu_{max}$ can also be obtained at the point where the
instabilities completely suppress. In Fig.\ref{fig:nu_max_vs_k_A} we have shown the behavior of 
$\nu_{max}/\omega_{p}^{LO}$ for A-mode with respect to
 $k/\omega_{p}^{LO}$  for different
$\xi$ ($\xi = 0.1, 0.2 ~\text{and}~ 0.3$) 
at $\theta_n = 0, \pi/6 ~\text{and}~\pi/3$. 
In a similar way, we have plotted Fig.\ref{fig:nu_max_vs_k_G1} for
G1-mode at $\theta_n = 0 ~\text{and}~ \pi/6$. Here we have found that
with the increase in $k/\omega_{p}^{LO}$,
$\nu_{max}/\omega_{p}^{LO}$ decreases and also it gives smaller numbers
with the increase in $\theta_n$. But with the increase in anisotropy($\xi$),
 it increases. Again the results 
for A- and G1-mode are same for $\theta_n = 0$ but for
higher $\theta_n$ ($\theta_n = \pi/6$), $\nu_{max}$
corresponding to G1-mode decreases faster.
In Fig.\ref{fig:nu_max_vs_th_A} and
\ref{fig:nu_max_vs_th_G1},  we have shown the plots of 
$\nu_{max}$ with respect to $Cos[\theta_n]$ for the
parameters mentioned in the figures. Here also 
as we move from $\theta_n = \pi/2$ to $\theta_n = 0$ or from $\theta_n = \pi/2$ to 
$\theta_n = \pi$ there is symmetry in values of $\nu_{max}/\omega_{p}^{LO}$ and hence there is a symmetry in the system.
Again the EoSs are following the similar pattern with
slightly different numbers.
 
\subsubsection{Maximum values of the $Cos[\theta_n]$ or $\theta_n$ at which instability completely suppresses}
\label{TM}

We again follow the similar procedure to obtain maximum values of $Cos[\theta_n]$ 
as we did for $\nu_{max}$ and $k_{max}$. In Fig. \ref{fig:th_max_vs_xi_A}
and \ref{fig:th_max_vs_xi_G1}, we have plotted maximum possible value 
of $Cos[\theta_{n}]$ with respect to $\xi$ for A- and G1-modes, respectively
with the parameters mentioned in the plots.
In both the cases, we have found that as the anisotropy($\xi$) increases, 
the value of $Cos[\theta_{n}]_{max}$ shifted to the smaller values positive and negative values( or the higher 
values of $\theta_n$ maximum). Thus, we can say that higher the value of $\xi$, higher will be the value of
maximum $\theta_n$({\it i.e.,} towards $\theta_n = \pi/2$) or lower will be the value of $Cos[\theta_n]_{max}$({\it i.e.,} 
towards $Cos[\theta_n]=0$), where the instabilities reach or exist.

\section{Summary and Conclusions}
\label{SC}
We have obtained the analytical results for the gluon self-energy in terms 
of structure functions in the presence of BGK collisional kernel for a 
hot anisotropic QCD medium in the small anisotropy limit. 
The dispersion equations for collective modes have been obtained regarding the structure functions of the gluon self-energy. 
We have studied the stable modes, which are found to have less affected by the anisotropy.
We have also investigated the unstable modes and found that the presence of anisotropy and the collision frequency
profoundly affects the instabilities of the system. We have incorporated the QCD medium interaction by exploiting the quasi-particle
description of the hot QCD equations of state concerning quark and gluon effective fugacities in their distribution functions. 
The EoSs include 3-loop HTLpt and a very recent Lattice  EoS along with ideal EoS.

It turns out that the results obtained 
for collective modes are very close (irrespective of any change in anisotropy parameter and collision frequency) for first two 
EoSs and differ significantly in numbers with the case of ideal EoS. This suggests us that the interactions affect the modes
significantly (temperature dependence). Hence the instabilities in QGP is found to have a high impact of collisional frequency and anisotropy
and also have directionality dependence.

To get a more closer picture to the experiments, one can also observe the behavior of 
collective mades while subsuming the non-local BGK kernel. This will be taken up in near future.

\section{Acknowledgements}
 VC would like to sincerely acknowledge DST, Govt. of India for Inspire Faculty Award -IFA13/PH-15
 and Early Career Research Award(ECRA/2016) Grant. We  would like to acknowledge Sukanya Mitra for
 providing numerical  help in effective description of hot QCD equations of state employed in the present work. 
  We further acknowledge Amit Reza, Soumen Roy, Chkresh Singh and 
 Manu George for their help in developing numerical understanding of the work. A. Kumar acknowledges 
 the hospitality of IIT Gandhinagar. We would like to acknowledge people of INDIA for 
 their generous support for the research in fundamental sciences in the country.
 
{}

\begin{thebibliography}{99}

\bibitem{expt_rhic}
J. Adams {\it et al.}  (STAR Collaboration), Nucl.  Phys.  A {\bf 757}, 102 (2005);
K. Adcox {\it et al.} PHENIX Collaboration, Nucl. Phys.  A {\bf 757}, 184 (2005);
B.B. Back {\it et al.} PHOBOS Collaboration, Nucl. Phys. A  {\bf 757}, 28 (2005);
I. Arsene {\it et al.} BRAHMS Collaboration, Nucl. Phys. A {\bf 757}, 1 (2005).

\bibitem{expt_lhc}
K. Aamodt {\it et al.} (The Alice Collaboration), Phys. Rev. Lett. {\bf 105}, 252302 (2010);
Phys. Rev.  Lett. {\bf 105}, 252301 (2010); Phys. Rev. Lett. {\bf 106}, 032301 (2011).

\bibitem{Ryu}
S. Ryu, J. F. Paquet, C. Shen, G. S. Denicol, B. Schenke, S. Jeon, C. Gale, Phys. Rev. Lett. {\bf 115}, 132301 (2015).

\bibitem{Denicol1}
G. Denicol, A. Monnai, B. Schenke, Phys. Rev. Lett. {\bf 116}, 212301 (2016).

\bibitem{Chu:1988wh}
M. C. Chu and T. Matsui, Phys. Rev. D {\bf 39}, 1892 (1989).
 
\bibitem{Landau:1984}
{L.D.Landau and E.M.Lifshitz}, {\it Electrodynamics of continuous media}, {Butterworth-Heinemann}, (1984).

\bibitem{Koike:1991mf}
Y.~Koike, AIP Conf.\ Proc.\ {\bf 243}, 916 (1992) .

\bibitem{Mrowczynski:1993qm}
S.~Mrowczynski, Phys.\ Lett.\ B {\bf 314}, 118 (1993).

\bibitem{Mrowczynski:1994xv}
S.~Mrowczynski, Phys.\ Rev.\ C {\bf 49}, 2191 (1994).

\bibitem{Mrowczynski:1996vh}
S.~Mrowczynski, Phys.\ Lett.\ B {\bf 393}, 26 (1997).

\bibitem{Jamal:2017dqs} 
M.~Y.~Jamal, S.~Mitra and V.~Chandra, Phys.\ Rev.\ D {\bf 95}, 094022 (2017)

\bibitem{avdhesh}
Avdhesh Kumar, Jitesh. R. Bhatt, Predhiman. K. Kaw, Phys. Letts.  {\bf B 757}, 317-323 (2016).

\bibitem{dm_rev1}
Daniel F. Litim, C. Manual, Phys. Rep. {\bf 364}, 451 (2002).

\bibitem{dm_rev2}
J.-P.Blaizot and E.Iancu, Phys.\ Rep.\ {\bf 359}, 355 (2002).

\bibitem{Mrowczynski:2000ed}
S.~Mrowczynski and M.~H.~Thoma, Phys.\ Rev.\ D {\bf 62}, 036011 (2000) .

\bibitem{Mrowczynski:2004kv}
S. Mrowczynski, A. Rebhan and M.~Strickland, Phys.\ Rev.\ D {\bf 70}, 025004 (2004).


\bibitem{Akamatsu:2013pjd}
Yukinao Akamatsu, Naoki Yamamoto, Phys. Rev. Lett.{\bf{111}}, 052002 (2013). 

\bibitem{Bhatnagar:1954} 
P. L. Bhatnagar, E. P. Gross, and M. Krook, Phys. Rev. {\bf 94}, 511 (1954).

\bibitem{Jiang:2016dkf}
Bing-feng Jiang, De-fu Hou and Jia-rong Li, Phys. Rev. D {\bf 94}, 074026 (2016).

\bibitem{Schenke:2006xu}
Bjoern Schenke, Michael Strickland, Carsten Greiner, Markus H. Thoma, Phys. Rev. D {\bf 73}, 125004 (2006).

\bibitem{Weibel:1959zz}
E. S. Weibel, Phys.\ Rev.\ Lett.\  {\bf 2}, 83 (1959).

\bibitem{Arnold:2003rq}
P. B. Arnold, J. Lenaghan and G. D. Moore, JHEP {\bf 0308}, 002 (2003).

\bibitem{Mrowczynski:2005ki}
S. Mrowczynski, Acta Phys.\ Polon.\ B {\bf 37}, 427 (2006).

\bibitem{Romatschke:2003ms}
P. Romatschke and M.Strickland, Phys. Rev. D {\bf 68}, 036004 (2003).

\bibitem{Romatschke:2004jh}
P. Romatschke and M.Strickland Phys. Rev. D {\bf 70}, 116006 (2004).

\bibitem{Schenke:2006yp}
B. Schenke and M. Strickland, Phys.\ Rev.\ D {\bf 76}, 025023 (2007).

\bibitem{Dumitru:2007hy}
A.~Dumitru, Y.~Guo and M.~Strickland, Phys.\ Lett.\ B {\bf 662}, 37 (2008).

\bibitem{Martinez:2008di}
M.~Martinez and M.~Strickland, Phys.\ Rev.\ C {\bf 78}, 034917 (2008).

\bibitem{Attems:2012js}
M.~Attems, A.~Rebhan and M.~Strickland, Phys.\ Rev.\ D {\bf 87}, no.2, 025010 (2013).

\bibitem{Florkowski:2012as}
W.~Florkowski, R.~Maj, R.~Ryblewski and M.~Strickland, Phys.\ Rev.\ C {\bf 87}, no.3, 034914 (2013).


\bibitem{Carrington:2004}
M. Carrington, T. Fugleberg, D. Pickering, and M. Thoma, Can. J. Phys. {\bf 82}, 671 (2004).

\bibitem{chandra_quasi1}
V. Chandra, R. Kumar, V. Ravishankar, Phys. Rev.  C {\bf 76}, 054909 (2007);
[Erratum: Phys. Rev. C {\bf 76}, 069904 (2007)];
V. Chandra, A. Ranjan, V. Ravishankar, Eur. Phys. J. A {\bf 40}, 109-117 (2009).
 
\bibitem{chandra_quasi2}
V. Chandra, V. Ravishankar, Phys. Rev.  D {\bf 84}, 074013 (2011). 

\bibitem{effmass1}
A. Peshier {\it et. al}, Phys. Lett. {\bf B 337}, 235 (1994); Phys.Rev. D {\bf 54}, 2399 (1996).

\bibitem{effmass2}
A. Peshier, B. K\"{a}mpfer, G. Soff, Phys. Rev. C {\bf 61},045203 (2000);
Phys. Rev.  D {\bf 66}, 094003 (2002); V. M. Bannur, Phys. Rev. C {\bf 75}, 044905 (2007);
{\it ibid}. C {\bf 78}, 045206 (2008), JHEP {\bf 0709}, 046 (2007);
A. Rebhan, P. Romatschke, Phys. Rev. D {\bf 68}, 0250022 (2003); 
M. A. Thaler, R. A. Scheider, W. Weise, Phys. Rev. C {\bf 69}, 035210 (2004);
K. K. Szabo, A. I. Toth, JHEP  {\bf 0306}, 008 (2003). 

\bibitem{polya}
M. D\'Elia, A. Di Giacomo, E. Meggiolaro, Phys. Lett. B {\bf 408}, 315 (1997);
Phys. Rev. D {\bf  67}, 114504 (2003); P. Castorina, M. Mannarelli, Phys. Rev. C {\bf 75}, 054901 (2007);
Phys.  Lett. B  {\bf 664}, 336 (2007); Paolo Alba {\it et al.}, Nucl. Phys. A {\bf 934}, 41-51 (2014).  

\bibitem{pnjl}
A. Dumitru, R. D. Pisarski, Phys. Lett. B  {\bf 525}, 95 (2002);
K. Fukushima, Phys.  Lett. B  {\bf 591}, 277 (2004);
S. K. Ghosh {\it et. al}, Phys.  Rev. D {\bf 73}, 114007 (2006);
H. Abuki, K. Fukushima, Phys. Lett. B  {\bf 676}, 57 (2006);
H. M. Tsai, B. M\"{u}ller,  J. Phys.  G  {\bf 36}, 075101 (2009).

\bibitem{flor}  W.  Florkowski, R. Ryblewski, Nan Su, K. Tywoniuk, 
Phys. Rev.  C {\bf 94} no.4, 044904 (2016) ; Acta  Phys. Polon. B {\bf 47}, 1833 (2016).

\bibitem{PJI}
P. Chakraborty, J. I. Kapusta , Phys.  Rev. C {\bf 83}, 014906 (2011).
 
\bibitem{Mkap}
M. Albright and J. I. Kapusta,  Phys. Rev. C {\bf 93}, 014903 (2016).

\bibitem{Mitra:2016zdw}
S.~Mitra and V.~Chandra, Phys.\ Rev.\ D {\bf 94}, no.3, 034025 (2016).


\bibitem{chandra_etazeta}
V. Chandra, Phys. Rev. D {\bf 86} 114008 (2012); {\it ibid.}, D {\bf 84}, 094025  (2011).

\bibitem{chandra_eta}
V. Chandra, V. Ravishankar, Eur. Phys. J. C  {\bf 64}, 63-72  (2009); {\it ibid.} C {\bf 59}, 705-714  (2009).

\bibitem{Bluhm}
M.~Bluhm, B.~Kampfer and K.~Redlich,Phys.\ Rev.\ C {\bf 84}, 025201 (2011).

\bibitem{Greco}
A. Puglisi, S. Plumari, V. Greco, Phys. Lett. B {\bf 751}, 326-330 (2015).


\bibitem{Das:2012ck}
S.~K.~Das, V.~Chandra and J.~e.~Alam, J.\ Phys.\ G {\bf 41}, 015102 (2013).  

\bibitem{Chandra:2015gma}
V.~Chandra and S.~K.~Das, Phys.\ Rev.\ D {\bf 93}, no.9, 094036 (2016).

\bibitem{Chandra:2010xg}
V.~Chandra and V.~Ravishankar, Nucl.\ Phys.\ A {\bf 848}, 330 (2010).
  
\bibitem{Agotiya:2016bqr}
V.~K.~Agotiya, V.~Chandra, M.~Y.~Jamal and I.~Nilima, Phys.\ Rev.\ D {\bf 94}, no.9, 094006 (2016)

\bibitem{Chandra:2015rdz}
V.~Chandra and V.~Sreekanth, Phys.\ Rev.\ D {\bf 92}, no.9, 094027 (2015).

\bibitem{Chandra:2016dwy}
V.~Chandra and V.~Sreekanth, arXiv:1602.07142 [nucl-th].


\bibitem{cheng}
M. Cheng {\it et. al}, Phys. Rev. D {\bf 77}, 014511 (2008). 

\bibitem{nhaque}
N. Haque, A.  Bandyopadhyay, J. O. Andersen, Munshi G. Mustafa, M. Strickland and Nan Su, 
JHEP {\bf 1405}, 027  (2014).


\bibitem{Andersen:2015eoa}
J. O. Andersen, N. Haque, M. G. Mustafa and M. Strickland,
Phys. Rev. D {\bf 93}, no.5,  054045  (2016)

\bibitem{bazabov2014} A. Bazabov {\it et, al.},  Phys.  Rev  D {\bf 90}, 094503 (2014).

\bibitem{fodor2014}
S.  Borsanyi, Z. Fodor, C.  Hoelbling, S. D. Katz, S. Krieg, Kalman K. Szabo, 
Phys. Lett.  B {\bf 370}, 99-104,  (2014).


\bibitem{dmass1}
P. F. Kelly  {\it et. al}, Phys. Rev. Lett. {\bf 72}, 3461 (1994); Phys. Rev. D {\bf 50}, 4209 (1995).

\bibitem{qcd_coupling}
M. Laine and Y. Sch\"{o}der, JHEP {\bf 0503}, 067 (2005).

\bibitem{Carrington:2014bla}
M.~E.~Carrington, K.~Deja and S.~Mrowczynski, Phys.\ Rev.\ C {\bf 90},  no.3, 034913 (2014).

\bibitem{Kobes:1990dc}
R.~Kobes, G.~Kunstatter and A.~Rebhan, Nucl.\ Phys.\ B {\bf 355}, 1 (1991).

\end{thebibliography}
\end{document}